\documentclass[reprint,amsmath,amssymb,aps]{revtex4-2}
\usepackage{physics}
\usepackage{graphicx}
\usepackage{dcolumn}
\usepackage{bm}
\usepackage[table]{xcolor}
\usepackage{amsthm}
\usepackage{dsfont}

\usepackage{import}
\usepackage{physics}

\usepackage[utf8]{inputenc}
\usepackage{lineno,hyperref}
\usepackage{graphicx}
\usepackage{comment}

\begin{document}

\title{
Generalized Kerr-Cat Qubit Codes}

\author{Alonso Viladomat$^1$}

\author{Shahram Dehdashti$^1$}

\author{Amin Kargarian$^2$ }

\author{Janis N\"{o}tzel$^1$}

\author{Peter  van Loock$^{3}$}

\affiliation{$^1$Emmy-Noether Gruppe Theoretisches Quantensystemdesign Lehrstuhl F\"{u}r Theoretische Informationstechnik Technische Universit\"{a}t M\"{u}nchen, Germany}
\affiliation{$^2$Electrical and Computer Engineering Department, Louisiana State University, Baton Rouge, 70803, USA.}
\affiliation{$^3$Johannes-Gutenberg University of Mainz, Institute of Physics, Staudingerweg 7, 55128 Mainz, Germany}

 \thanks{A. V, and S. D  contributed equally and share co-first authorship.}

\begin{abstract}
We present a systematic study of Schr\"{o}dinger cat codes constructed
from Kerr-type coherent states, including displaced Kerr coherent states
and Barut--Girardello Kerr coherent states, each admitting two distinct
families determined by the sign of the Kerr nonlinearity. By tuning the
Kerr parameter and coherent-state amplitude, these states interpolate
between $\mathfrak{su}(2)$, $\mathfrak{su}(1,1)$ coherent states, providing a
unified and versatile foundation for this type of bosonic quantum error correction.
Unlike standard two-component Schr\"{o}dinger cat codes, where a single photon-loss
event induces an uncorrectable bit-flip, the nonlinear phase-space
structure of Kerr cat states enables simultaneous detection and correction
of both photon-loss and dephasing errors within a unified recovery
framework, with optimal recovery operations determined via convex
optimization. We demonstrate that Kerr cat encodings significantly
outperform conventional cat codes under
combined loss and dephasing noise, and that judicious parameter
optimization can suppress both error channels to a level that reduces the
overhead of additional error correction layers. We further show that
Kerr-deformed coherent-state manifolds under engineered two-photon driving
emerge as effective steady states of driven-dissipative dynamics, with
single-photon decoherence strongly suppressed and leakage outside the
protected manifold appearing only as higher-order corrections in the
deformation strength.
Our extended formalism identifies generalized Kerr Schr\"{o}dinger cat codes as
promising candidates for fault-tolerant bosonic quantum computation in 
experimental platforms such as nonlinear photonics.

\end{abstract}

\maketitle

\section{Introduction}
Quantum computing is a promising tool for solving complex computational
problems far beyond the reach of classical computers. This advantage over
classical computation emerges through the use of superposition properties
of quantum states as part of its computational
process~\cite{Preskill2018quantumcomputingin,arute2019quantum}. The gain
of quantum computing translates not only into speed
efficiency~\cite{madsen2022quantum}, but also into energy
efficiency~\cite{fellous2023optimizing}. However, achieving reliable
quantum computation has remained a formidable challenge due to the
detrimental effects of decoherence and environmental
noise~\cite{Weber2024Construction,Georgopoulos2021Modeling,Hann2021Resilience}.
Quantum noise limits computational accuracy and scalability by impacting
both the longevity of quantum states and the accuracy of gates applied
during computation~\cite{Weber2024Construction}. It is therefore necessary
to employ effective quantum error correction (QEC) techniques in order to
achieve fault-tolerant quantum computers~\cite{548464,PhysRevX.8.031027}.

QEC strategies involve encoding logical states in multiple quantum physical
systems~\cite{shor1995scheme,PhysRevLett.77.793,PhysRevA.54.1862}.
This technique exploits quantum entanglement between many physical systems
in a way that defines a code subspace contained in the total Hilbert space
of the composite system~\cite{gottesman2001encoding,fujii2015quantum}.
Logical states are then prepared to store and process quantum information
in this code subspace~\cite{nielsen2000quantum}. Due to highly entangled
correlations, greater fault tolerance, robustness, and longevity against
decoherent errors can be achieved compared to storing information directly
in a single physical system. Since code spaces are entirely abstract, QEC
strategies have the freedom to be applicable to any kind of physical
system, though the choice of physical system may induce noise adherent to
its own nature, which can sometimes degrade rather than enhance fault
tolerance~\cite{Preskill2018quantumcomputingin,kitaev2003fault,raussendorf2007fault,terhal2015quantum}.
A commonly present type of error in qubit spin systems is a local bit-flip
or phase-flip error, which can be corrected with the quantum repetition
code by embedding a logical qubit into many physical
qubits~\cite{nielsen2000quantum,knill1998resilient,lidar2013quantum}.

Extensive theoretical research on QEC has been carried out for many-body
qubit encodings~\cite{devitt2013quantum,fowler2012surface,Hastings2021dynamically,Andreasson2019quantumerror}.
Shor's code, for example, is capable of correcting an arbitrary
single-qubit Pauli error in the
system~\cite{shor1995scheme,gottesman2010introduction}. Topological
surface codes offer the possibility of locally correcting noise errors
within a physical subsystem neighborhood without substantially altering the
logical embedded state in an irreversible
way~\cite{fowler2012surface,bravyi1998quantum,preskill1998fault}. Most
many-body qubit QEC codes have been demonstrated experimentally with
trapped ions or spins~\cite{schindler2011experimental,reed2012realization},
yet achieving such codes demands highly entangled many-body states and
sophisticated cooling methods, creating a bottleneck that prevents en masse
implementation of reliable quantum computing chips.

Bosonic codes propose an alternative candidate for QEC
strategies~\cite{leghtas2013hardware}. One of the greatest advantages of
bosonic QEC encodings is that they encode a qubit by employing a
superposition over all Fock basis elements, inherently providing a
scalable method for many-body
systems~\cite{matsuura2024continuous,PRXQuantum.5.020355,PhysRevA.108.022428,ofek2016extending}.
These codes also provide mechanisms for storing and processing quantum
information while being robust to certain types of
noise~\cite{cai2021bosonic,brady2023advances,noh2019fault}. In particular,
bosonic codes based on photonic systems are promising for overcoming the
challenge of cooling the quantum system, as they can operate at higher
temperatures~\cite{noh2020encoding,knill2001scheme,krastanov2021room},
making them attractive for scalable quantum technologies. Experimental
advances have demonstrated the feasibility of creating and manipulating
such bosonic encodings in laboratory
settings~\cite{sychev2018generating,motes2017encoding}. Among the
different types of bosonic codes, cat states and
Gottesman-Kitaev-Preskill (GKP) codes have emerged as leading
candidates~\cite{ofek2016extending,PRXQuantum.6.010330}.

The GKP code~\cite{gottesman2001encoding} encodes a logical qubit in the
phase space of the harmonic oscillator by forming a periodic lattice in
phase space. Small displacement errors can be detected and corrected as
long as the displacement is smaller than the lattice period, enabling
detection and correction of photon gain or loss induced by ladder operator
noise~\cite{PhysRevA.108.052413,albert2018performance,michael2016new}. GKP
codes have been implemented in photonic
experiments \cite{konno2024logical,aghaee2025scaling}, trapped
ions~\cite{fluhmann2019encoding}, and superconducting
circuits~\cite{campagne2020quantum}. However, GKP codes are not robust
against dephasing noise, which shifts code states in phase space by an
arbitrarily large amount, translating into a logical
error~\cite{PRXQuantum.2.020101}.

Another prominent class of bosonic QEC codes involves Schrödinger cat states (SCSs) \cite{cochrane1999macroscopically,mirrahimi2014dynamically,bergmann2016quantum,PhysRevA.72.022320,yang2025minute,PRXQuantum.6.010321,li2017cat}. In this approach, logical qubits are encoded into superpositions of coherent states and their phase-inverted counterparts. In the simplest, two-component construction \cite{cochrane1999macroscopically}, the logical zero is restricted to the even photon-number subspace of the Fock basis, while the logical one resides in the odd subspace. This two-component encoding is intrinsically vulnerable to photon loss: a single photon-loss event maps the logical zero to the logical one, effectively inducing a bit-flip error in the encoded qubit. To overcome this limitation, the four-component cat code was introduced \cite{mirrahimi2014dynamically}, which encodes logical states as superpositions of four coherent states equally spaced in phase and is capable of correcting a single photon-loss error. This construction has since been extended to multi-component cat codes involving arbitrarily many coherent-state constituents \cite{bergmann2016quantum}, which can in principle correct higher orders of photon loss without requiring a Kerr nonlinearity or other special resources. 
These considerations highlight a general principle: engineering the geometry of bosonic codewords in phase space, through the choice of encoding, nonlinearity, or number of coherent-state components, is a powerful strategy for improving robustness against photon-loss errors \cite{hastrup2022all,PhysRevX.15.011070,schlegel2022quantum,xu2023autonomous,PhysRevA.110.012460}.

In this context, Kerr nonlinearity~\cite{miranowicz2013two}, or more
generally anharmonicity in quantum
systems~\cite{garcia2020quantum}, provides a natural and physically
well-motivated route toward generalized bosonic encodings. Although
Kerr-type nonlinearities are often suppressed or treated perturbatively in
quantum device engineering, their controlled incorporation can
fundamentally modify the structure of oscillator states. More broadly,
nonlinear coherent states provide a natural generalization of standard
coherent-state constructions. Kerr coherent states belong to this broader
class and are defined via deformed annihilation and creation operators of
the form
\begin{equation*}
  \hat{A} = f(\hat{n})\,\hat{a}, \qquad
  \hat{A}^{\dagger} = \hat{a}^{\dagger}\,f^{\dagger}(\hat{n}),
\end{equation*}
which replace the standard bosonic
operators~\cite{Dodonov2003,dehdashti2015decoherence,dehdashti2013decoherence,dehdashti2025quantum}.
The nonlinear function $f(\hat{n})$ encodes the algebraic deformation and
determines the properties of the resulting coherent states. Depending on the choice of deformation, these states can interpolate between different algebraic structures such as $\mathfrak{su}(2)$ and  $\mathfrak{su}(1, 1)$, providing a unified framework for generating a wide class of nonclassical states.
Introducing Kerr-type anharmonicity gives rise to two distinct families of generalized coherent states: the displaced Kerr coherent states \cite{dehdashti2025quantum} (DKCSs) and the Barut-Girardello Kerr coherent states (BGKCSs).
These states extend canonical coherent states by explicitly incorporating nonlinear effects and exhibit close connections to the algebraic structures mentioned above.
Importantly, such nonlinear coherent states are not purely formal constructs; they can, in principle, be realized in a range of experimental platforms, including trapped-ion systems and cavity-QED architectures, where engineered light--matter interactions enable controlled access to anharmonic dynamics.
Photonic implementations based on engineered waveguide lattices further provide classical analogues that capture essential features of the underlying displacement mechanisms, offering an intuitive framework for visualizing these states.

In recent years, engineered dissipation and nonlinear driven quantum
systems have emerged as powerful approaches for the preparation and
stabilization of nonclassical states of
light~\cite{puri2017engineering,hajr2024high,grimm2020stabilization,xu2022engineering,qing2026quantum}.
In conventional driven-dissipative cat-state platforms, however,
single-photon loss typically induces decoherence that gradually destroys
quantum coherence between the coherent-state components, ultimately driving
the system toward an incoherent statistical mixture. Consequently,
suppressing decoherence while maintaining the nonclassical structure of the
cat manifold remains one of the central challenges in dissipative
quantum-state engineering.

Motivated by these developments, we investigate Schr\"{o}dinger cat
encodings constructed from Kerr nonlinear coherent states. In particular,
we consider Kerr coherent states, displaced Kerr coherent states, and
Barut-Girardello Kerr coherent states, each of which admits two distinct
families depending on the sign of the Kerr parameter.
From these building blocks, we construct \emph{four} distinct Kerr
cat-state encodings and analyze their performance under dominant bosonic
noise channels.

We model decoherence using a Markovian open quantum system framework
described by the Lindblad master equation in terms of the standard bosonic
annihilation and creation operators $\hat{a}$ and $\hat{a}^{\dagger}$. In
addition, we comment on an effective nonlinear extension of the
environmental dynamics incorporating Kerr-type contributions. While such
nonlinear modifications alter the microscopic structure of the dynamics,
we find that their qualitative impact on error processes remains comparable
to the standard dissipative model, indicating the robustness of the
conclusions across different physical regimes.

We also investigate Kerr-deformed coherent-state (KDCS) manifolds
generated through nonlinear algebraic structures and analyze their
associated dissipative dynamics under engineered two-photon driving. By
explicitly analyzing the action of nonlinear jump operators on the KDCS
manifold, we show that, in the weak-deformation regime, the dominant
dissipative evolution remains effectively confined to the degenerate
subspace spanned by these states. We further demonstrate that, in contrast
to conventional cat-state stabilization schemes, the interplay between
Kerr deformation and appropriately tuned two-photon driving can strongly
suppress decoherence induced by single-photon loss. In this regime, the Kerr cat states (KCSs)
emerge as effective steady states of the driven-dissipative dynamics,
while leakage outside the protected manifold appears only as higher-order
corrections in the deformation strength.
This driven-dissipative stabilization realizes an autonomous form of QEC, in which error correction is built directly into the system dynamics rather than relying on explicit syndrome measurements and classical feedback.

The Kerr-deformed structure introduces an additional tunable parameter that
controls the phase-space geometry of the codewords. This tunability enables
improved separation of logical components and enhanced distinguishability
of error-induced states. In contrast to standard SCS codes, Kerr-based cat
states inherit a nonlinear structure that enables the engineering of a
noise-biased logical subspace. In the case of Kerr cat states, whether
DKCSs or BGKCSs, standard particle loss and dephasing processes do not
transform the logically encoded states into either orthogonal states or the
states themselves, respectively. This behavior differs qualitatively from
conventional two-component cat codes, where photon loss primarily induces
phase flips without intrinsic correction capability. As a result, both
particle-loss and dephasing errors can, in principle, be detected via
appropriate error-syndrome measurements and corrected using suitable
recovery operations within a unified recovery framework, where the Kerr
parameter plays a central role in optimizing the trade-off between
different noise channels. In particular, the Kerr-induced structure allows
for a partial mitigation of photon-loss effects at the level of the code
space itself, rather than relying solely on external redundancy.

By carefully selecting the amplitude of the Kerr coherent states, the Kerr
parameter, and the ratio of the optical frequency to the Kerr parameter,
both dephasing and particle-loss errors can be simultaneously and
effectively suppressed to a negligible level.
This identifies a parameter regime in which Kerr-based cat codes go beyond mere error detection and approach intrinsic error suppression, resembling the behavior of loss-correcting bosonic codes, but achieved here through Kerr nonlinearity alone.
The effectiveness of this error mitigation strategy depends critically on the amplitude of the Kerr coherent states and the Kerr parameter.
Although Kerr cat states are intrinsically non-Gaussian, this
non-Gaussianity \emph{alone} does not guarantee fault tolerance. Instead,
it serves as a resource that enables controllable phase-space deformation
and interference structure, which can be exploited to enhance error
suppression and recovery performance at the logical-qubit level.

We quantify performance using both the Knill-Laflamme (KL) conditions and
quantum channel fidelity under Lindblad dynamics.
We demonstrate that Kerr cat encodings can significantly outperform conventional cat codes in experimentally relevant parameter regimes, particularly in the presence of combined loss and dephasing noise.
Moreover, the Kerr parameter provides a flexible control knob for
optimizing code performance without introducing additional nonlinear
control operations. These results establish a direct connection between
nonlinear algebraic deformations, engineered dissipation, and the
stabilization of protected bosonic quantum states, suggesting that
Kerr-deformed coherent-state manifolds provide a flexible and robust
platform for dissipative quantum-state engineering and for implementing
noise-resilient bosonic encodings in nonlinear photonic and
superconducting architectures.

The structure of this paper is as follows.
In Sec.~\ref{sec:Kerr_Cat_States}, we briefly introduce the displaced Kerr
coherent states and Barut-Girardello Kerr coherent states and construct
their associated cat states.
In Sec.~\ref{sec:Dynamics_of_KCSs}, we analyze the fidelity of the DKCSs,
BGKCSs, and standard Schr\"{o}dinger cat states under both dephasing and
particle-loss noise channels.
Section~\ref{sec:Errors_and_their_corrections} focuses on optimizing
quantum error correction by numerically determining the optimal recovery
operations using convex optimization methods.
In Sec.~\ref{engineering}, we investigate Kerr-deformed coherent-state
manifolds generated through nonlinear algebraic structures and their
associated dissipative dynamics under engineered two-photon driving, and
demonstrate the stabilization of displaced Kerr coherent states.
Finally, Sec.~\ref{conclusion} provides concluding remarks and outlines
future directions.

\section{Kerr Cat States}\label{sec:Kerr_Cat_States}
The Kerr Hamiltonian is expressed as
\begin{eqnarray}
\hat{H} = \omega \hat{n} + \frac{\lambda}{2} \hat{n}^{2},
\end{eqnarray}
where $\hat{n} = \hat{a}^{\dagger}\hat{a}$ denotes the number operator, and $\lambda$ is referred to as the Kerr parameter \cite{scully1997quantum}. This Hamiltonian is widely encountered in quantum optics and condensed matter physics, modeling systems such as Kerr cavities \cite{miranowicz2013two} and transmon superconducting qubits \cite{garcia2020quantum}. It introduces anharmonicity in the energy spectrum, resulting in unequal spacing between adjacent energy levels.
In Kerr cavities, the Kerr nonlinearity characterizes the strength of photon-photon interactions and is proportional to the real part of the third-order nonlinear susceptibility, $\Re(\chi^{(3)})$. In this context, the Kerr parameter $\lambda$ is typically positive. In contrast, for superconducting circuits such as transmons, the parameter is commonly referred to as the anharmonicity and can be tuned to take either positive or negative values.

In the remainder of this work, we distinguish between positive and negative values of the Kerr parameter and rewrite the Hamiltonian in the form
\begin{eqnarray}
\hat{H} = \omega \hat{n} + \frac{\lambda}{2} \hat{n}^{2}
= \hat{A}_{\pm}^{\dagger}\hat{A}_{\pm},
\end{eqnarray}
where $\hat{n} = \hat{a}^{\dagger}\hat{a}$ denotes the number operator and $\lambda \in \mathbb{R}$ is the Kerr nonlinearity parameter.

To account for both signs of $\lambda$, we define the Kerr annihilation operators as
\begin{eqnarray}
\hat{A}_{\pm}
=
\sqrt{\frac{|\lambda|}{2}}\,
\hat{a}\,
\sqrt{2j \mp 1 \pm \hat{n}},
\label{anh_pos}
\end{eqnarray}
where $\hat{a}$ and $\hat{a}^{\dagger}$ are the bosonic annihilation and creation operators, respectively, and $ j := \frac{\omega}{\lambda} \pm \frac{1}{2}$. The corresponding Kerr creation operators $\hat{A}_{\pm}^{\dagger}$ are given by the Hermitian conjugates of $\hat{A}_{\pm}$ \cite{dehdashti2025quantum}.

The following briefly introduces the displaced Kerr cat states and Barut-Girardello Kerr cat states.  
\subsection{ Displaced Kerr cat  states}
A displaced Kerr coherent state  is defined as \cite{ dehdashti2015realization, dehdashti2013coherent}:
\begin{eqnarray}\label{eq:DKCS_from_displacement_op}
\ket{\alpha;j,\lambda_{\pm}}=D(\alpha; j, \lambda_{\pm})\ket{0},\ \alpha=re^{i\theta} \in \mathbb{C}
\end{eqnarray}
in which the displaced operator, using the Kerr annihilation  operators (\ref{anh_pos}), is defined as
\begin{align}\label{dis}
    D(\alpha; j, \lambda_{\pm})=\exp\left[\alpha \hat{A}_{\pm}^{\dagger}-\alpha^{\ast}\hat{A}_{\pm}\right] 
\end{align}
The positive  displaced Kerr coherent state can be expanded in terms of the Fock state basis 
as \cite{dehdashti2024enhancing}
\begin{eqnarray}
\ket{\alpha;j,\lambda_{+}}&=&\cosh^{-2j}\left[\sqrt{\frac{\lambda}{2}} |\alpha| \right] 
\sum_{n=0}^{\infty} \sqrt{\frac{\Gamma(2j+n)}{\Gamma(2j) n!}}\nonumber\\
&\times&e^{-i n\phi}\tanh^{n}\left[\sqrt{\frac{\lambda}{2}} |\alpha| \right] \ket{n},
\end{eqnarray}
in which $j$ can be an integer or a half-integer. It is crucial to highlight that if the parameter $\lambda$ is assigned the value of $2$, the resulting state is a $\mathfrak{su}(1,1)$-coherent state. 

\begin{figure*}[t]
    \centering
\includegraphics[width=4.3cm]{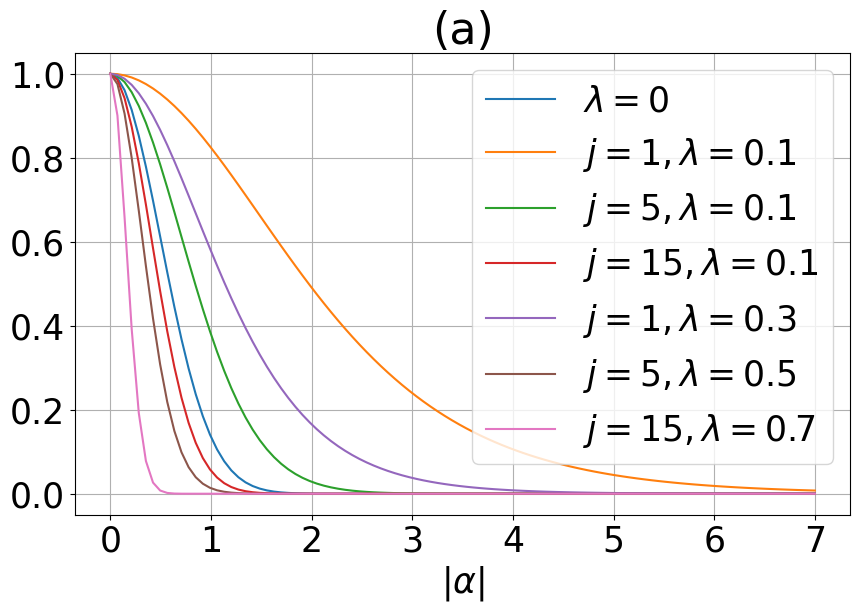}
\includegraphics[width=4.3cm]{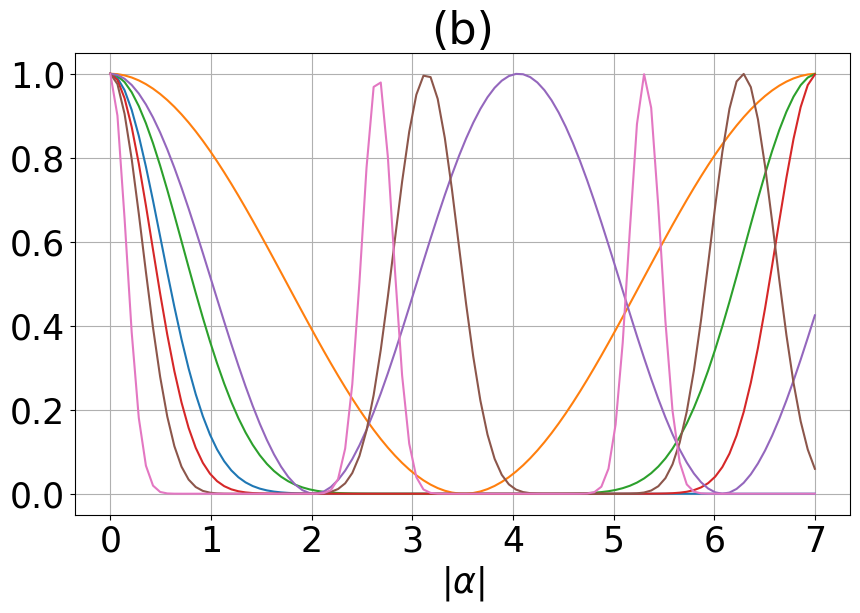}
\includegraphics[width=4.3cm]{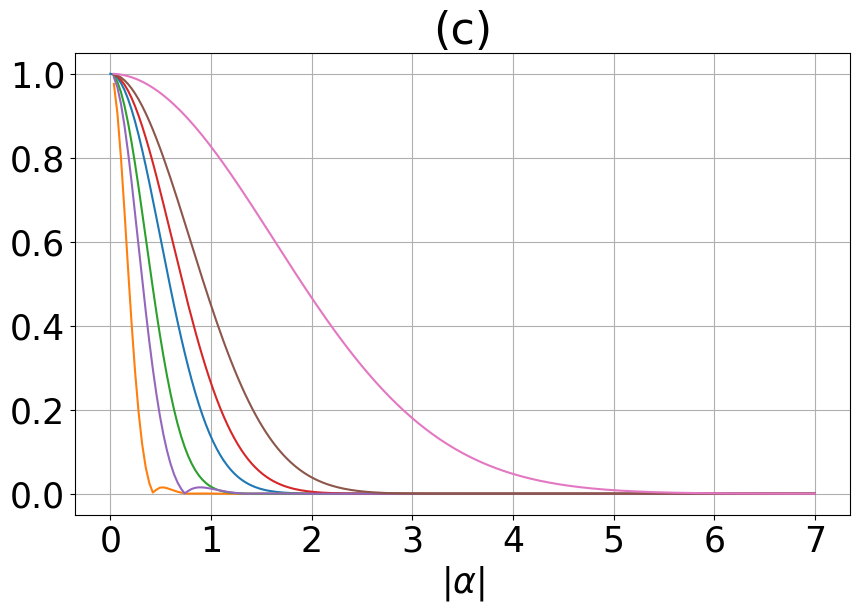}
\includegraphics[width=4.3cm]{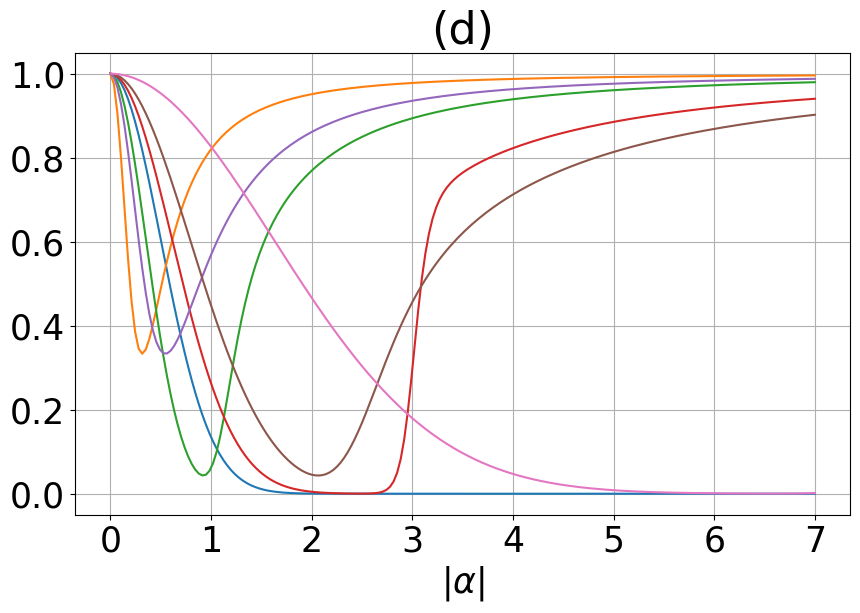}
    \caption{Overlap of the positive and negative KCSs are depicted in plots (a) and (b), respectively. Plots (c) and (d) illustrate the overlap for the BGKCSs, considering both positive and negative Kerr parameters. }
    \label{fig_inner}
\end{figure*}
For negative values of the  Kerr parameter $\lambda$,
the negative displaced Kerr coherent state $\ket{\alpha; j,\lambda_{-}}$
is given by 
\begin{eqnarray}
\ket{\alpha;j,\lambda_{-}}&=&\cos^{2j}\left[\sqrt{\frac{|\lambda|}{2}} |\alpha| \right] 
\sum_{n=0}^{2j} \sqrt{\frac{(2j)!}{(2j-n)!n!}}\nonumber\\
&\times&e^{-in\phi}\tan^{n}\left[\sqrt{\frac{|\lambda|}{2}} |\alpha| \right] \ket{n},
\end{eqnarray}
Note that for the $\lambda=-2$, the negative DKCS represents a $\mathfrak{su}(2)$-coherent state.\\
\indent  Two-component  Kerr cat states (KCS) are defined as the even and odd superposition of the displaced Kerr coherent state with opposite displacement, i.e., \begin{eqnarray}\label{cat_dis}
\ket{\mathsf{C}^{\pm};j, \lambda_{\pm}}_{D}=\frac{1}{\mathsf{N}^{\pm}}\left(\ket{\alpha; j,\lambda_{\pm}}\pm \ket{-\alpha;j, \lambda_{\pm}}\right)
    \end{eqnarray}
in which $\mathsf{N}^{\pm}$ is the normalization constant. The normalization constant is given by
\begin{eqnarray}
\mathsf{N}^{\pm}&=&\sqrt{2\left(1\pm\braket{\alpha; j,\lambda_{\pm} }{ -\alpha; j,\lambda_{\pm}}\right)}
\end{eqnarray} 
The quasi-orthogonality -referring to the property that the overlap becomes negligible for sufficiently large values of $\alpha$-  of the displaced Kerr coherent states occurs for positive values of the Kerr parameter, that is 
\begin{eqnarray}
\braket{\alpha; j, \lambda_{+}}{ -\alpha; j, \lambda_{+}}=\cosh^{-2j}\left[\sqrt{2\lambda}|\alpha|\right]
\end{eqnarray}
Notably, the Kerr parameter can be adjusted for any intensity of light, i.e., $|\alpha|^{2}$, where the degree of overlap approaches zero with high precision, as illustrated in Fig. \ref{fig_inner}-(a). However, for the negative value of the Kerr parameter $\lambda$, the inner  product of the KCSs with the opposite phase is given by 
\begin{align}\label{overlab_neg}
\braket{\alpha;j, \lambda_{-} }{ -\alpha;j, \lambda_{-}}=\cos^{2j}\left[\sqrt{2|\lambda|}|\alpha|\right]
\end{align}
Specifically,  for $\sqrt{2|\lambda|}|\alpha|=n\pi/2$, $n\in \mathbb{N}$, the two KCSs are orthogonal, i.e., $\braket{\alpha;j, \lambda_{-} }{ -\alpha;j, \lambda_{-}}=0$, Fig. \ref{fig_inner}-(b) illustrates the behavior of the overlap (\ref{overlab_neg}) as a function of light intensity for various values of the Kerr parameter.
\subsection{Barut–Girardello Kerr cat state}
Following Barut and Girardello \cite{barut1971new}, we define  a Barut-Girardello Kerr coherent state   as the  eigenstates of the Kerr annihilation operators  (\ref{anh_pos}),  
\begin{eqnarray}\label{BG_def}
\hat{A}_{\pm}\ket{\alpha;j,\lambda_{\pm}}_{BG}&=&\alpha\ket{\alpha;j,\lambda_{\pm}}_{BG}
\end{eqnarray}
In the Fock state basis, we can represent the 
Barut–Girardello Kerr coherent state, for the positive Kerr parameter, as follows: 
\begin{eqnarray}\label{GK-def-pos}
\ket{\alpha;j,\lambda_{+}}_{BG}=\mathsf{N}^{-1/2}\sum_{n=0}^{\infty} \sqrt{\frac{\Gamma(2j)}{\Gamma(2j+n) n!}} \left[\sqrt{\frac{2}{\lambda}}\alpha\right]^{n}\ket{n}\nonumber\\
\end{eqnarray}
  The normalization $\mathsf{N}$
is given by
\begin{eqnarray}
\mathsf{N}=\frac{\Gamma(2j)}{\left[|\alpha|\sqrt{\frac{8}{\lambda}}\right]^{2j-1}}I_{2j-1}\left(\sqrt{\frac{8}{\lambda}}|\alpha|\right),
\end{eqnarray}
where $I_{n}(x)$ is the $n^{\text{th}}$-order modified Bessel function.
For the negative value of the Kerr parameter, the Fock representation of the  Barut–Girardello Kerr coherent state is given by
\begin{eqnarray}\label{GK-def-neg}
\ket{\alpha;j,\lambda_{-}}_{BG}=\mathsf{N}^{-1/2}\sum_{n=0}^{2j}\sqrt{\frac{(2j-n)!}{(2j)!n!}}\left[\sqrt{\frac{2}{|\lambda|}}\ \alpha\right]^{n}\ket{n}\nonumber\\
\end{eqnarray}
in which the normalization is given by
\begin{eqnarray} \mathsf{N}=\sum_{n=0}^{2j}\frac{(2j-n)!}{(2j)!n!}\left[\sqrt{\frac{2}{|\lambda|}}\ |\alpha|\right]^{2n}
\end{eqnarray}
\indent Two-component Barut-Girardello Kerr cat states (BGKCSs) are defined as follows,
 \begin{eqnarray}\label{cat_BG}
\ket{\mathcal{C}^{\pm},\lambda_{\pm}}_{BG}=
\frac{1}{\mathsf{N}^{\pm}}\left(\ket{\alpha;\lambda_{\pm}}\pm\ket{-\alpha;\lambda_{\pm}}\right)
\end{eqnarray}
where
\begin{eqnarray}
\mathsf{N}_{\alpha}^{\pm}&=&\sqrt{2\left(1\pm\braket{\alpha,\lambda }{ -\alpha,\lambda}_{BG}\right)}
\end{eqnarray} 
\begin{figure*}[t!]
    \centering 
\includegraphics[width=18cm]{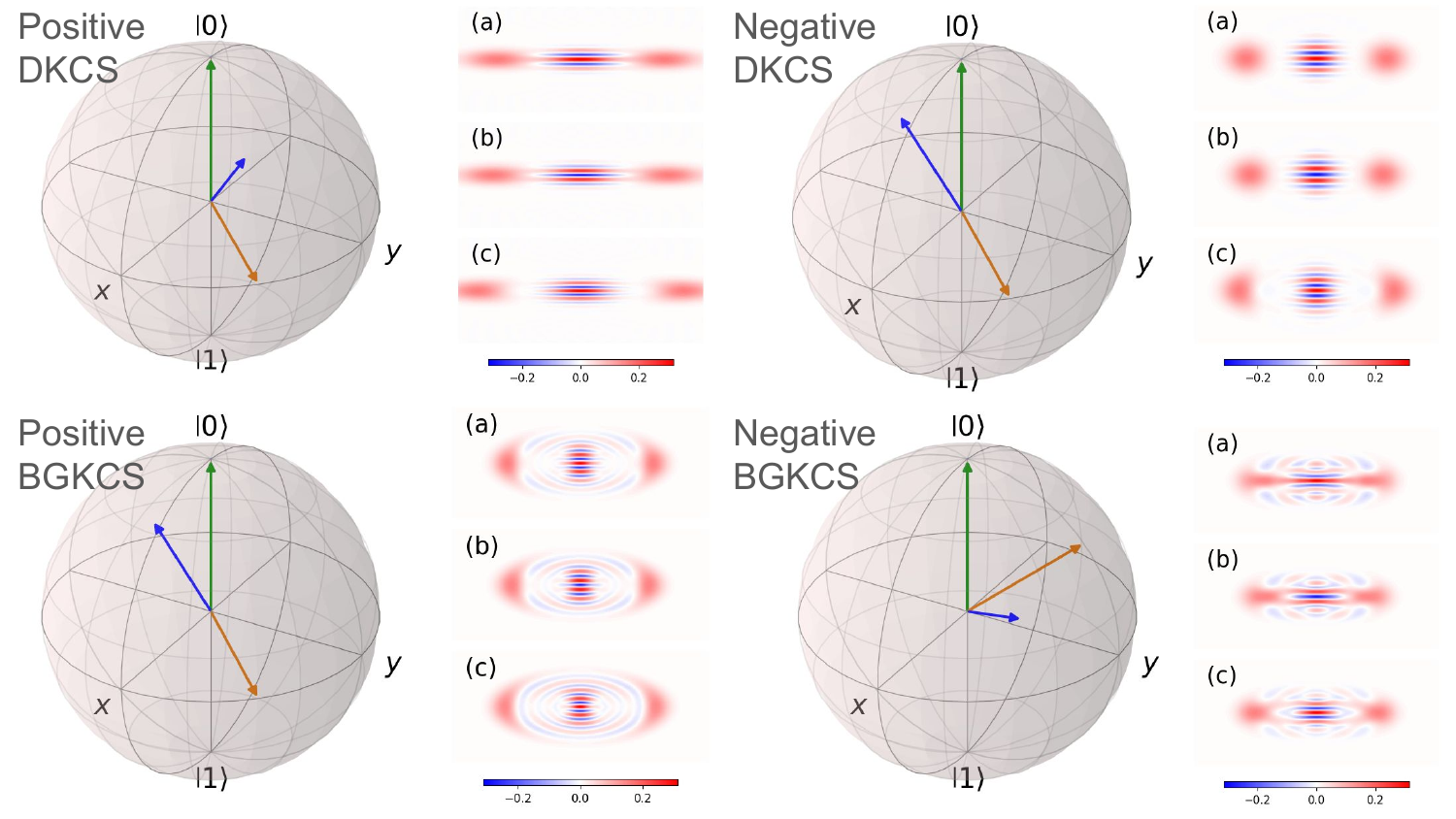}

    \caption{
The green vector represents the cat state, while the orange and blue vectors depict the effects of the standard loss and dephasing operators, respectively, on the Bloch sphere. The Wigner quasi-probability distribution function $ W(X, Y)$ for the cat state is shown in plot (a), while plots (b) and (c) illustrate the Wigner functions corresponding to the influence of the loss and dephasing operators. For all plots, the parameters are set as follows: $ \lambda = \pm 0.5 $ for DKCSs, and $ \lambda = \pm 0.2$, $ \alpha = 1.5 $ for the Wigner function visualizations.
}\label{fig_wigner_dis}
\end{figure*}
The quasi-orthogonality of BGKCS occurs for positive values of the Kerr parameter, that is
\begin{eqnarray}\label{over_BG_pos}
\braket{\alpha; j,\lambda_{+} }{ -\alpha;j,\lambda_{+}}_{BG}=\frac{J_{2j-1}\left(\sqrt{\frac{8}{\lambda}}|\alpha|\right)}{I_{2j-1}\left(\sqrt{\frac{8}{\lambda}}|\alpha|\right)}\nonumber\\
\end{eqnarray}
in which $J_{n}(x)$ is the Bessel function of the first kind. For the negative values of the Kerr parameters, the normalization is given by
\begin{eqnarray}\label{over_BG_neg}
\braket{\alpha; j,\lambda_{-} }{ -\alpha;j,\lambda_{-}}_{BG}&=&\sum_{n,m=0}^{2j}\frac{(2j-n)!m!}{(2j-m)!n!} (-1)^{n} \nonumber\\
&\times &\left[\frac{2}{|\lambda|} |\alpha|^{2}\right]^{n-m}
\end{eqnarray}

In Figs. (\ref{fig_inner})-(c) and -(d) display the overlap functions (\ref{over_BG_pos}) and (\ref{over_BG_neg}) as a function of light intensity for various values of the Kerr parameter. 

The DKCSs and BGKCSs are a superposition of only even and odd number states, so they are orthogonal. As a consequence, they are eigenstates of the parity operator $\Pi=\exp(i\pi\hat{a}^{\dagger}\hat{a})$ with eigenvalues $\pm 1$.

Therefore, quantum information can be stored in the two deformed-cat BQC, where the cat states $\ket{\mathsf{C}^{\pm},\lambda}_{D}$ and $\ket{\mathsf{C}^{\pm},\lambda}_{BG}$ encode the basis states $\ket{0}$ and $\ket{1}$ of the logical qubit.

\section{Dynamics of  Kerr cat states}\label{sec:Dynamics_of_KCSs}
In the following, we examine two distinct operators that model the environment. First, we consider the environment represented by the annihilation  $\hat{a}$ and the creation operator $\hat{a}^{\dagger}$. Second, we model the environment using a set of Kerr-modified annihilation and creation operators (\ref{anh_pos}). 

We describe the bosonic mode as a (weakly) anharmonic oscillator within the framework of the Gorini-Kossakowski-Sudarshan-Lindblad master equation,
\begin{equation}\label{eq:dynamics}
\partial_t \hat{\rho} = \mathcal{L}\hat{\rho}.
\end{equation}
The Liouvillian superoperator is given by
\begin{equation}\label{eq:Lindbladian_standard}
\mathcal{L}\hat{\rho}
= \kappa_{l}\,\mathcal{D}[\hat{a}]\,\hat{\rho}
+ \kappa_{d}\,\mathcal{D}[\hat{a}^{\dagger}\hat{a}]\,\hat{\rho},
\end{equation}
where $\hat{a}$ denotes the bosonic annihilation operator. In a generalized (dressed) representation, the dynamics can equivalently be expressed as
\begin{equation}\label{eq:Lindbladian_modified}
\mathcal{L}\hat{\rho}
= \kappa_{l}\,\mathcal{D}[\hat{A}]\,\hat{\rho}
+ \kappa_{d}\,\mathcal{D}[\hat{A}^{\dagger}\hat{A}]\,\hat{\rho},
\end{equation}
with $\hat{A}$ an effective annihilation operator incorporating nonlinearities or engineered dissipative processes. The dissipator is defined as
\begin{equation}
\mathcal{D}[\hat{C}]\,\hat{\rho}
= \hat{C}\hat{\rho}\hat{C}^{\dagger}
- \frac{1}{2}\left\{\hat{C}^{\dagger}\hat{C}, \hat{\rho}\right\},
\end{equation}
for an arbitrary operator $\hat{C}$. The terms proportional to $\kappa_{l}$ and $\kappa_{d}$ describe single-particle loss and pure dephasing processes, respectively.

\indent The noise channel  gives the evolution of an initial state $\hat{\rho}$  for the interval time $\tau$,
\begin{eqnarray}\label{qe4}
\mathcal{N}_{\kappa_{l},\kappa_{d}}(\hat{\rho}(0))&=&e^{\mathcal{L}\tau}\hat{\rho}(0)\nonumber\\
&=&\sum_{j}^{\infty}\hat{K}_{j}\hat{\rho}(0)\hat{K}_{j}^{\dagger}
\end{eqnarray}
where  $\hat{K}_{j}$s are Kraus operators.  To leading order in $\kappa_{l,d}\tau \ll 1$, the  channel (\ref{qe4}) can be written as \cite{schlegel2022quantum}
\begin{eqnarray}\label{eq:noise_channel}
\mathcal{N}_{\kappa_{l},\kappa_{d}}(\hat{\rho}(0))\approx \sum_{j=0}^{2} \hat{K}_{j}\hat{\rho}(0)\hat{K}_{j}^{\dagger}
\end{eqnarray}
where 
\begin{eqnarray}\label{eq:Kraus_operators}
\hat{K_{0}}&=&\mathbb{I}-\frac{\kappa_{l}\tau}{2}\hat{C}^{\dagger}\hat{C}-\frac{\kappa_{d}\tau}{2}\left(\hat{C}^{\dagger}\hat{C}\right)^{2}\\
\hat{K_{1}}&=&\sqrt{\kappa_{d}\tau} \hat{C}^{\dagger}\hat{C}\\
\hat{K_{2}}&=& \sqrt{\kappa_{l}\tau} \hat{C}
\end{eqnarray}
where $\hat{C}$ represents either the linear operator $\hat{a}$ or the nonlinear operator $\hat{A}$, depending on whether the noise is linear or nonlinear in nature. Thus, based on the environmental model, the following sets of  errors can be considered:
\begin{eqnarray}
\{\hat{E}_{m}\}&=&\{\mathbb{I},\hat{a},\hat{a}^{\dagger}\hat{a},(\hat{a}^{\dagger}\hat{a})^{2}\},\label{eq:error_matrices_sta}
\\
\{\hat{E}_{m}\}&=&\{\mathbb{I},\hat{A},\hat{A}^{\dagger}\hat{A},(\hat{A}^{\dagger}\hat{A})^{2}\},\label{eq:error_matrices_mod}
\end{eqnarray}

\indent The correction capability of a bosonic quantum code for a finite set of errors can be characterized by the Knill–Laflamme (KL) conditions \cite{ng2010simple,schlegel2022quantum,klesse2007approximate}. Given two distinct errors $\hat{E}^{m}$ and $\hat{E}^{m^{\prime}}$, belonging to either the set of errors (\ref{eq:error_matrices_sta}) or (\ref{eq:error_matrices_mod}),

 and codewords $\{\ket{\mathcal{C}^{\pm}}_{f}\}$, where $f= D, BG$, i.e., the cat state defined by the  relations (\ref{cat_dis}) and (\ref{cat_BG}), the KL conditions state that errors $\hat{E}^{m}$ and $\hat{E}^{m^{\prime}}$ can be distinguished and corrected if and only if
\begin{eqnarray}\label{eq-KL}
\bra{\psi_{i}}\hat{E}_{m}\hat{E}_{m^{\prime}}\ket{\psi_{j}}=\delta_{ij} \alpha_{m,m^{\prime}}
\end{eqnarray}
in which $\ket{\psi_{k}}$, where $k= i, j $ is one of the above-mentioned Kerr cat states  $\{\ket{\mathcal{C}^{\pm}}_{f}\}$, $\alpha_{m,m^{\prime}}$ is a Hermitian matrix, independent of $i$ and $j$. We aim to consider a BQC capable of approximately correcting errors and assess how much a code is correctable to the action of the noise channel $\mathcal{N}_{\kappa_{l,d}}$ defined in equation (\ref{qe4}). The  KL condition can be fulfilled  the following relation:
\begin{eqnarray}\label{qeeq10}
0\leq KL(\hat{E}_{m},\hat{E}_{m^{\prime}})\leq 1 
\end{eqnarray}
which 
\begin{eqnarray}\label{eq_KL}
KL(\hat{E}_{m},\hat{E}_{m^{\prime}})=\frac{|\bra{\psi_{i}}\hat{E}_{m}^{\dagger}\hat{E}_{m^{\prime}}\ket{\psi_{j}}|}{\sqrt{\bra{\psi_{j}}\hat{E}_{m}^{\dagger}\hat{E}_{m^{\prime}}\hat{E}_{m}^{\dagger}\hat{E}_{m^{\prime}}\ket{\psi_{j}}}}
\end{eqnarray}
In the KL condition zero corresponds to perfect correction of errors if $i\neq j$, and $1$ corresponds to a maximum violation of KL conditions.

In light of the observation that Kerr/standard annihilation operator induces a shift in each Fock state within the Fock representation, the effect of the error operators $\hat{a}$ and $\hat{A}$ on the DKCSs can be expressed as follows:
\begin{eqnarray*}
    \hat{a} \ket{\mathcal{C}^{\pm},\lambda}_{D} &=& c_{1}\ket{\mathcal{C}^{\mp},\lambda}_{D} + d_{1} \ket{\Tilde{\mathcal{C}}^{\mp},\lambda}_{D}\\
    \hat{A} \ket{\mathcal{C}^{\pm},\lambda}_{D} &=& c_{2}\ket{\mathcal{C}^{\mp},\lambda}_{D} + d_{2} \ket{\Tilde{\mathcal{C}}^{\mp},\lambda}_{D}
\end{eqnarray*}
Here, the states $\ket{\tilde{\mathcal{C}}^{\mp},\lambda}_{D}$ form an error space orthogonal to the code space. Notably, in a scenario where the cat states are regarded as BGKCSs, as defined in (\ref{BG_def}), and the environmental impact is modeled by associated Kerr operators, i.e., $\hat{A}_{\lambda_{\pm}}\ket{\mathcal{C}^{\pm},\lambda_{\pm}}_{BG}\propto \ket{\mathcal{C}^{\mp},\lambda_{\pm}}_{BG}$, it is crucial to highlight that no syndrome will be able to detect the occurrence of such an error, similar to normal coherent states.

Fig. \ref{fig_wigner_dis} illustrates the impact of the annihilation and number operators on the cat states defined in Eqs. (\ref{cat_dis}) and (\ref{cat_BG}). In the Bloch sphere representation, the green vector corresponds to the states $\ket{\mathcal{C}^{+}, \lambda_{\pm}}_{D}$ and $\ket{\mathcal{C}^{+}, \lambda_{\pm}}_{BG}$, which define the logical code space. Plot~(a) displays the corresponding Wigner function.
The blue and orange vectors on the Bloch sphere represent the action of the annihilation and number operators on the cat state, i.e., $\hat{a}\ket{\mathcal{C}^{+}, \lambda_{\pm}}_{f}$ and $\hat{a}^{\dagger}\hat{a}\ket{\mathcal{C}^{+}, \lambda_{\pm}}_{f}$, where $f \in \{D, BG\}$. These operations correspond to photon loss and dephasing processes in the system, respectively.
Plot~(b) demonstrates the effect of photon loss on various cat states, while Plots~(d) depicts the influence of dephasing.
It is noteworthy that, although the annihilation operator maps the standard cat state $\ket{\mathcal{C}^{+}}$ (logical $\ket{0}$) to $\ket{\mathcal{C}^{-}}$ (logical $\ket{1}$), and the number operator preserves $\ket{\mathcal{C}^{+}}$ (logical $\ket{0}$), the effects of the annihilation and number operators on the Kerr cat states correspond to distinct locations on the Bloch sphere. These positions vary depending on the type of coherent state, as well as the values of the Kerr parameter $\lambda$ and the light intensity $\alpha$.

\section{Errors and their Corrections Using Kerr Bosonic Codes}\label{sec:Errors_and_their_corrections}
\subsection{KL condition for the Kerr cat states}
Let us consider a cat state $\ket{\mathcal{C}^{\pm},\lambda_{\pm}}_{D}$ and $\ket{\mathcal{C}^{\pm},\lambda_{\pm}}_{BG}$ subject to a particle loss event. We consider two different scenarios in which either an annihilation operator $\hat{a}$ or a Kerr annihilation operator $\hat{A}$ are responsible for interacting with light. 

KL condition \eqref{eq_KL} provides a quantitative criterion for evaluating the correctability of errors in bosonic quantum codes. For standard cat states, the KL condition associated with single-photon loss is given by \cite{schlegel2022quantum}
\begin{equation}
\frac{\langle \mathcal{C}^{\mp} | \hat{a} | \mathcal{C}^{\pm} \rangle}
{\sqrt{\langle \mathcal{C}^{\pm} | \hat{a}^{\dagger}\hat{a} | \mathcal{C}^{\pm} \rangle}}
= \frac{\alpha}{|\alpha|},
\end{equation}
which shows that the action of the photon-loss operator is independent of the light intensity and depends only on the phase of the coherent amplitude.

In contrast, for DKCSs, the KL condition depends explicitly on the coherent amplitude, the Kerr nonlinearity, and $j$. For DKCSs with positive and negative Kerr parameters $\lambda_{\pm}$, one obtains (see Appendix \ref{appendix_B})
\begin{align}
{}_D\langle \mathcal{C}^{\mp}, \lambda_{+} | \hat{A} | \mathcal{C}^{\pm}, \lambda_{+} \rangle_D
&= \frac{1+\sech^{2j+1}\left[\sqrt{2\lambda}\, |\alpha|\right]}{1-\sech^{4j}\left[\sqrt{2\lambda}\, |\alpha|\right]}\nonumber\\
&\times j\sqrt{\frac{\lambda}{2}} \sinh\left[\sqrt{2\lambda}\, |\alpha|\right], 
\label{error_KL_DKC_pos} \\
{}_D\langle \mathcal{C}^{\mp}, \lambda_{-} | \hat{A} | \mathcal{C}^{\pm}, \lambda_{-} \rangle_D
&=  \frac{1+\cos^{2j+1}\left[\sqrt{2\lambda}\, |\alpha|\right]}{1-\cos^{4j}\left[\sqrt{2\lambda}\, |\alpha|\right]}\nonumber\\
&\times j\sqrt{\frac{\lambda}{2}} \sin\left[\sqrt{2\lambda}\, |\alpha|\right]
\label{error_KL_DKC_neg}
\end{align}
In the case of $j \gg 1$, the above relations can be reduced to
\begin{eqnarray}
{}_D\langle \mathcal{C}^{\mp}, \lambda_{+} | \hat{A} | \mathcal{C}^{\pm}, \lambda_{+} \rangle_D 
&\approx&
j\sqrt{\frac{\lambda}{2}} \, \sinh\!\left[\sqrt{2\lambda}|\alpha|\right],
\label{eq_approx_pos}
\\
{}_D\langle \mathcal{C}^{\mp}, \lambda_{-} | \hat{A} | \mathcal{C}^{\pm}, \lambda_{-} \rangle_D 
&\approx&
j\sqrt{\frac{\lambda}{2}} \, \sin\!\left[\sqrt{2\lambda}|\alpha|\right].
\label{eq_approx_neg}
\end{eqnarray}
Relations~\eqref{eq_approx_pos} and~\eqref{eq_approx_neg} show that, in both the positive and negative Kerr-parameter cases, the logical error induced by the Kerr loss operator becomes increasingly correctable. The case of a negative Kerr parameter is particularly intriguing due to the periodicity of Eq.~\eqref{eq_approx_neg}. In fact, by appropriately adjusting the value of the Kerr parameter, the loss in the system can be corrected for arbitrary values of $\alpha$, representing a capability beyond that of standard  cat states. Moreover, for small values of the Kerr and light-intensity parameters, both equations scale linearly, and the loss error can be corrected by increasing any of the parameters $j$, $\lambda$, or $|\alpha|$, provided that $\sqrt{\lambda}|\alpha| \ll 1$. 
\begin{figure}[t!]
\includegraphics[width=4.2cm]{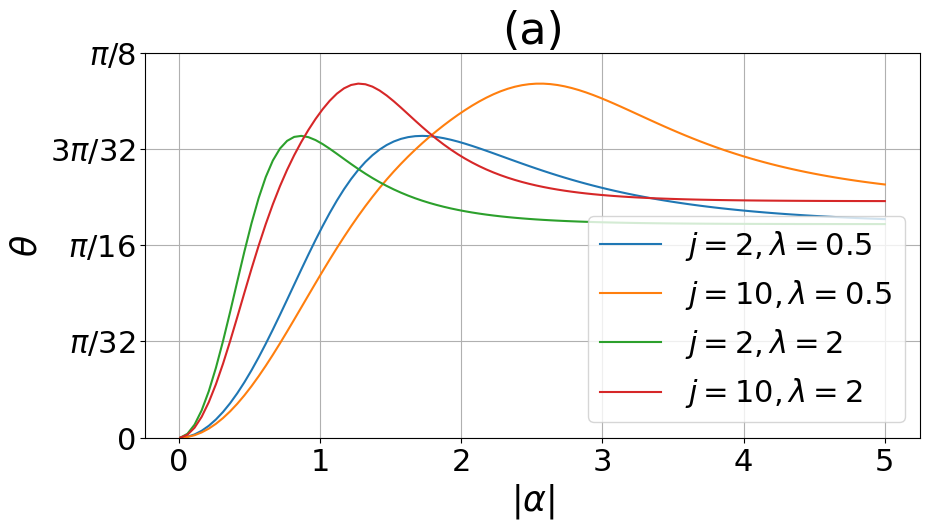}   \includegraphics[width=4.2cm]{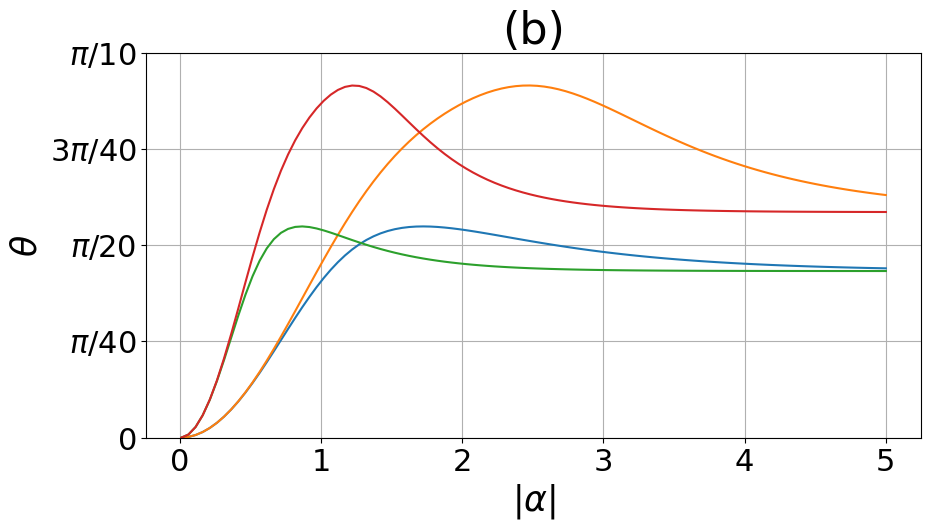}
\includegraphics[width=4.2cm]{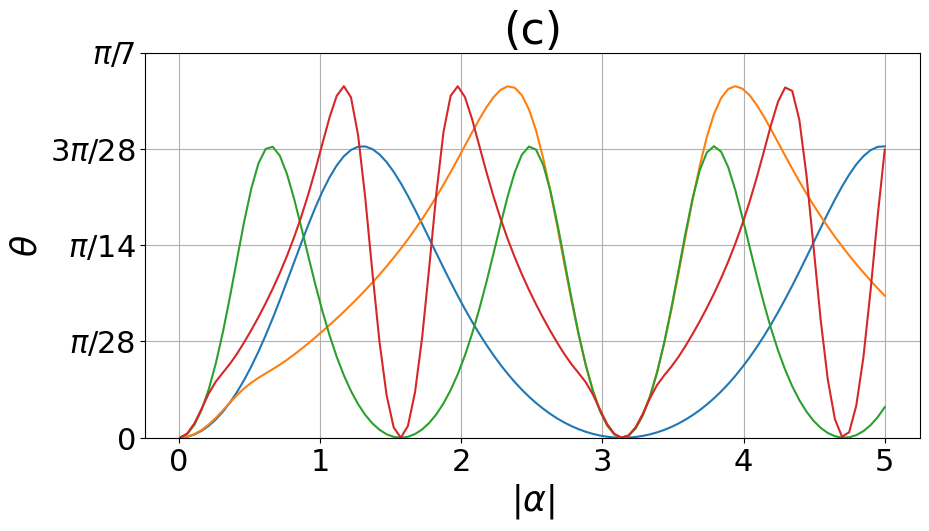}   \includegraphics[width=4.2cm]{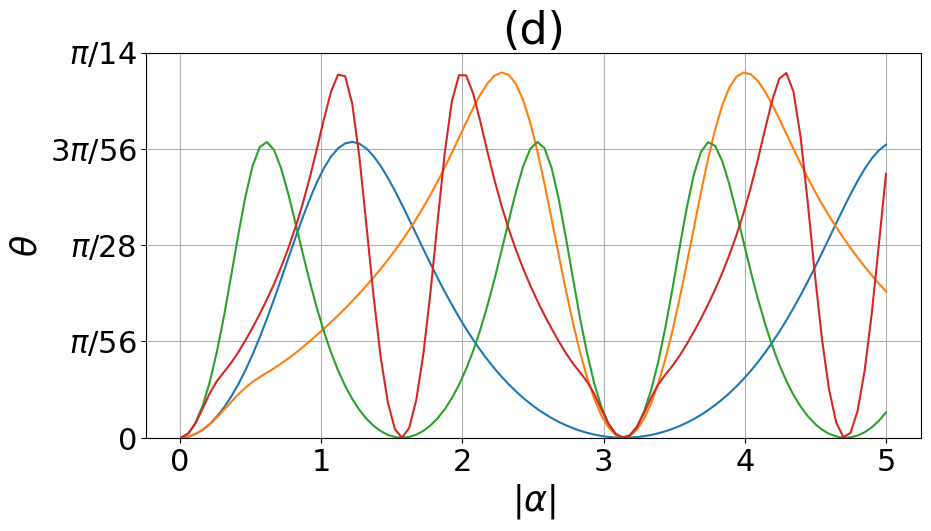}

    \caption{
The angle $\vartheta$, denoted as $\arccos{KL}$, is measured between the states $\ket{\mathcal{C}^{\pm},\lambda}$ and $\hat{A}\ket{\mathcal{C}^{\mp},\lambda}$, as well as between the states $\ket{\mathcal{C}^{\pm},\lambda}$ and $\hat{a}\ket{\mathcal{C}^{\mp},\lambda}$, as a function of $|\alpha|$ for positive Kerr parameters, depicted in plots (a) and (b). Similarly, plots (c) and (d) illustrate the same curves for negative Kerr parameters. }\label{fig_angle_anhi_DKCS}
\end{figure}

Moreover, although no closed-form analytical expression exists for the action of the standard loss operator, the lower bound is given by (see \eqref{B_17} and \eqref{B_18}), 
\begin{align}
{}_{D}\langle \mathcal{C}^{\mp}, \lambda_{\pm} \vert \hat{a} \vert \mathcal{C}^{\pm}, \lambda_{\pm} \rangle_{D}
\ge \sqrt{2j} \tanh^{2}\left[\sqrt{\frac{\lambda}{2}} |\alpha|\right].
\end{align}
Consequently, increasing the parameter $j$ enhances the correctability of the system.

In addition, by considering the KL condition for (Kerr-type) loss processes, we define
\begin{align}
\cos{\vartheta}=
\frac{{}_{D}\bra{\mathcal{C}^{\mp},\lambda_{\pm}}\hat{C}   \ket{\mathcal{C}^{\pm},\lambda_{\pm}}_{D}}{\sqrt{{}_{D}\bra{\mathcal{C}^{\pm},\lambda_{\pm}}
\hat{C}^{\dagger}\hat{C}\ket{\mathcal{C}^{\pm},\lambda_{\pm}}_{D}}},
\end{align}
where $\hat{C}$ denotes either the standard annihilation operator $\hat{a}$ or its nonlinear counterpart $\hat{A}$. 
The parameter $\vartheta$ thus quantifies the angle between the original code state $\ket{\mathcal{C}^{\pm},\lambda_{\pm}}_{D}$ and the state obtained after the action of the error operator, $\hat{C}\ket{\mathcal{C}^{\pm},\lambda_{\pm}}_{D}$, thereby providing a geometric measure of the distinguishability induced by the loss channel.
We analyze its impact numerically, as shown in Fig.~\ref{fig_angle_anhi_DKCS}. Fig  illustrates the variation of the angle $\vartheta = \cos^{-1}(KL)$ as a function of the light intensity $|\alpha|$, for different values of the parameters $j$ and $\lambda$. Plots~(a) and~(b) depict the impact of the Kerr-modified annihilation operator and the standard annihilation operator, respectively, on the distinguishability of the DKCSs under the action of the loss operator. Similarly, plots~(c) and~(d) show the corresponding effects for the negative Kerr parameters.
Plot (a) indicates that the distinguishability increases with the Kerr parameter $\lambda$ when the light intensity $|\alpha|$ is relatively small. However, in the regime where $|\alpha| $ is relatively large, the influence of the Kerr parameter on distinguishability becomes negligible. A comparison between plots (a) and (b) reveals that the Kerr-modified annihilation operator leads to greater distinguishability compared to the standard annihilation operator.
Plots~(c) and~(d) exhibit a periodic behavior that depends on the parameters $\alpha$, $\lambda$, and $j$. Although this behavior is observed in both cases, the Kerr-modified annihilation operator results in greater distinguishability.

Furthermore, in the case of the positive DKCSs, the states $\ket{\mathsf{C}^{\mp},\lambda}$, $\hat{A}^{\dagger}\hat{A}\ket{\mathsf{C}^{\mp},\lambda}$, and $\hat{a}^{\dagger}\hat{a}\ket{\mathsf{C}^{\mp},\lambda}$ remain distinguishable, i.e.,
\begin{align}
{}_{D}\bra{\mathsf{C}^{\mp},\lambda_{\pm}}\hat{A}^{\dagger}\hat{A}\ket{\mathsf{C}^{\mp},\lambda_{\pm}}_{D}&=0\\
{}_{D}\bra{\mathsf{C}^{\mp},\lambda_{\pm}}\hat{a}^{\dagger}\hat{a}\ket{\mathsf{C}^{\mp},\lambda_{\pm}}_{D}&=0
\end{align}
under the condition $j\gg 1$, which fulfill  the following:
\begin{align}\nonumber
{}_{D}\bra{\mathsf{C}^{+},\lambda_{\pm}}\hat{A}^{\dagger}\hat{A}\ket{\mathsf{C}^{+},\lambda_{\pm}}_{D}]\approx {}_{D}\bra{\mathsf{C}^{-},\lambda_{\pm}}\hat{A}^{\dagger}\hat{A}\ket{\mathsf{C}^{-},\lambda_{\pm}}_{D}\\
{}_{D}\bra{\mathsf{C}^{+},\lambda_{\pm}}\hat{a}^{\dagger}\hat{a}\ket{\mathsf{C}^{+},\lambda_{\pm}}_{D}]\approx {}_{D}\bra{\mathsf{C}^{-},\lambda_{\pm}}\hat{A}^{\dagger}\hat{A}\ket{\mathsf{C}^{-},\lambda_{\pm}}_{D}
\end{align}

On the other hand, using the eigenvalue relation
$\hat{A}\ket{\alpha,\lambda_{\pm}}_{BG}
= \alpha \ket{\alpha,\lambda_{\pm}}_{BG}$,
the KL condition for the Kerr-loss operator in the case of BGKCSs  becomes
\begin{align}
\label{eq_pos_neg_KL}
\frac{|{}_{BG}\bra{\mathcal{C}^{\mp},\lambda_{\pm}}
\hat{A}
\ket{\mathcal{C}^{\pm},\lambda_{\pm}}_{BG}|}
{\sqrt{
{}_{BG}\bra{\mathcal{C}^{\pm},\lambda_{\pm}}
\hat{A}^{\dagger}\hat{A}
\ket{\mathcal{C}^{\mp},\lambda_{\pm}}_{BG}
}}
= 1,
\end{align}
which corresponds to a maximal violation of the KL condition.

\begin{figure}[t]
\includegraphics[width=4.2cm]{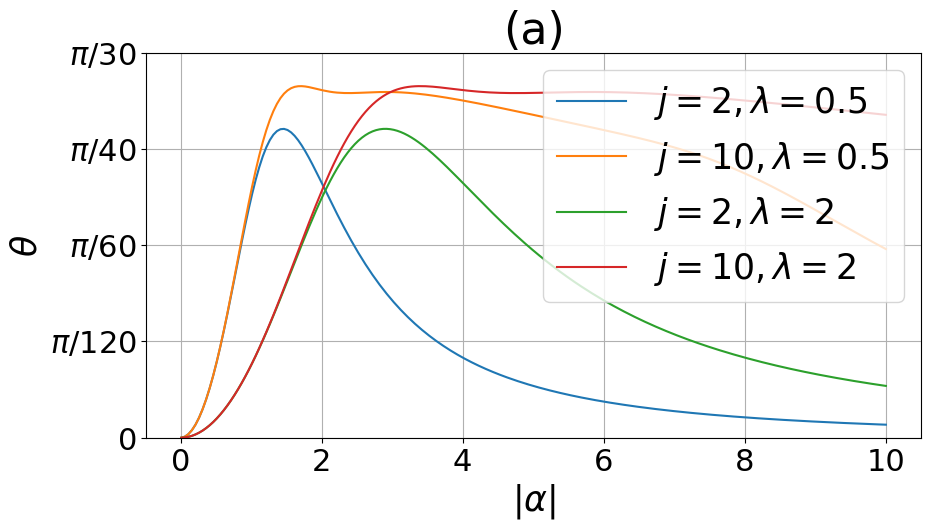}   \includegraphics[width=4.2cm]{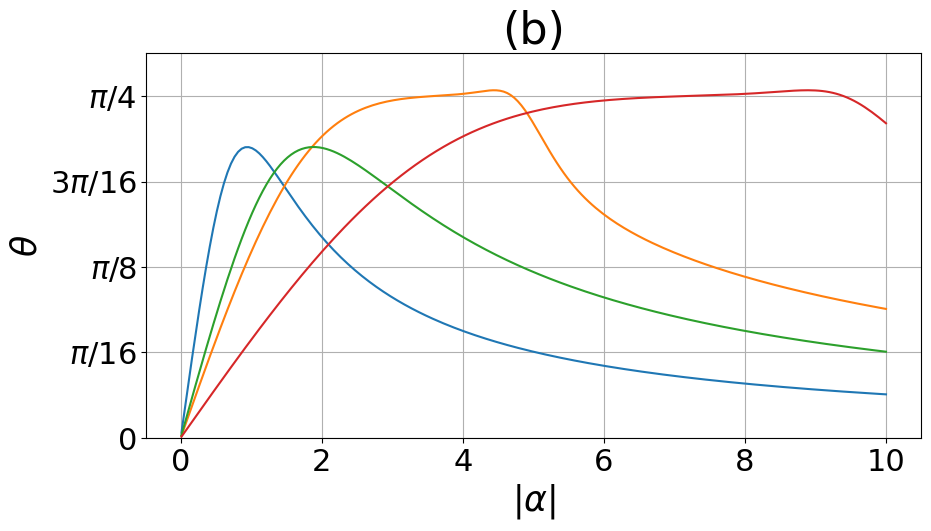}
\caption{The angle $\vartheta = \arccos(\mathrm{KL})$ between the states $\ket{\mathcal{C}^{\pm},\lambda}_{\mathrm{BG}}$ and $\hat{a}\ket{\mathcal{C}^{\mp},\lambda}_{\mathrm{BG}}$ as a function of the light intensity $|\alpha|$, for both positive and negative values of the Kerr parameter  are shown in plots~(a) and~(b),  respectively.  }\label{fig_angle_anhi_KBG}
\end{figure}

However, since
\(
\hat{a}\ket{\alpha,\lambda_{\pm}}_{BG}
\neq
\alpha \ket{\alpha,\lambda_{\pm}}_{BG},
\)
Eq.~\eqref{eq_pos_neg_KL} does not hold for the standard particle-loss operator.
Nevertheless, for $j \gg 1$, a upper bound can be obtained as
\begin{align}
{}_{BG}\bra{\mathcal{C}^{\mp},\lambda_{+}}
\hat{a}
\ket{\mathcal{C}^{\pm},\lambda_{+}}_{BG}
\ge
\frac{\sqrt{2}}{j}
\left[
1+\frac{|\alpha|^2}{2j|\lambda|}
 \right]
\end{align}
This shows that either large values of the Kerr parameter or small values of  $|\alpha|$ render the logical code approximately correctable.
  Figs. \ref{fig_angle_anhi_KBG}-(a) and (b) illustrate the angle $\vartheta=\arccos KL$ between $\hat{a}\ket{
\mathcal{C}^{\pm},\lambda_{+}}_{BG}$, and $\ket{\mathcal{C}^{\mp},\lambda_{\pm}}_{BG}$, i.e., positive and negative Kerr parameter,   respectively, for different values of the Kerr parameters with respect to the amplitude of light $|\alpha|$.

BGKCSs are able to suppress dephasing errors, as follows from the KL conditions in Eq.~(\ref{eq-KL}), independently of the dephasing model and of the sign of the Kerr parameter, namely
\begin{eqnarray*}
{}_{BG}\bra{\mathsf{C}^{\pm},\lambda}
\hat{A}^{\dagger}\hat{A}
\ket{\mathsf{C}^{\mp},\lambda}_{BG} &=& 0, \\
{}_{BG}\bra{\mathsf{C}^{\pm},\lambda}
\hat{a}^{\dagger}\hat{a}
\ket{\mathsf{C}^{\mp},\lambda}_{BG} &=& 0.
\end{eqnarray*}

However, the KL conditions further require
\begin{align}\label{eq_45_condition_deformed_number}
{}_{BG}\bra{\mathsf{C}^{+},\lambda_{+}}
\hat{A}^{\dagger}\hat{A}
\ket{\mathsf{C}^{+},\lambda_{+}}_{BG}
\approx
{}_{BG}\bra{\mathsf{C}^{-},\lambda_{+}}
\hat{A}^{\dagger}\hat{A}
\ket{\mathsf{C}^{-},\lambda_{+}}_{BG},\nonumber\\
\end{align}
which imposes  
\begin{align}
{}_{BG}\braket{\alpha,\lambda_{+}}{-\alpha,\lambda_{+}}_{BG}={}_{0}F_{1}(,2j,-\frac{2}{|\lambda|}|\alpha|^{2})\approx 0
\end{align}
In the case of standard dephasing error, the following condition should be satisfied: 
\begin{align}\label{eq_45_condition_number}
{}_{BG}\bra{\mathsf{C}^{+},\lambda_{+}}
\hat{a}^{\dagger}\hat{a}
\ket{\mathsf{C}^{+},\lambda_{+}}_{BG}
\approx
{}_{BG}\bra{\mathsf{C}^{-},\lambda_{+}}
\hat{a}^{\dagger}\hat{a}
\ket{\mathsf{C}^{-},\lambda_{+}}_{BG}
\end{align}
which applies to  scenarios in which Kerr-induced dephasing and standard dephasing are applied to the cat states, respectively.
The relation \eqref{eq_45_condition_number} is fulfilled when
$
|\alpha| \approx \sqrt{\frac{2}{\lambda}}\, j $, for $j \gg 1$.

In the case of a negative Kerr parameter, the condition
\begin{align}\nonumber
{}_{BG}\bra{\mathsf{C}^{+},\lambda_{-}}
\hat{A}^{\dagger}\hat{A}
\ket{\mathsf{C}^{+},\lambda_{-}}_{BG}
\approx
{}_{BG}\bra{\mathsf{C}^{-},\lambda_{-}}
\hat{A}^{\dagger}\hat{A}
\ket{\mathsf{C}^{-},\lambda_{-}}_{BG}
\end{align}
requires 
\begin{align}
{}_{BG}\braket{\alpha,\lambda_{-}}{-\alpha,\lambda_{-}}_{BG}={}_{0}F_{1}(,-2j,\frac{2}{|\lambda|}|\alpha|^{2})\approx 0
\end{align}
Considering standard dephasing error,
\begin{align}\nonumber
{}_{BG}\bra{\mathsf{C}^{+},\lambda_{-}}
\hat{a}^{\dagger}\hat{a}
\ket{\mathsf{C}^{+},\lambda_{-}}_{BG}
\approx
{}_{BG}\bra{\mathsf{C}^{-},\lambda_{-}}
\hat{a}^{\dagger}\hat{a}
\ket{\mathsf{C}^{-},\lambda_{-}}_{BG}
\end{align}
which means
$|\alpha|^{2}\approx 2j|\lambda|$, for $j\gg 1$. In addition, by assuming either $|\lambda|\ll 1$ or $|\alpha|\gg 1$, regardless of the value of the Kerr parameter, the condition fulfills.

\subsection{Error Corrections}
By employing the channel fidelity approach, one can periodically monitor and recover a quantum state at discrete time intervals $\tau$~\cite{kosut2009quantum}. Specifically, the Lindblad master equation governs the evolution of the system under the influence of a noise channel $\mathcal{N}_{\kappa_{l},\kappa_{d}}$ for a duration $\tau$. During this interval, a projective measurement onto the orthogonal code and error subspaces detects the error syndrome $\mathcal{S}$, followed by the application of a correction operation $\mathcal{C}$ that projects the system back onto the code space~\cite{kosut2009quantum,schlegel2022quantum}. 
Thus, the combined effect of the syndrome measurement and correction can be represented as a recovery map $\mathcal{R} = \mathcal{C} \circ \mathcal{S}$. The evolution of the quantum state over the interval $\tau$ is then described by:
\begin{equation}
    \hat{\rho}(t + \tau) = \mathcal{R} \left( e^{\mathcal{L} \tau} \hat{\rho}(t) \right) = \mathcal{R} \circ \mathcal{N}_{\kappa_{l},\kappa_{d}} \left( \hat{\rho}(t) \right),
\end{equation}
where $\mathcal{L}$ is the Liouvillian superoperator associated with the noise dynamics.

Hence, the whole dynamical process  can be written by the Kraus operators:
\begin{eqnarray}
\mathcal{E}(\hat{\rho}(\tau))=\sum_{j=0}^{\infty}\hat{S}_{j}\hat{\rho}(0)\hat{S}_{j}^{\dagger}
\end{eqnarray}
in which
\begin{eqnarray}
\hat{S}_{j}=\sum_{l,r=0}^{\infty}R_{l}K_{r}
\end{eqnarray}
where $K_{r}$ are the Kraus operators associated with the noise channel $\mathcal{N}_{\kappa_{l},\kappa_{d}}$, i.e., the equation (\ref{eq:noise_channel}), and $R_{l}$ are the Kraus operators associated with the recovery $\mathcal{R}$. \\
\indent The average channel fidelity, defined as \cite{albert2018performance,schumacher1996sending}
\begin{eqnarray}\label{eq:average_fidelity}
\mathcal{F}_{avg}=\frac{1}{d^{2}}\sum_{j=0}^{\infty}|\Tr(\hat{S}_{j})|^{2}=\frac{1}{d^{2}}\sum_{l,k=0}^{\infty}|\Tr(\hat{R}_{l}\hat{K}_{r})|^{2}
\end{eqnarray}
quantifies the average performance of a certain recovery operation $\mathcal{R}$.
We should  find the best recovery operation, i.e. the $\mathcal{R}$ which maximizes the average channel fidelity. 
To optimize the recovery operation $\mathcal{R}$, we can express the operators $\hat{R}_{l}$ in terms of a set of basis operators $\hat{R}_{i}=\sum x_{il}\hat{B}_{l}$, in which the set of basis operators $\{\hat{B}_{i}\}$ is orthogonal, i.e., $\Tr\left[\hat{B}_{i}\hat{B}_{j}\right]=\delta_{ij}$.\\
\indent The optimization procedure aims to find the coefficients $x_{il}$ which maximize the channel fidelity $\mathcal{F}_{avg}$, for a given set $\{\hat{B}_{i}\}$. By considering  components of the recovery matrix $\mathbf{X}=[X_{ij}]$ and the process matrix $\mathbf{W}=[W_{ij}]$ as 
\begin{eqnarray}
X_{ij}&=& \sum_{l}x_{il}x_{lj}^{\ast}\\
W_{ij}&=& \sum_{l} \Tr{B_{i}K_{l}} \Tr{B_{j}K_{l}}^{\ast}\nonumber\\
&=& \Tr{\mathcal{N}_{\kappa_{d},\kappa_{l}}\left(\hat{B}_{i}\otimes\hat{B}_{j}\right)}
\end{eqnarray}

One can obtain the process matrix as
\begin{eqnarray}
W_{ij}=\Tr \left[e^{\mathcal{L \tau}}\left( \hat{B}_{i}\otimes\hat{B}_{j}\right)\right]= \Tr\left[\mathcal{N}_{\kappa_{d},\kappa_{l}}\left(\hat{B}_{i}\otimes\hat{B}_{j}\right)\right]\nonumber\\
\end{eqnarray}
where $\Tr[\cdot]$ in this case denotes the trace of the super-operators.
We follow Refs. \cite{schlegel2022quantum,kosut2009quantum}, in which 
the recovery matrix is assumed as a  positive semidefinite operator.  Therefore, we should maximize 
\begin{eqnarray}
\frac{1}{d^{2}}\sum_{ij}X_{ij} W_{ij}=\frac{1}{d^{2}}\Tr\left[XW\right]
\label{eq:recovery_channel}
\end{eqnarray}
under the following conditions:
\begin{eqnarray}
&\mathbf{X}\succcurlyeq 0\\
&\sum_{ij} 
X_{ij}B^{\dagger}_{i}B_{j}=
\mathds{1}
\end{eqnarray}
which enforce positive semidefiniteness of the recovery matrices and trace preservation, respectively.

By defining the non-orthogonal subspaces using the error sets~(\ref{eq:error_matrices_sta}) or~(\ref{eq:error_matrices_mod}), which are spanned by the states $\ket{\psi^{\pm}_{i}}$, one can obtain:
\begin{eqnarray}
\ket{\psi^{\pm}_{i}}=\hat{E}_{m}\ket{\mathcal{C}_{f}^{\pm}},
\end{eqnarray}
Hence, the recovery operations can be defined as follows:
\begin{eqnarray}
P_{m}^{(0)}&=&\ket{\mathcal{C}_{f}^{+}}\bra{\psi_{m}^{+}}+\ket{\mathcal{C}_{f}^{-}}\bra{\psi_{m}^{+}}\label{recov_1}\\
P_{m}^{(1)}&=&\ket{\mathcal{C}_{f}^{+}}\bra{\psi_{m}^{-}}+\ket{\mathcal{C}_{f}^{-}}\bra{\psi_{m}^{+}}\label{recov_2}\\
P_{m}^{(2)}&=&\ket{\mathcal{C}_{f}^{+}}\bra{\psi_{m}^{-}}-i\ket{\mathcal{C}_{f}^{-}}\bra{\psi_{m}^{+}}\label{recov_3}\\
P_{m}^{(3)}&=&\ket{\mathcal{C}_{f}^{+}}\bra{\psi_{m}^{+}}+\ket{\mathcal{C}_{f}^{-}}\bra{\psi_{m}^{-}}\label{recov_4}
\end{eqnarray}
where $m = 0, 1, 2, 3$. Recovery operators \eqref{recov_1}-\eqref{recov_4} are mutually orthogonal and characterize the action of the recovery operation, which projects $\ket{\psi_{m}^{\pm}}$ back onto the initial states. The symbol $f$ denotes the type of Kerr cat states used, with $f \in \{D, BG\}$ representing DKCSs and BGKCSs, respectively.
\subsection{Channel fidelity for dephasing and loss}\label{subsec:channel_fidelities}
\begin{figure}
\includegraphics[width=\linewidth,keepaspectratio]{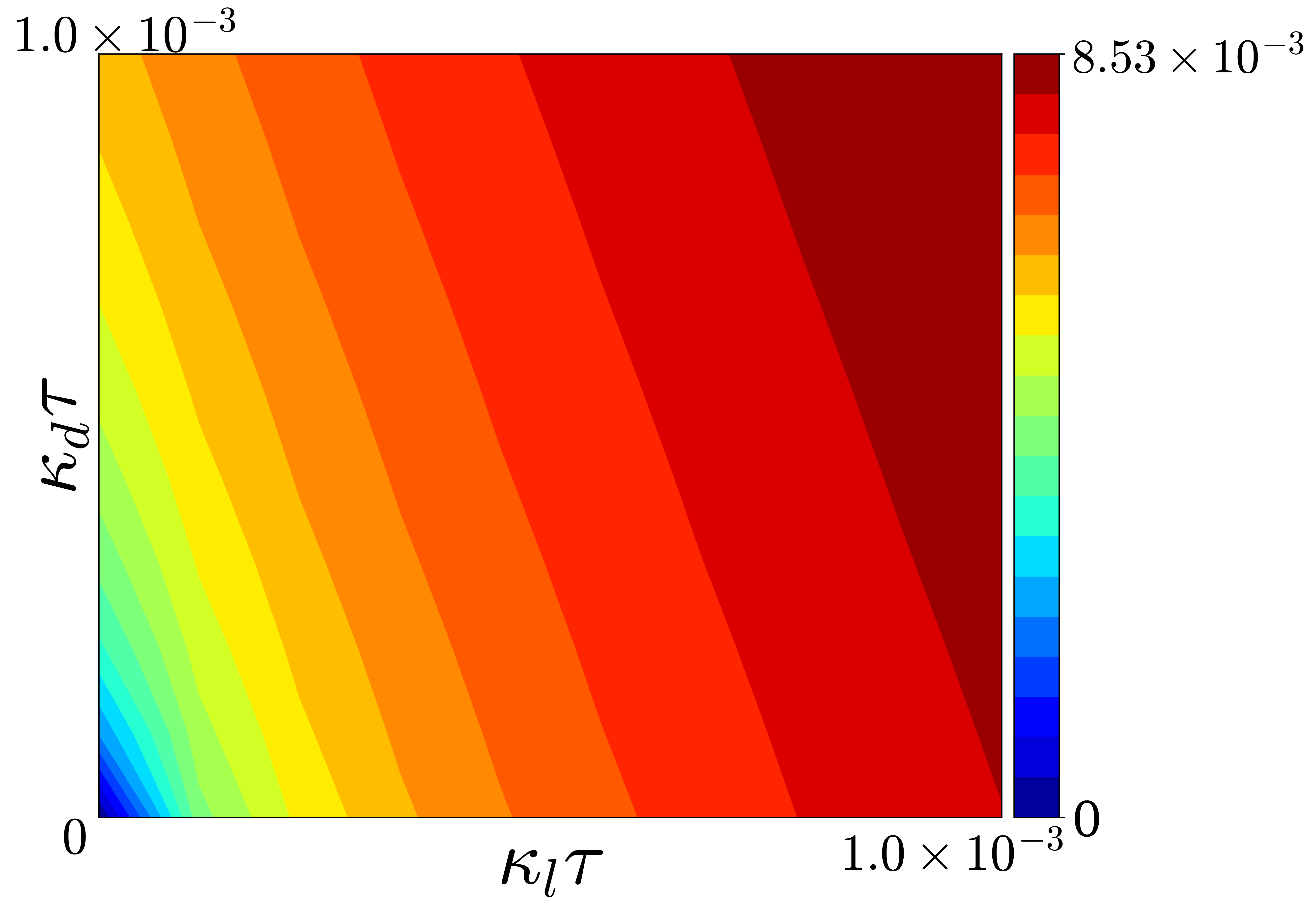}
    \caption{Heatmap of the average infidelity (logarithmic scale) for SCS under the combined effect of dephasing $\kappa_d \tau$ and photon loss $\kappa_l \tau$ errors. The coherent state corresponds to $|\alpha| = 1.5$.}
    \label{fig:heatmaps_SCS}
\end{figure}
We compute the optimal channel infidelity $1 - \mathcal{F}$, defined via Eq.~(\ref{eq:average_fidelity}), as a function of the dimensionless times $\kappa_{l} \tau$ and $\kappa_{d} \tau$, by applying the optimization procedure described above. Here, $\tau$ denotes the interval between successive recoveries; thus, larger values of $\kappa_{l} \tau$ or $\kappa_{d} \tau$ can be interpreted as either stronger dissipation or longer recovery intervals.

In the case of standard cat states (SCSs), the  infidelity as a function of the loss parameter $\kappa_{l}\tau$ and the dephasing parameter $\kappa_{d}\tau$ is shown in Fig.~\ref{fig:heatmaps_SCS}.
The figure illustrates that  the effect of particle loss begins to slightly dominate.

\begin{figure*}
\includegraphics[width=\linewidth,keepaspectratio]{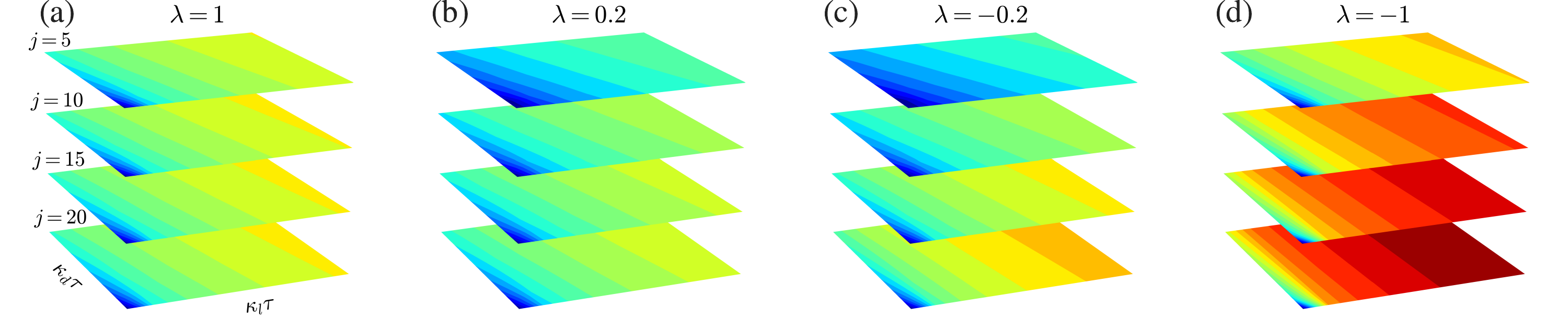}
\\
\includegraphics[width=0.7\linewidth]{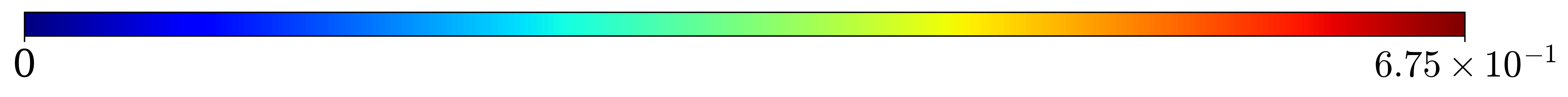}
    \caption{Stacked heatmaps of the average infidelity (logarithmic scale) for DKCS under combined dephasing and photon-loss channels. The error rates $\kappa_d \tau$ and $\kappa_l \tau$ are varied within $[0, 10^{-3}]$ and $|\alpha| = 1.5$.}
    \label{fig:heatmaps_DKCS}
\end{figure*}

\begin{figure*}
\includegraphics[width=\linewidth,keepaspectratio]{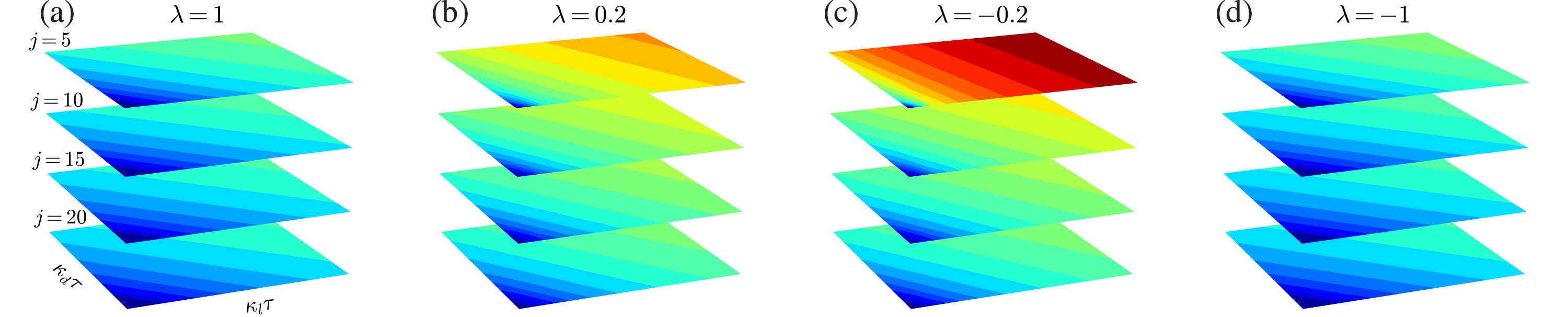}
\\
\includegraphics[width=0.7\linewidth]{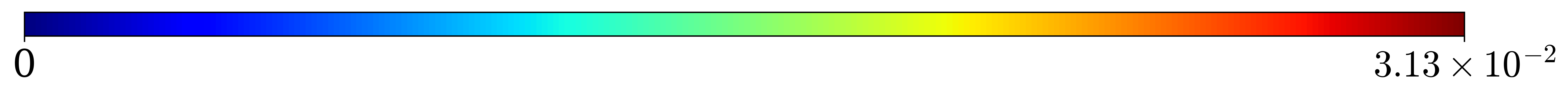}
    \caption{Stacked heatmaps of the average infidelity (logarithmic scale) for BGKCS under the effect of dephasing and photon loss, with error rates $\kappa_d \tau$, $\kappa_l \tau \in [0,10^{-3}]$ respectively. The state corresponds to $|\alpha| = 1.5$.}
    \label{fig:heatmaps_BGKCS}
\end{figure*}

However, according to the KL conditions, the performance of the code depends on both the light amplitude $|\alpha|$ and the Kerr nonlinearity parameter $\lambda$. 
Furthermore, as a direct consequence of Eq.~(\ref{anh_pos}), the preparation of DKCSs and  BGCSs introduces an additional degree of freedom, i.e., $j=\omega/\lambda\pm 1/2$. The impact of varying $j$, $\lambda$, and $\alpha$ on the average infidelity is illustrated in Figs.~\ref{fig:heatmaps_DKCS} and \ref{fig:heatmaps_BGKCS} for DKCSs and BGKCSs, respectively.

Figure~\ref{fig:heatmaps_DKCS} presents contour plots of the infidelity as a function of $\kappa_{l} \tau$ and $\kappa_{d} \tau$ for DKCSs with various values of the Kerr parameter $\lambda$. Plot (a) shows that increasing $\kappa_{l} \tau$ leads to a corresponding rise in infidelity. This trend is consistently observed across all DKCSs, regardless of the specific values of the parameters $j$ and $\lambda$. In particular, as $\lambda$ increases, a slight increase in infidelity is observed, as illustrated in plot (b). A similar behavior is evident for negative values of the Kerr parameter, as shown in plots (c) and (d). Furthermore, a comparison between plots (a) and (c), as well as between (b) and (d), reveals that the influence of the loss parameter is more pronounced in negatively deformed DKCSs.

Fig. \ref{fig:heatmaps_BGKCS} displays contour plots of the infidelity as a function of $\kappa_{l} \tau$ and $\kappa_{d} \tau$ for BGKCSs with various values of the Kerr parameter $\lambda$. Plot (a) indicates that both $\kappa_{l} \tau$ and $\kappa_{d} \tau$ contribute similarly to the increase in infidelity. This trend is consistently observed across all BGKCSs, independent of the specific values of $j$ and $\lambda$. Interestingly, as $\lambda$ increases, a slight \emph{decrease} in infidelity is observed, as shown in plot (b), in contrast to the behavior observed for DKCSs. A similar trend is seen for negative values of the Kerr parameter in plots (c) and (d). Additionally, a comparison between plots (a) and (c), as well as between (b) and (d), indicates that the effect of the loss parameter is more pronounced in negatively deformed BGKCSs.

\subsection{Dynamics: state fidelity evolution}\label{subsec:Dynamics}
As previously mentioned, the average fidelity is independent of the input state. Nevertheless, the dependence of individual quantum trajectories on the input state is a well-established phenomenon. Furthermore, the modeling of the noise channel, i.e., whether standard or Kerr operators,  constitutes another critical aspect that warrants detailed investigation.

To illustrate the performance of QEC using the Kerr cat state, we define the deviation from the initial state in terms of the overlap fidelity as
\begin{equation}\label{eq:fidelity}
    P(t) = 1 - \bra{\mathcal{C}_{f}^{+}} \hat{\rho}(t) \ket{\mathcal{C}_{f}^{+}},
\end{equation}
where $\hat{\rho}(t)$ denotes the density matrix of the system at time $t$, evolving according to the master equation~(\ref{eq:dynamics}).

Assuming the system is initially prepared in the state $\ket{\mathcal{C}_{f}^{+}}$, we consider the application of a loss operator at a specific time. The overlap fidelity defined in Eq.~(\ref{eq:fidelity}) thus quantifies the effect of the particle loss as well as the subsequent recovery process on the cat code.
 
\begin{figure*}
    \includegraphics[width=\linewidth,keepaspectratio]{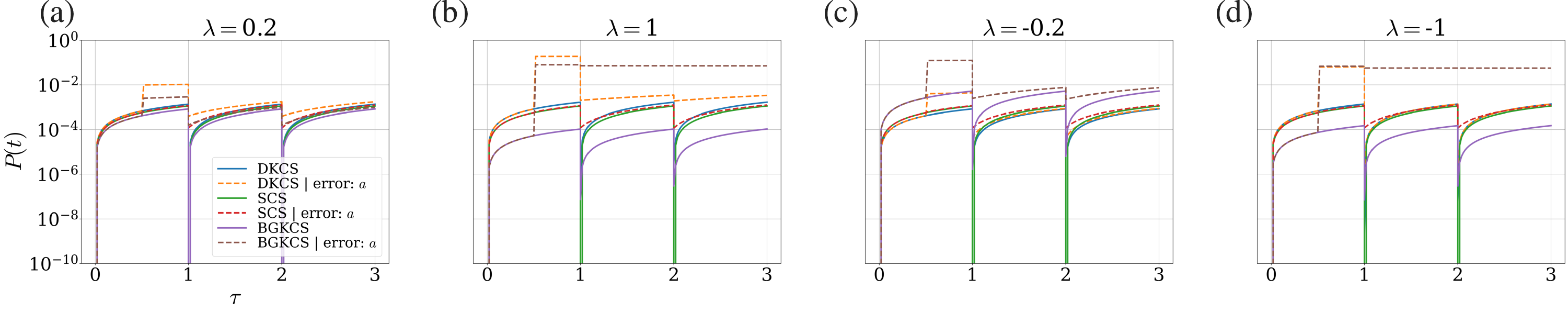}
    \caption{Evolution plot of the fidelity (logarithmic scale) of DKCS, BGKCS and SCS for values $j=5$ (not applicable to SCS) and $|\alpha|=1.5$. The initial logical state $1/2\left( \ket{\mathcal{C}^{+}}_{f}\bra{\mathcal{C}^{+}}_{f} + \ket{\mathcal{C}^{-}}_{f}\bra{\mathcal{C}^{-}}_{f} \right)$ is prepared and evolved under the noise channel of Eq. (\ref{eq:noise_channel}) with loss and dephasing rates $\kappa_l, \kappa_d = 10^{-3}$. In all of the evolution trajectories the optimized recovery channel of Eq. (\ref{eq:recovery_channel}) is applied to the state at time $\tau \in \{1, 2\}$ with $\tau = t/\kappa $. In the  dotted-line evolution plots a photon loss error is induced at $\tau= 0.5$, then measured and corrected at $\tau= 1$.}
    \label{fig:evolution_standard_noise}
\end{figure*}

Fig. \ref{fig:evolution_standard_noise} illustrates the time evolution of the quantity $P(t)$ for various cat states—namely, the DKCSs, SCSs, and BGKCSs—under different signs and magnitudes of the Kerr parameter. A recovery operation is applied at times $\tau = 2$ and $\tau = 3$, while a particle loss operator is applied at $\tau = 0.5$.
As shown in plot (a), for small positive values of the Kerr parameter, the dynamics of all cat states exhibit similar behavior, regardless of whether the system is subjected to particle loss followed by a combined dephasing and loss channel, or only to the combined dephasing and loss channel. However, as the magnitude of the Kerr parameter increases, the recovery operation plays a more significant role in mitigating quantum noise. In particular, the BGKCS demonstrates superior robustness, whether the recovery channel is applied following the combined dephasing and loss noise, or if only the loss and dephasing channels are present.
As shown in plot (c), the performance of the recovery operation under both scenarios closely resembles that observed in plot (a), with the exception that the DKCS exhibits a slightly improved fidelity. Plot (d) presents behavior similar to that of Plot (b), but for a negative value of the Kerr parameter.
However, when the Kerr nonlinearity is used to model the environment, the effectiveness of the recovery channel improves significantly. For further details, see Appendix~\ref{sec:Broader_scope}.

\section{Effective Dynamics in the Kerr-Displaced Frame}\label{engineering}

In this section we present the main results of the effective dynamics
analysis in the Kerr-displaced frame; the complete derivations and
technical details underlying all calculations presented here are given in
Appendix~\ref{app_C}. Our starting point is the two-photon--driven
Kerr-nonlinear resonator Hamiltonian, expressed in a frame rotating at the
resonator frequency and in the presence of single-photon loss,
\begin{align}
    H &= \Omega \hat{A}_{\pm}^{\dagger} \hat{A}_{\pm}
    + \varepsilon_{p}^{\ast} \hat{A}_{\pm}^{2}
    + \varepsilon_{p} \hat{A}_{\pm}^{\dagger\, 2}
    + \mathcal{D}[\hat{A}_{\pm}].
\end{align}
The corresponding Lindblad master equation takes the form
\begin{align}\label{dynamic_a}
    \dot{\rho}
    = -i\left(\hat{H}_{eff}\hat{\rho} - \hat{\rho}\hat{H}_{eff}^{\dagger}\right)
    + \frac{\kappa_{l}}{2} \hat{A}_{\pm}^{\dagger} \hat{\rho} \hat{A}_{\pm},
\end{align}
with the effective non-Hermitian Hamiltonian
\begin{align}\label{Hamiltonian_eff}
    \hat{H}_{eff}
    = \left(\Omega + i\frac{\kappa_{l}}{2} \right)\hat{A}_{\pm}^{\dagger}\hat{A}_{\pm}
    + \varepsilon_{p}^{\ast} \hat{A}_{\pm}^{2}
    + \varepsilon_{p} \hat{A}_{\pm}^{\dagger\, 2}.
\end{align}
Shifting to the Kerr-displaced frame via the relation~\eqref{dis},
\begin{align}
    \hat{H}^{\prime \, \pm}_{\mathrm{eff}}
    =
    D^{\dagger}(\alpha; j,\lambda_{\pm})
    \hat{H}_{\mathrm{eff}}
    D(\alpha; j,\lambda_{\pm}),
\end{align}
the transformed effective Hamiltonian becomes
\begin{align}
\hat{H}^{\prime \, \pm}_{\mathrm{eff}}
&=
K_{A_{\pm}^{\dagger}A_{\pm}}
\,\hat{A}_{\pm}^{\dagger}\hat{A}_{\pm}
+
K_{A_{\pm}^{2}}
\,\hat{A}_{\pm}^{2}
+
K_{A_{\pm}^{\dagger 2}}
\,\hat{A}_{\pm}^{\dagger 2}
\nonumber\\
&\quad
+
K_{A_{\pm}M_{\pm}}
\,\hat{A}_{\pm}\hat{M}_{\pm}
+
K_{A_{\pm}^{\dagger}M_{\pm}}
\,\hat{A}_{\pm}^{\dagger}\hat{M}_{\pm}
\nonumber\\
&\quad
+
K_{M_{\pm}A_{\pm}}
\,\hat{M}_{\pm}\hat{A}_{\pm}
+
K_{M_{\pm}A_{\pm}^{\dagger}}
\,\hat{M}_{\pm}\hat{A}_{\pm}^{\dagger}
\nonumber\\
&\quad
+
K_{M^{2}}
\,\hat{M}_{\pm}^{2}.
\end{align}
Here we have omitted the constant contribution associated with the Casimir
operator, since it produces only an overall energy shift of the
non-Hermitian effective Hamiltonian. Retaining terms up to second order in
the deformation parameter,
$\mathcal{O}(\lambda^{2})$, and choosing the displacement parameter
$\alpha$ such that all single-photon driving terms vanish, we obtain the
following approximate solutions.

For positive Kerr parameters, the displacement amplitude satisfies
\begin{align}
r_{+}^{2}
\simeq
\frac{1}{\lambda}
\frac{
\displaystyle
\frac{\varepsilon}{|\tilde{\Omega}|}
-
\frac{j}{2j+1}
}{
\displaystyle
\frac{1}{2}
-
\frac{\varepsilon \Omega^{2}}{|\tilde{\Omega}|^{3/2}}
},
\end{align}
with the corresponding phase relation
\begin{align}
\tan(2\theta_{+}+\phi)
\simeq
-\frac{\kappa_{l}}{2\Omega}
\left[
1
+
\frac{
\displaystyle
\frac{\varepsilon}{|\tilde{\Omega}|}
-
\frac{j}{2j+1}
}{
\displaystyle
\frac{1}{2}
-
\frac{\varepsilon \Omega^{2}}{|\tilde{\Omega}|^{3/2}}
}
\right].
\end{align}
For negative Kerr parameters, the analogous expressions are
\begin{align}
r_{-}^{2}
&\simeq
\frac{
(2j-1)^{2}\varepsilon^{2}
-
j^{2}\tilde{\Omega}^{2}
}{
\lambda
\left(
\frac{j^{2}\kappa_{l}^{2}}{2}
-
j\tilde{\Omega}^{2}
\right)
},
\label{eq_r2-}
\end{align}
and
\begin{align}
\tan(2\theta_{-}+\phi)
&\simeq
-\frac{\kappa_{l}}{2\Omega}
\left[
1
+
\left(
\frac{
(2j-1)^{2}\varepsilon^{2}
-
j^{2}\tilde{\Omega}^{2}
}{
\frac{j^{2}\kappa_{l}^{2}}{2}
-
j\tilde{\Omega}^{2}
}
\right)
\right],
\label{eq_phase2}
\end{align}
where $\tilde{\Omega}=\Omega+i\kappa_{l}/2$.

With these choices, the transformed effective Hamiltonian reduces to
\begin{align}\label{H_eff}
\hat{H}^{\prime\,\pm}_{\mathrm{eff}}
&=
\tilde{\Omega}_{\mathrm{eff}}
\hat{A}_{\pm}^{\dagger}\hat{A}_{\pm}
+
\varepsilon_{\mathrm{eff}}
\hat{A}_{\pm}^{\dagger 2}
+
\varepsilon_{\mathrm{eff}}^{\ast}
\hat{A}_{\pm}^{2}
\nonumber\\
&\quad-
\frac{\lambda^{2}r_{\pm}}{2}
\left[
\tilde{\Omega}e^{i\phi}
+
\cos(\phi+2\theta_{\pm})
\right]
\hat{A}_{\pm}^{\dagger}\hat{a}^{\dagger}\hat{a}
\nonumber\\
&\quad-
\frac{\lambda^{2}r_{\pm}}{2}
\left[
\tilde{\Omega}e^{-i\phi}
+
\cos(\phi+2\theta_{\pm})
\right]
\hat{a}^{\dagger}\hat{a}\hat{A}_{\pm}.
\end{align}
Here the first line contains an effective detuning,
\begin{align}
\tilde{\Omega}_{\mathrm{eff}}
&=
\tilde{\Omega}(1\pm\lambda r_{\pm}^{2})
+
2\lambda r_{\pm}^{2}\varepsilon
\cos(\phi+2\theta_{\pm}),
\end{align}
and an effective two-photon parametric drive with amplitude
\begin{align}
\varepsilon_{\mathrm{eff}}
&\simeq
\varepsilon e^{-i\phi}
(1\pm\lambda r_{\pm}^{2})
+
\frac{\lambda r_{\pm}^{2}}{2}
\tilde{\Omega}e^{-2i\theta_{\pm}},
\end{align}
while the remaining terms represent residual nonlinear corrections that
are second order in the deformation parameter.

A key advantage of the Kerr-deformed framework is that the effective
two-photon parametric drive can be completely eliminated by choosing the
external drive parameter $\varepsilon_{p}=\varepsilon_{c}$, with
\begin{align}
\varepsilon_{c}
=\frac{\lambda r^{2}}
{1-\lambda r^{2}}|\tilde{\Omega}|,
\end{align}
together with the phase condition
\begin{align}
\tan\phi_{c}
=
\frac{
\kappa_{l}
+
2\Omega\tan(2\theta)
}{
\Omega
-
2\kappa_{l}\tan(2\theta)
}.
\end{align}
Under these conditions the effective parametric coefficient
$\varepsilon_{\mathrm{eff}}$ vanishes identically, so that the vacuum
state $\ket{0}$ remains an eigenstate in the Kerr-displaced frame even in
the presence of finite single-photon loss. This is in sharp contrast to
the conventional non-deformed Kerr-displacement framework, where the
analogous cancellation condition imposes the restrictive constraint
$\kappa_{l}\ll 8\lambda r^{2}$ (see Ref.~\cite{puri2017engineering}),
effectively requiring the single-photon dissipation rate to be negligibly
small and therefore limiting applicability to near-ideal, weakly
dissipative environments. In the Kerr-deformed framework, no such
constraint arises: the conditions on $\varepsilon_{c}$ and $\phi_{c}$
derived above hold for \emph{arbitrary} values of $\kappa_{l}$, meaning
that the effective parametric drive can be suppressed regardless of how
strongly the system couples to its environment. The present scheme is
therefore applicable across the full range of dissipative regimes,
from weakly damped resonators to strongly dissipative platforms, without
any fine-tuning of the loss rate. Equivalently, in the laboratory frame,
the coherent states $\ket{\pm\alpha_{0}, \lambda_{\pm}}_D$ constitute
degenerate eigenstates of the effective Hamiltonian for any value of
$\kappa_{l}$.

The $\mathbb{Z}_{2}$ symmetry of the system is preserved under the
Kerr-deformed transformation. Specifically, the generalized parity
operator
\begin{align}
\mathbb{Z}_{p}
=
\exp\!\left(
i\pi \hat{a}^{\dagger}\hat{a}/p
\right)
\end{align}
commutes with the effective Hamiltonian, and for the physically relevant
case $p=2$ one obtains
\begin{align}
[\mathbb{Z}_{2},\hat{H}_{\mathrm{eff}}]=0.
\end{align}
The steady states of the Hamiltonian are therefore Kerr-deformed cat
states, i.e., coherent superpositions of the form
\begin{align}
\ket{\mathcal{C}_{\pm},\lambda_{-}}
\propto
\ket{\alpha,\lambda_{-}}_{D}
\pm
\ket{-\alpha,\lambda_{-}}_{D}.
\end{align}

We now analyze the action of the quantum jump operators on the KDCSs,
which determines the structure of the dissipative dynamics within the
cat-state manifold. Explicit calculation yields
\begin{align}\label{eq_apos_impact}
\hat{A}_{+}\ket{\alpha,\lambda_{+}}_{D}
&=
2j\sqrt{\frac{\lambda}{2}}
\tanh\!\left[\sqrt{\frac{\lambda}{2}}|\alpha|\right]
e^{i\theta}
\ket{\alpha,\lambda_{+}}_{D}
\nonumber\\
&\quad+
e^{i\theta}
\sqrt{\frac{\lambda}{2}}
\tanh\!\left[\sqrt{\frac{\lambda}{2}}|\alpha|\right]
\sech^{j}\!\left[\sqrt{\frac{\lambda}{2}}|\alpha|\right]
\nonumber\\
&\quad\times
\sum_{n=0}^{\infty}
\sqrt{
\frac{\Gamma(2j+n)}
{\Gamma(2j)n!}
}\,
n\,
\tanh^{n}\!\left[\sqrt{\frac{\lambda}{2}}|\alpha|\right]
e^{in\theta}
\ket{n},
\end{align}
and
\begin{align}\label{eq_aneg_impact}
\hat{A}_{-}\ket{\alpha,\lambda_{-}}_{D}
&=
2j\sqrt{\frac{\lambda}{2}}
\tan\!\left[\sqrt{\frac{\lambda}{2}}|\alpha|\right]
e^{i\theta}
\ket{\alpha,\lambda_{-}}_{D}
\nonumber\\
&\quad+
e^{i\theta}
\sqrt{\frac{\lambda}{2}}
\tan\!\left[\sqrt{\frac{\lambda}{2}}|\alpha|\right]
\sec^{j}\!\left[\sqrt{\frac{\lambda}{2}}|\alpha|\right]
\nonumber\\
&\quad\times
\sum_{n=0}^{2j}
\sqrt{
\frac{(2j+n)!}
{(2j)!n!}
}\,
n\,
\tan^{n}\!\left[\sqrt{\frac{\lambda}{2}}|\alpha|\right]
e^{in\theta}
\ket{n}.
\end{align}
Each expression consists of two contributions: a leading term proportional
to the original KDCS and a subleading summation over Fock states. In the
weak-Kerr regime $\lambda\ll 1$, or equivalently for small amplitudes
$|\alpha|\ll 1$, the hyperbolic and circular functions admit the
expansions
\begin{align}
\tanh\!\left[\sqrt{\frac{\lambda}{2}}|\alpha|\right]
&=
\sqrt{\frac{\lambda}{2}}|\alpha|
-\frac{1}{3}
\left(
\sqrt{\frac{\lambda}{2}}|\alpha|
\right)^3
+\mathcal{O}(\lambda^{5/2}),
\\
\tan\!\left[\sqrt{\frac{\lambda}{2}}|\alpha|\right]
&=
\sqrt{\frac{\lambda}{2}}|\alpha|
+\frac{1}{3}
\left(
\sqrt{\frac{\lambda}{2}}|\alpha|
\right)^3
+\mathcal{O}(\lambda^{5/2}).
\end{align}
It follows that the subleading summation terms in
Eqs.~\eqref{eq_apos_impact} and~\eqref{eq_aneg_impact} scale as
$\mathcal{O}(\lambda^{3/2})$, whereas the leading terms proportional to
the KDCSs scale as $\mathcal{O}(\lambda)$. Consequently, to leading order
in $\lambda$ the subleading contributions may be safely neglected, giving
the approximate eigenvalue-like relations
\begin{align}
\hat{A}_{+}\ket{\alpha,\lambda_{+}}_{D}
&\simeq
2j\sqrt{\frac{\lambda}{2}}
\tanh\!\left[\sqrt{\frac{\lambda}{2}}|\alpha|\right]
e^{i\theta}
\ket{\alpha,\lambda_{+}}_{D},
\\
\hat{A}_{-}\ket{\alpha,\lambda_{-}}_{D}
&\simeq
2j\sqrt{\frac{\lambda}{2}}
\tan\!\left[\sqrt{\frac{\lambda}{2}}|\alpha|\right]
e^{i\theta}
\ket{\alpha,\lambda_{-}}_{D}.
\end{align}
These relations imply that the quantum jump term
$\kappa_l \hat{A}_{\pm}\hat{\rho} \hat{A}_{\pm}^{\dagger}$
in the master equation induces nondeterministic jumps between the
even- and odd-parity Kerr cat states,
$\ket{\mathcal{C}^{+}_{\alpha_{0}},\lambda_{\pm}}_{D}$ and
$\ket{\mathcal{C}^{-}_{\alpha_{0}},\lambda_{\pm}}_{D}$.
This process produces decoherence \emph{within} the Kerr cat-state
manifold, while the dynamics remain confined to the degenerate subspace
$\mathrm{span}\{\ket{\mathcal{C}^{\pm}_{\alpha_{0}},\lambda_{\pm}}_{D}\}$,
thereby preventing leakage out of the protected subspace.

In summary, we have established two complementary results. First, the
action of the nonlinear quantum jump operators on the KDCSs preserves the
underlying coherent-state structure in the weak-deformation regime: for
$\lambda\ll1$, the dominant contribution to
$\hat{A}_{\pm}\ket{\alpha,\lambda_{\pm}}_{D}$ remains proportional to the
original KDCS, while all corrections are parametrically suppressed as
$\mathcal{O}(\lambda^{3/2})$. As a consequence, single-photon loss induces
only intra-manifold transitions between even- and odd-parity Kerr cat
states, while leakage outside the protected subspace is suppressed to
higher order in the deformation parameter. Second, and crucially, the
conditions required to eliminate the effective parametric drive impose no
constraint on the single-photon dissipation rate $\kappa_{l}$, in sharp
contrast to the standard non-deformed framework where $\kappa_{l}\ll
8\lambda r^{2}$ is required. Together, these two results constitute the
central mechanism by which Kerr-deformed cat-state manifolds achieve
robust dissipative stabilization under engineered two-photon driving:
the protected subspace is preserved for arbitrary dissipation strength,
and coherence within the manifold is maintained to leading order in the
deformation parameter.

\section{Conclusion}\label{conclusion}
In this work, we have constructed and analyzed a family of
Schr\"{o}dinger cat codes derived from Kerr-type coherent states,
including displacement Kerr coherent states (DKCSs) and Barut-Girardello
Kerr coherent states (BGKCSs). By exploring both signs of the Kerr
nonlinearity, we revealed two distinct and tunable families of cat codes
within each class, yielding four generalized Kerr cat-state encodings in
total. These constructions recover important limiting cases, such as
$\mathfrak{su}(2)$ and  $\mathfrak{su}(1,1)$ coherent states, thereby establishing a unified
and versatile framework for encoding logical qubits in a single bosonic
mode.

We have demonstrated that Kerr-based cat codes exhibit superior robustness
to both photon loss and dephasing noise compared to conventional
two-component Schr\"{o}dinger cat codes. This resilience arises from the
nontrivial structure of the code space under the action of typical error
channels. Unlike standard cat codes, in which a single photon-loss event
maps the logical states to orthogonal states without intrinsic correction
capability, the nonlinear phase-space geometry of Kerr cat states ensures
that neither particle-loss nor dephasing errors transform the logical
codewords into themselves or into orthogonal states. As a result, both
error channels remain simultaneously detectable and correctable via
optimized recovery protocols. By employing a master-equation-based model
and evaluating both the Knill-Laflamme (KL) conditions and channel
fidelity under Lindblad dynamics, we have established the effectiveness of
a semi-autonomous QEC strategy that applies recovery operations
periodically and unconditionally. We further provided a practical method
to compute the optimal recovery map for each class of Kerr cat states using
convex optimization, highlighting the critical role of system
parameters, particularly the coherent-state amplitude, Kerr strength, and
the ratio of optical frequency to Kerr nonlinearity $j = \omega/\lambda
\pm 1/2$, in simultaneously suppressing both error channels to a negligible
level.

We have also investigated Kerr-deformed coherent-state (KDCS) manifolds
under engineered two-photon driving in the presence of single-photon loss.
By analyzing the action of nonlinear jump operators on the KDCS manifold,
we showed that in the weak-deformation regime, the dominant dissipative evolution remains effectively confined to the degenerate subspace spanned
by these states, and that the interplay between Kerr deformation and
appropriately tuned two-photon driving can strongly suppress single-photon
decoherence. Consequently, the KDCSs emerge as effective steady states of
the driven-dissipative dynamics, with leakage outside the protected
manifold appearing only as higher-order corrections in the deformation
strength. These results establish a direct connection between nonlinear
algebraic deformations, engineered dissipation, and the stabilization of
protected bosonic quantum states—complementing the error-correction
analysis and further demonstrating the robustness of the Kerr framework
across different physical regimes.

To further connect our theoretical framework with physically relevant
settings, we carried out an explicit variational analysis within the model
of a Kerr nonlinear resonator (KNR). By evaluating the expectation value
of the KNR Hamiltonian with respect to both BGKCSs and DKCSs and
minimizing over the complex amplitude, we identified the configurations
that extremize the energy for each cat-state type. This treatment yields
analytical expressions for the optimal amplitudes for different signs of
the Kerr nonlinearity, clarifying how the structure of the emergent cat
states depends on the system parameters (see Appendix \ref{KNR}).

From an algebraic perspective, the present construction can be placed within a broader coherent-state framework. The coherent states considered here arise from a nonlinear deformation of the underlying ladder-operator algebra and may be viewed as a generalization of both $\mathfrak{su}(2)$ and $\mathfrak{su}(1,1)$ coherent-state formalisms. In particular, within the standard bosonic realization of $\mathfrak{su}(1,1)$, the minimal representation ($j=1/2$) is directly associated with squeezed-state constructions generated by Gaussian squeezing operators. In this sense, Gaussian coherent and squeezed-state constructions arise as special limiting cases within the broader algebraic hierarchy. The deformed annihilation and creation operators, together with the associated nonlinear displacement operator $D_f(\alpha; j, \lambda_{\pm})$, extend this hierarchy by introducing a nonlinear deformation of the underlying ladder-operator algebra rather than a linear canonical (Gaussian) transformation. Consequently, Gaussian squeezed-cat encodings appear only as embedded limiting cases of the formalism, whereas the generic regime explored in this work corresponds to intrinsically non-Gaussian coherent-state manifolds that extend beyond conventional $\mathfrak{su}(2)$ and $\mathfrak{su}(1,1)$ coherent-state structures.

From the perspective of quantum error correction, the present family of cat-code constructions offers a broader parameter landscape for satisfying the KL conditions. Whereas conventional bosonic encodings typically require operation within relatively constrained parameter regimes, the KDCSs and BGKCSs considered here achieve approximate error-correcting conditions through different physical mechanisms and in distinct regions of parameter space. In particular, displaced Kerr-cat codes exhibit favorable error-correcting properties for comparatively small Kerr nonlinearities and moderate state amplitudes, while BG Kerr-cat codes satisfy the KL criteria in regimes of larger coherent amplitudes and stronger nonlinear deformations. This flexibility highlights the versatility of the algebraic framework and provides multiple experimentally relevant routes toward robust bosonic quantum-information encoding. Consequently, within the parameter regimes considered here and commonly employed in the literature, the proposed Kerr-cat constructions can exhibit improved performance relative to existing bosonic cat-code encodings in certain experimentally relevant regimes. We emphasize, however, that a comprehensive optimization over all encoding parameters has not been performed, and such an analysis could further refine the quantitative comparison.

Although the primary focus of this work is theoretical, the nonlinear
coherent states studied here have well-established routes toward
experimental realization. Laser-cooling and ion-trapping techniques
provide a natural platform in which the internal electronic states of a
trapped ion couple to its quantized vibrational
motion~\cite{deMatosFilho1996}. By appropriately tuning the driving laser
fields, one can engineer the vibrational dynamics to generate nonlinear
coherent states, with a key advantage being the extremely weak coupling of
vibrational modes to the external environment, enabling high-stability
preparation and observation~\cite{deMatosFilho1996,Vogel1995,dehdashti2024enhancing}.
In a complementary setting, Kerr coherent states may arise in cavity-QED
architectures, where a two-level atom interacts with a single-mode
quantized field via an intensity-dependent Jaynes-Cummings model while
being driven by a strong classical field. Under suitable conditions, the
system evolution produces superpositions of Kerr coherent states, and
conditional measurements on the atomic state can project the cavity field
onto the desired class of states~\cite{dehdashti2024enhancing}. These
platforms are closely related to the physical settings considered in
Kerr-cat qubit proposals, where a combination of nonlinear interactions
and coherent driving stabilizes cat-like states in a controllable manner.
Beyond these quantum platforms, photonic lattices composed of evanescently
coupled waveguides with engineered coupling profiles provide a classical
optical analogue for visualizing the underlying physics~\cite{PerezLeija2010,Christodoulides2003}.
Within such Glauber-Fock lattices, the excitation of individual
waveguides maps onto Fock states, while the spatial propagation of light
reproduces the probability amplitudes associated with Kerr coherent states.
Although this does not constitute a direct realization of the quantum
states, it offers an intuitive and experimentally accessible framework for
probing the key mechanisms underlying our
construction~\cite{dehdashti2024enhancing}.

These findings identify the central mechanism underlying the robust dissipative stabilization of Kerr-deformed cat-state manifolds under engineered two-photon driving: the protected subspace remains preserved for arbitrary dissipation strengths, while coherence within the manifold is maintained to leading order in the deformation parameter. Taken together, our results pave the way toward experimentally feasible, noise-resilient bosonic quantum codes. The Kerr parameter provides a flexible control knob for optimizing code performance, while the tunability of Kerr-based constructions enables a unified framework interpolating between $\mathfrak{su}(2)$ and  $\mathfrak{su}(1,1)$-coherent state limits. Several important directions remain open for future work, including the extension of Kerr cat encodings to multi-qubit architectures, the development of fault-tolerant logical gate sets, and the experimental realization of the stabilization schemes proposed here.

\begin{comment}

\end{comment}

\begin{acknowledgments}
The authors acknowledge the financial support by the Federal Ministry of Education and Research of Germany in the programme of “Souverän. Digital. Vernetzt.”. Joint project 6G-life, project identification number: 16KISK002 and via projects 16KISQ039, 16KISQ093, 16KIS0948, 16KIS1598K, and the DFG Emmy-Noether program under grant number NO 1129/2-1.
PvL further acknowledges support from the EU/BMBF via CLUSTEC and from the BMFTR (former BMBF) in Germany through QR.N, PhotonQ, and QuKuK.\\
This work was supported by the Munich Quantum Valley, which is supported by the Bavarian state government with funds from the Hightech Agenda Bayern Plus.
\end{acknowledgments}

\bibliography{Refs}

\appendix

\section{ System Hamiltonian}\label{KNR}
The starting point for engineering quantum states of light in a Kerr–nonlinear resonator is the two-photon–driven Kerr nonlinear resonator (KNR) Hamiltonian,
\begin{equation}\label{KNR_Hamilton}
\hat{H} = \Delta a^{\dagger} a
\pm K a^{\dagger 2} a^{2}
+ \left( \varepsilon_{2} a^{\dagger 2}
+ \varepsilon_{2}^{\ast} a^{2} \right),
\end{equation}
where $\Delta$ denotes the detuning between the resonator and the rotating frame, $a$ and $a^{\dagger}$ are the annihilation and creation operators of the resonator mode, $K$ is the Kerr nonlinearity strength, and $\varepsilon_{2}$ represents the amplitude of the two-photon drive. Most previous studies have focused on more specific parameter regimes where analytic simplifications and approximate descriptions become possible~\cite{goto2016universal,puri2017engineering,hajr2024high,grimm2020stabilization,xu2022engineering,qing2026quantum}.

In the following, we relax this commonly assumed condition and consider a more general parameter regime. In this regime, the energy spectrum becomes strongly anharmonic, and the spacing between successive energy levels is no longer uniform.

To account for this structure, we introduce Kerr-modified annihilation and creation operators defined as
\begin{align}
\hat{A}_{\pm}
&= \sqrt{\frac{|\lambda|}{2}} \,
\hat{a}\sqrt{2j \mp 1 \pm \hat{n}}, \\
\hat{A}_{\pm}^{\dagger}
&= \sqrt{\frac{|\lambda|}{2}} \,
\sqrt{2j \pm 1 \mp \hat{n}} \, \hat{a}^{\dagger},
\end{align}
where $\hat{n}=a^{\dagger}a$ is the number operator, and the operators $\hat{A}_{+}$ and $\hat{A}_{-}$ correspond respectively to positive and negative values of the Kerr parameter.

In terms of these operators, the KNR Hamiltonian can be written in the compact form
\begin{align}
\hat{H}
= \Omega \hat{A}_{\pm}^{\dagger}\hat{A}_{\pm}
+ 
\varepsilon_{2} \hat{a}_{\pm}^{\dagger 2}
+ \varepsilon_{2}^{\ast} \hat{a}_{\pm}^{2},
\end{align}
where $\Omega$ denotes the effective mode frequency in the Kerr-modified representation, $\Delta=\Omega \, j |\lambda|$, and $2K=\Omega\, |\lambda|$.

The cat qubit spanned by the BG coherent states 
$\ket{\pm \alpha,\lambda}_{BG}$ 
can be analyzed within a variational framework for a Kerr oscillator
subject to a parametric two-photon drive.
Specifically, we evaluate the expectation value of the Hamiltonian
with respect to the BGCSs and minimize it with respect to the
complex amplitude $\alpha$.

Writing
\begin{equation}
\alpha = r e^{i\vartheta}, 
\qquad 
\varepsilon_{2} = |\varepsilon_{2}| e^{i\phi},
\end{equation}
and expanding up to fourth order in the small parameter $\alpha$,
the minimization with respect to the phase yields the phase-locking condition
\begin{equation}
2\vartheta = \phi + 2n\pi, 
\qquad n \in \mathbb{Z}.
\end{equation}

Minimization with respect to the amplitude then gives the optimal values of $\alpha$.
For positive Kerr parameter ($\lambda>0$), one obtains
\begin{align}
\alpha=\pm \sqrt{
\frac{
j \sqrt{2j+1}\, \lambda^2
}{4|\varepsilon_2|
\left(
\frac{1}{\sqrt{2j}} - \frac{1}{\sqrt{2j+2}}
\right)}
\left[
\Omega+
\frac{4|\varepsilon_2|}
{\lambda \sqrt{(2j)(2j+1)}}
\right]}.
\end{align}

For negative Kerr parameter ($\lambda<0$), the minimization instead yields
\begin{equation}
\alpha=\pm\sqrt{
\frac{
j^{3/2}\lambda^{2}
}{
2^{3/2}|\varepsilon_{2}|
\left(
\frac{4j^{2}}{\sqrt{2j-1}}
-\frac{1}{\sqrt{2j+1}}
\right)
}\left[
\Omega-
\frac{4|\varepsilon_2|}
{|\lambda|\sqrt{(2j)(2j+1)}}
\right]}.
\end{equation}
Consequently, coherent superpositions of the BGCSs, namely BG cat states,
emerge as energy-minimizing configurations of the Kerr nonlinear resonator
within appropriate parameter regimes.

In close analogy, the cat qubit spanned by the DK coherent states
$\ket{\pm \alpha,\lambda}_{D}$
can be treated within the same variational procedure.
By minimizing the expectation value of the Hamiltonian
with respect to $\alpha$, one determines the optimal amplitudes.

For positive Kerr parameter ($\lambda>0$), one finds
\begin{align}
\alpha
=\pm
\sqrt{\frac{2}{\lambda}}\,
\sinh^{-1}\!\left(
\sqrt{
\frac{
2(2j+1)\lvert \varepsilon_{2}\rvert
-
j\,\Omega\lambda
}{
(2j+1)\big(\Omega\lambda-4\lvert \varepsilon_{2}\rvert\big)
}
}
\right).
\end{align}

For negative Kerr parameter ($\lambda<0$), one obtains
\begin{equation}
\alpha
=
\pm
\sqrt{\frac{2}{|\lambda|}}\,
\sin^{-1}\!\left(
\sqrt{
\frac{
j\,\Omega|\lambda|
-
2(2j-1)\lvert \varepsilon_{2}\rvert
}{
(2j-1)\big(\Omega|\lambda|-4\lvert \varepsilon_{2}\rvert\big)
}
}
\right).
\end{equation}
Accordingly, coherent superpositions of the DKCSs 
also arise as energy-minimizing configurations of the Kerr nonlinear resonator
for suitable choices of the system parameters.
\section{ Bogoliubov transformation of the Kerr operators}
The goal of this subsection is to  calculate the impact of the Kerr displacement operator on the Kerr annihilation and creation operator, i.e., 
\begin{align}\label{d1}
    \hat{A}_{\pm}(\alpha)=\hat{D}_{\pm}(\alpha) \hat{A}_{\pm}  \hat{D}_{\pm}(\alpha)
\end{align}
in which  the Kerr displacement operator is defined by the relation \eqref{dis}.
The first step in evaluating the right side of this equation is to define the Kerr 
generator 
\begin{align}
    G(\alpha)=-i \left[\alpha \hat{A}_{\pm}^{\dagger}-\alpha^{\ast}\hat{A}_{\pm}\right]
\end{align}
Then, we are able to rewrite the equation \eqref{d1}, using ancilla parmaeter $\tau \in [0,1]$, as
\begin{align}
    \hat{A}_{\pm}= e^{i\tau\hat{G}_{\pm}(\alpha)} \hat{A}_{\pm} e^{-i\tau\hat{G}_{\pm}(\alpha)}
\end{align}
The interpolation formula has the form of a time evolution with Hamiltonian $\hat{G}_{\pm}$, so the interpolating operators satisfy the
Heisenberg-like equations:
\begin{equation}
\frac{d\hat{A}_{\pm}}{d\tau}= i[\hat{A}_{\pm},\hat{G}_{\pm}]
\end{equation}
Using the commutation relations \cite{dehdashti2013coherent,dehdashti2025quantum}
\begin{equation}\label{eq_algebraic}
[\hat{A}_{\pm}, \hat{A}_{\pm}^\dagger] = \mp 2\hat{M}_{\pm}, \qquad
[\hat{A}_{\pm}, \hat{M}_{\pm}] = \frac{|\lambda|}{2}\hat{A}_{\pm},
\end{equation}
we obtain a closed system of coupled differential equations:
\begin{align}
\frac{d}{d\tau}\hat{A}_{\pm}(\tau) &= - 2\alpha \, \hat{M}_{\pm}(\tau), \\
\frac{d}{d\tau}\hat{A}_{\pm}^\dagger(\tau) &= - 2\alpha^\ast \, \hat{M}_{\pm}(\tau), \\
\frac{d}{d\tau}\hat{M}_{\pm}(\tau) &= \mp \frac{|\lambda|}{2}
\left( \alpha \hat{A}_{\pm}^\dagger(\tau) + \alpha^\ast \hat{A}_{\pm}(\tau) \right).
\end{align}
We define $\omega = \sqrt{2|\lambda|}\, |\alpha|$ and $\hat{X}_{\pm}=\alpha \hat{A}_{\pm}^\dagger(\tau) + \alpha^\ast \hat{A}_{\pm}(\tau)$. Now, for the positive Kerr parameter, we find  
\begin{align}
\hat{M}_{+}(\tau) =  \cosh(\omega \tau)\hat{M}_{+} -\frac{\lambda}{2\omega}  \sinh(\omega \tau) \hat{X}_{+}\\
\hat{X}_{+}(\tau) = \hat{X}_{+} \cosh(\omega \tau) - \frac{2\omega}{\lambda}  \sinh(\omega \tau) \hat{M}_{+}
\end{align}
Therefore, we have 
\begin{align}\label{dis_tr}
\hat{D}_{+}^{\dagger}(\alpha)\,\hat{A}_{+}\,\hat{D}_{+}(\alpha)
&=\cosh^{2}(\sqrt{\frac{\lambda}{2}}|\alpha|) \hat{A}_{+}\nonumber\\
&+\frac{\alpha^{2}}{|\alpha|^{2}}\sinh^{2}(\sqrt{\frac{\lambda}{2}}|\alpha|) \hat{A}_{+}^{\dagger}\nonumber\\
&-\sqrt{\frac{\lambda}{2}}\,\frac{\alpha}{|\alpha|}\sinh(\sqrt{2\lambda}|\alpha|) \hat{M}_{+}.
\end{align}
For negative values of the Kerr parameter, we have 
\begin{align}
\hat{M}_{-}(\tau) = \hat{M}_{-} \cos(\omega \tau) + \frac{|\lambda|}{2\omega}  \hat{X}_{-} \sin(\omega \tau)\\
\hat{X}_{-}(\tau) = \hat{X}_{-} \cos(\omega \tau) - \frac{2\omega}{|\lambda|} \hat{M}_{-} \sin(\omega \tau)
\end{align}
and therefore, the transformation of the Kerr annihilation is given by
\begin{align}\label{dis_tr_neg}
\hat{D}_{-}^{\dagger}(\alpha)\,\hat{A}_{-}\,\hat{D}_{-}(\alpha)
&=\cos^{2}(\sqrt{\frac{\lambda}{2}}|\alpha|) \hat{A}_{-}\nonumber\\
&+\frac{\alpha^{2}}{|\alpha|^{2}}\sin^{2}(\sqrt{\frac{\lambda}{2}}|\alpha|) \hat{A}_{-}^{\dagger}\nonumber\\
&-\sqrt{\frac{\lambda}{2}}\,\frac{\alpha}{|\alpha|}\sin(\sqrt{2\lambda}|\alpha|) \hat{M}_{-}.
\end{align}

\section{Derivation of the Kerr-Displaced Effective Hamiltonian}\label{app_C}
We begin with the Kerr Hamiltonian in the presence of single-photon loss,
\begin{align}
    H &= \Omega \hat{A}_{\pm}^{\dagger} \hat{A}_{\pm}
    + \varepsilon_{p}^{\ast} \hat{A}_{\pm}^{2}
    + \varepsilon_{p} \hat{A}_{\pm}^{\dagger\, 2}
    + \mathcal{D}[\hat{A}_{\pm}].
\end{align}

The corresponding Lindblad equation can be written as
\begin{align}\label{dynamic_b}
    \dot{\rho}
    = -i\left(\hat{H}_{eff}\hat{\rho} - \hat{\rho}\hat{H}_{eff}^{\dagger}\right)
    + \frac{\kappa_{l}}{2} \hat{A}_{\pm}^{\dagger} \hat{\rho} \hat{A}_{\pm},
\end{align}
where the effective non-Hermitian Hamiltonian is given by
\begin{align}\label{Hamiltonian_eff_b}
    \hat{H}_{eff}
    = \left(\Omega + i\frac{\kappa_{l}}{2} \right)\hat{A}_{\pm}^{\dagger}\hat{A}_{\pm}
    + \varepsilon_{p}^{\ast} \hat{A}_{\pm}^{2}
    + \varepsilon_{p} \hat{A}_{\pm}^{\dagger\, 2}.
\end{align}

Using Eqs. \eqref{dis_tr} and \eqref{dis_tr_neg}, i.e., the Kerr displacement transformation, and neglecting constant energy shifts, the effective Hamiltonian becomes
\begin{align}
\hat{H}^{\prime}_{\mathrm{eff}} &=
K_{A_{\pm}^{\dagger}A_{\pm}}\,\hat{A}_{\pm}^\dagger \hat{A}_{\pm}
+ K_{A_{\pm}^2}\,\hat{A}_{\pm}^2
+ K_{A_{\pm}^{\dagger 2}}\,\hat{A}_{\pm}^{\dagger 2} \nonumber\\
& + K_{A_{\pm}M_{\pm}}\,\hat{A}_{\pm}
\hat{M}_{\pm}
+ K_{A_{\pm}^{\dagger} M_{\pm}}\,\hat{A}_{\pm}^\dagger \hat{M}_{\pm} \nonumber\\
& + K_{M_{\pm}A_{\pm}}\,\hat{M}_{\pm}\hat{A}_{\pm}
+ K_{M_{\pm} A_{\pm}^{\dagger}}\,\hat{M}_{\pm}\hat{A}_{\pm}^{\dagger}.
\end{align}
Now, using \eqref{eq_algebraic}
and 
\begin{align}
\hat{M}_{\pm} = \frac{|\lambda|}{2}\,(\hat{a}^{\dag}\hat{a} \mp j)
\end{align}
we rewrite 
\begin{align}
&K_{A_{\pm}^{\dag}M_{\pm}}\hat{A}_{\pm}^{\dagger}\hat{M}_{\pm}+ K_{M_{\pm}A_{\pm}^{\dag}}\hat{M}_{\pm}\hat{A}_{\pm}^{\dagger}=\mp \frac{|\lambda|}{2} \Big[j K_{A_{\pm}^{\dag}M_{\pm}}\nonumber\\
&\quad+(j\pm 1)K_{M_{\pm}A_{\pm}^{\dag}}\Big]\hat{A}_{\pm}^{\dagger}\nonumber\\
&\quad\pm\frac{|\lambda|}{2} \Big[ K_{A_{\pm}^{\dag}M_{\pm}}+K_{M_{\pm}A_{\pm}^{\dag}}\Big]\hat{A}_{\pm}^{\dagger}\hat{a}^{\dag}\hat{a}
\end{align}
In the following subsection, we take $\alpha_{0}$ satisfy the coefficient $F_{j}=0$, where $F_{j}=j K_{A_{\pm}^{\dag}M_{\pm}}+(1 \mp j)K_{M_{\pm}A_{\pm}^{\dag}}$.

\subsection{Positive Kerr parameters}
Now, by considering the coefficients,
\begin{align}
K_{M_{+}A_{+}^{\dag}}&=-\left[ \sqrt{\frac{\lambda}{2}}\sinh(\sqrt{2\lambda}|\alpha|) \right]\nonumber\\
&\times\Bigg[\left(\Omega + i\frac{\kappa_{l}}{2} \right)  \frac{\alpha}{|\alpha|} \sinh^2(\sqrt{\frac{\lambda}{2}}|\alpha|) \nonumber\\
&+ \varepsilon_{p}^*  \frac{\alpha^3}{|\alpha|^3}\sinh^2(\sqrt{\frac{\lambda}{2}}|\alpha|)
+ \varepsilon_{p} \frac{\alpha^*}{|\alpha|} \cosh^2(\sqrt{\frac{\lambda}{2}}|\alpha|) \Bigg]
\end{align}
and 
\begin{align}
K_{A_{+}^{\dag}M_{+}}&=-\left[ \sqrt{\frac{\lambda}{2}}\sinh(\sqrt{2\lambda}|\alpha|) \right]\nonumber\\
&\times\Bigg[\left(\Omega + i\frac{\kappa_{l}}{2} \right)  \cosh^{2}(\sqrt{\frac{\lambda}{2}}|\alpha|) \,\frac{\alpha}{|\alpha|}\nonumber\\
&+ \varepsilon_{p}^*  \frac{\alpha^{3}}{|\alpha|^{3}}\sinh^{2}(\sqrt{\frac{\lambda}{2}}|\alpha|) + \varepsilon_{p}  \cosh^{2}(\sqrt{\frac{\lambda}{2}}|\alpha|) \,\frac{\alpha^*}{|\alpha|}\Bigg]
\end{align}
and setting $\alpha = r e^{i\theta}$, and $\varepsilon_{p} = \varepsilon e^{-i\phi}$, the condition
$F_{j} = j K_{A_{+}^{\dag}M_{+}}+(1-j)K_{M_{+}A_{+}^{\dag}}=0$ yields
\begin{align}
&\left(\Omega + i\frac{\kappa_l}{2}\right)\Big[j 
+ (2j+1)\sinh^2\!\left(\sqrt{\frac{\lambda}{2}}r\right)\Big] \nonumber\\
&\quad + (2j+1)\,\varepsilon\left[e^{i\psi}\sinh^2\!\left(\sqrt{\frac{\lambda}{2}}r\right) 
+ e^{-i\psi}\cosh^2\!\left(\sqrt{\frac{\lambda}{2}}r\right)\right]=0
\end{align}
where $\psi=2\theta-\phi$. Separating the real and imaginary parts of the above equation, one obtains
\begin{align}
 &\Omega \left( j + (2j+1) \sinh^{2}\left(\sqrt{\frac{\lambda}{2}}r\right) \right) \nonumber\\
 &+ (2j+1) \varepsilon \cosh\left(\sqrt{2\lambda}r\right) \cos(\phi + 2\theta) =0
\end{align}
and
\begin{align}
\frac{\kappa_l}{2} \left( j + (2j+1) \sinh^{2}\left(\sqrt{\frac{\lambda}{2}}r\right) \right) - (2j+1) \varepsilon \sin(\phi + 2\theta) =0
\end{align}
By neglecting terms of $\mathcal{O}(\lambda^{2})$, we find  
\begin{align}\label{r_eq}
r^2 \simeq \frac{1}{\lambda}  \frac{\frac{\varepsilon}{|\tilde{\Omega}|} - \frac{j}{2j+1}}{  \frac{1}{2} - \frac{\varepsilon \Omega^2}{|\tilde{\Omega}|^{3/2}}}
\end{align}
and 
\begin{align}\label{phse_eq}
\tan(\phi + 2\theta) \simeq -\frac{\kappa_l}{2\Omega} \left( 1 +  \frac{\frac{\varepsilon}{|\tilde{\Omega}|} - \frac{j}{2j+1}}{ \left[ \frac{1}{2} - \frac{\varepsilon \Omega^2}{|\tilde{\Omega}|^{3/2}} \right]} \right)
\end{align}
where $\tilde{\Omega}=\Omega+i\frac{\kappa_l}{2}$. In addition, by considering $j\gg 1$, we can obtain 
\begin{align}
r^{2} &\simeq 
 \frac{1}{\lambda}  \frac{\frac{\varepsilon}{|\tilde{\Omega}|} - \frac{1}{2}}{ \left[ \frac{1}{2} - \frac{\varepsilon \Omega^2}{|\tilde{\Omega}|^{3/2}} \right]} \\
\tan(2\theta - \phi) &\simeq -\frac{\kappa_l}{2\Omega} \left[ 1+ \frac{\frac{\varepsilon}{|\tilde{\Omega}|} - \frac{1}{2}}{ \left[ \frac{1}{2} - \frac{\varepsilon \Omega^2}{|\tilde{\Omega}|^{3/2}} \right]} \right]\label{eq_phase21}
\end{align}
Since the phase is determined from a tangent functions \eqref{phse_eq} and \eqref{eq_phase21}, it is defined modulo $2\pi$, i.e.,
$2\theta - \phi \equiv 2\theta - \phi + 2\pi$, 
which implies an equivalence between coherent amplitudes $\alpha$ and $-\alpha$.  Therefore, the effective Hamiltonian now reads
\begin{align}
    H^{\prime}_{eff}&= \left[\tilde{\Omega}(1+\lambda r^2)+2
\lambda r^2\varepsilon\cos(\phi-2\theta)\right]\hat{A}^{\dag}\hat{A}\nonumber\\
    &\quad+\left[\varepsilon e^{i\phi}(1+\lambda r^2)+
\frac{\lambda r^2}{2}\tilde{\Omega} e^{2i\theta}\right]\hat{A}^{\dag}\hat{A}^{\dag}
\nonumber\\
    &\quad+\left[\varepsilon e^{-i\phi}(1+\lambda r^2)+
\frac{\lambda r^2}{2}\tilde{\Omega} e^{-2i\theta}\right]\hat{A}\hat{A}\nonumber\\
    &\quad- \frac{\lambda^{2} r}{2} \left[\tilde{\Omega}e^{i\phi}+\cos(\phi-2\theta)\right] \hat{A}^{\dag}\hat{a}^{\dag}\hat{a}\nonumber\\
    &\quad- \frac{\lambda^{2} r}{2} \left[\tilde{\Omega}e^{-i\phi}+\cos(\phi-2\theta)\right] \hat{a}^{\dag}\hat{a} \hat{A}
\end{align}
where $r^{2}$ and $\theta$ are given by \eqref{r_eq} and \eqref{phse_eq}. 
Note that it is always possible to choose the near-resonant parametric drive amplitude in the form
\begin{align}
\mathcal{E}=\varepsilon_c e^{i\phi_c},
\end{align}
such that
\begin{align}
\mathcal{E}(1+\lambda r^2)
+\frac{\lambda r^2}{2}\tilde{\Omega}e^{2i\theta}=0,
\end{align}
without requiring the single-photon loss rate to vanish, i.e., $\kappa_l\neq 0$. Solving the above complex constraint yields the critical drive strength
\begin{align}
\varepsilon_c=
\frac{\lambda r^2 |\tilde{\Omega}|}
{1+\lambda r^2},
\end{align}
together with the phase condition
\begin{align}
\tan\phi_c=
\frac{\kappa_l+2\Omega\tan(2\theta)}
{\Omega-2\kappa_l\tan(2\theta)}.
\end{align}
where $r$ and $\theta$ are given by \eqref{r_eq} and \eqref{phse_eq}, respectively.
It follows that, even in the presence of finite single-photon dissipation, one can always tune the complex parametric drive such that the resonant condition is exactly satisfied.

Therefore, the vacuum state $\ket{0}$ remains an eigenstate in the Kerr-displaced frame. Consequently, in the laboratory frame, $\ket{\pm r e^{i\theta},\lambda_{+}}_{D}$ constitute degenerate eigenstates of the effective Hamiltonian \eqref{Hamiltonian_eff_b}. 
Furthermore, the effective Hamiltonian exhibits a strong symmetry. It is implied that by applying the generalized parity operator
$\mathbb{Z}_p = \exp\!\left(i\pi \hat{a}^{\dagger}\hat{a}/p\right)$ \cite{Albert2014,Cattaneo2020},
in particular, for $p=2$, the Hamiltonian commutes with the parity operator, i.e.,
$[\mathbb{Z}_2, \hat{H}_{eff}] = 0$.
As a result, the steady states of Eq.~\eqref{dynamic_b} are Kerr cat states,
\begin{equation}
\ket{\mathcal{C}_{\pm},\lambda_{+}} \propto \ket{\alpha,\lambda_{+}}_D \pm \ket{-\alpha,\lambda_{+}}_D.
\end{equation}
\subsection{Negative Kerr parameters}
By considering the coefficients,
\begin{align}
K_{M_{-}A_{-}^{\dag}}&=-\left[ \sqrt{\frac{\lambda}{2}}\sin(\sqrt{2\lambda}|\alpha|) \right]\nonumber\\
&\times\Bigg[\left(\Omega + i\frac{\kappa_{l}}{2} \right)  \frac{\alpha}{|\alpha|} \sin^2(\sqrt{\frac{\lambda}{2}}|\alpha|) \nonumber\\
&+ \varepsilon_{p}^{\ast}  \frac{\alpha^3}{|\alpha|^3}\sin^2(\sqrt{\frac{\lambda}{2}}|\alpha|)
+ \varepsilon_{p} \frac{\alpha^{\ast}}{|\alpha|} \cos^2(\sqrt{\frac{\lambda}{2}}|\alpha|) \Bigg]
\end{align}
and 
\begin{align}
K_{A_{-}^{\dag}M_{-}}&=-\left[ \sqrt{\frac{\lambda}{2}}\sin(\sqrt{2\lambda}|\alpha|) \right]\nonumber\\
&\times\Bigg[\left(\Omega + i\frac{\kappa_{l}}{2} \right)  \cos^{2}(\sqrt{\frac{\lambda}{2}}|\alpha|) \,\frac{\alpha}{|\alpha|}\nonumber\\
&+ \varepsilon_{p}^{\ast}  \frac{\alpha^{3}}{|\alpha|^{3}}\sin^{2}(\sqrt{\frac{\lambda}{2}}|\alpha|) + \varepsilon_{p}  \cos^{2}(\sqrt{\frac{\lambda}{2}}|\alpha|) \,\frac{\alpha^{\ast}}{|\alpha|}\Bigg]
\end{align}
and setting $\alpha = r e^{i\theta}$, and $\varepsilon_{p} = \varepsilon e^{-i\phi}$, the condition
$F_{j} =j K_{A_{-}^{\dag}M_{-}}+(j-1)K_{M_{-}A_{-}^{\dag}}= 0$ yields
\begin{align}
&\left(\Omega + i\frac{\kappa_l}{2}\right)\left[j - \sin^2\left(\sqrt{\frac{\lambda}{2}}r\right)\right] \nonumber\\
&\quad+ (2j-1)\varepsilon \left[ e^{i\psi}\sin^2\left(\sqrt{\frac{\lambda}{2}}r\right) + e^{-i\psi}\cos^2\left(\sqrt{\frac{\lambda}{2}}r\right)\right] = 0 
\end{align}
where $\psi=2\theta+\phi$. Separating the real and imaginary parts of the above equation, one obtains
\begin{align}
&\Omega \left[ j - \sin^2\left(\sqrt{\frac{\lambda}{2}}r\right) \right] + (2j-1)\varepsilon \cos\psi = 0 \nonumber\\
&\frac{\kappa_l}{2} \left[ j - \sin^2\left(\sqrt{\frac{\lambda}{2}}r\right) \right] - (2j-1)\varepsilon \sin\psi \cos\left(\sqrt{2\lambda}r\right) = 0
\end{align}
By neglecting terms of $\mathcal{O}(\lambda^{2})$, we can find  
\begin{align}
r^2 &\simeq \frac{(2j-1)^2\varepsilon^2 - j^2\tilde{\Omega}^2}{\lambda \left[ \frac{j^2 \kappa_l^2}{2} - j\tilde{\Omega}^2 \right]}\label{eq_r2}\\
\tan(2\theta + \phi) &= -\frac{\kappa_l}{2\Omega \cos(\sqrt{2\lambda}r)}\nonumber\\
&\simeq  -\frac{\kappa_l}{2\Omega} \left[ 1 + \left( \frac{(2j-1)^2\varepsilon^2 - j^2\tilde{\Omega}^2}{\frac{j^2 \kappa_l^2}{2} - j\tilde{\Omega}^2} \right)  \right]
\label{eq_phase2_appendix}
\end{align}
in which
$\tilde{\Omega}^{2}=\Omega^{2} + \frac{\kappa_{l}^{2}}{4}$.

In addition, by considering $j\gg 1$, we can obtain  
\begin{align}
r^2 &\simeq \frac{2}{\lambda } \frac{4\varepsilon^2 - \tilde{\Omega}^2 }{\kappa_l^2}\nonumber\\
\tan(2\theta + \phi) &\simeq -\frac{\kappa_l}{2\Omega} \left( \frac{1}{2} + 2\frac{4\varepsilon^2-\Omega^2}{\kappa_{l}^{2}}\right)
\end{align}
For negative Kerr nonlinearity, i.e., $\lambda < 0$, an analogous analysis applies. 
The phase is again determined through tangent relations and is therefore defined modulo $2\pi$, 
leading to the same equivalence between coherent amplitudes $\alpha$ and $-\alpha$. As a result, the effective Hamiltonian now is given by
\begin{align}
    H'_{eff}&= \left[\tilde{\Omega}(1-\lambda r^2)+2
\lambda r^2\varepsilon\cos(\phi-2\theta)\right]\hat{A}^{\dag}\hat{A}\nonumber\\
    &\quad+\left[\varepsilon e^{i\phi}(1-\lambda r^2)+
\frac{\lambda r^2}{2}\tilde{\Omega} e^{2i\theta}\right]\hat{A}^{\dag 2}
\nonumber\\
    &\quad+\left[\varepsilon e^{-i\phi}(1-\lambda r^2)+
\frac{\lambda r^2}{2}\tilde{\Omega} e^{-2i\theta}\right]\hat{A}^{2}\nonumber\\
    &\quad- \frac{\lambda^{2} r}{2} \left[\tilde{\Omega}e^{i\phi}+\cos(\phi-2\theta)\right] \hat{A}^{\dag}\hat{a}^{\dag}\hat{a}\nonumber\\
    &\quad- \frac{\lambda^{2} r}{2} \left[\tilde{\Omega}e^{-i\phi}+\cos(\phi-2\theta)\right] \hat{a}^{\dag}\hat{a} \hat{A}
\end{align}
where $\tilde{\Omega}=\Omega+i\frac{\kappa_l}{2}$ and $r^{2}$ and $\theta$ are given by \eqref{eq_r2} and \eqref{eq_phase2}. 
Note that it is always possible to choose the near-resonant parametric drive amplitude in the form
\begin{align}
\mathcal{E}=\varepsilon_c e^{i\phi_c},
\end{align}
such that
\begin{align}
\mathcal{E}(1-\lambda r^2)
+\frac{\lambda r^2}{2}\tilde{\Omega}e^{2i\theta}=0,
\end{align}
without requiring the single-photon loss rate to vanish, i.e., $\kappa_l\neq 0$. Solving the above complex constraint yields the critical drive strength
\begin{align}
\varepsilon_c=
\frac{\lambda r^2 |\tilde{\Omega}|}
{1-\lambda r^2},
\end{align}
together with the phase condition
\begin{align}
\tan\phi_c=
\frac{\kappa_l+2\Omega\tan(2\theta)}
{\Omega-2\kappa_l\tan(2\theta)}.
\end{align}
Consequently, the vacuum state $\ket{0}$ remains an eigenstate in the corresponding Kerr-displaced frame.

In the laboratory frame, this implies that the states $\ket{\pm r e^{i\theta}, \lambda_{-}}_{D}$ 
again form a pair of degenerate eigenstates of the effective Hamiltonian~\eqref{Hamiltonian_eff_b}.

Moreover, the symmetry properties of the Hamiltonian are preserved: the generalized parity operator
\begin{equation}
\mathbb{Z}_p = \exp\!\left(i\pi \hat{a}^\dagger \hat{a}/p\right)
\end{equation}
still commutes with the effective Hamiltonian. In particular, for $p=2$, one has
\begin{equation}
[\mathbb{Z}_2, \hat{H}_{\mathrm{eff}}] = 0,
\end{equation}
indicating a $\mathbb{Z}_2$ symmetry.

As a result, the steady states of Eq.~\eqref{dynamic_b} are again Kerr cat states, i.e., 
coherent superpositions of the form
\begin{equation}
\ket{\mathcal{C}_{\pm},\lambda_{-}} \propto \ket{\alpha,\lambda_{-}}_D \pm \ket{-\alpha,\lambda_{-}}_D.
\end{equation}

\section{Broader scope of DKCS, BGKCS and SCS under photon loss and dephasing noise}\label{sec:Broader_scope}
In this Appendix section we present complementary analysis and results to Sections~\ref{subsec:channel_fidelities} and~\ref{subsec:Dynamics}. 

\subsection{Dynamics analysis of DKCS, BGKCS and SCS}
In this Section we extend the analysis of Section~\ref{subsec:Dynamics} to a lower photon count regime, $|\alpha| = 0.6$, and consider the dynamics of the infidelity for the input states;
\begin{eqnarray}
    \hat{\rho}(0)=&\ketbra{\mathcal{C}_{f}^{+}}{\mathcal{C}_{f}^{+}} \label{eq:initial_cat_plus}\\
    \hat{\rho}(0)=&\ketbra{\mathcal{C}_{f}^{-}}{\mathcal{C}_{f}^{-}} \label{eq:initial_cat_minus}\\
    \hat{\rho}(0)=&\frac{1}{2}I_l \label{eq:initial_max_mixed}\\
    \hat{\rho}(0)=&\ketbra{\psi}{\psi} \label{eq:initial_pure_state}
    \;,
\end{eqnarray}
where
$\frac{1}{2}I_l = \frac{1}{2}(\ketbra{\mathcal{C}_{f}^{+}}{\mathcal{C}_{f}^{+}} + \ketbra{\mathcal{C}_{f}^{-}}{\mathcal{C}_{f}^{-}})$ is the maximally mixed state in the logical subspace and the pure state $\ket{\psi} = \frac{1}{\sqrt{2}}(\ket{\mathcal{C}_{f}^{+}} + \ket{\mathcal{C}_{f}^{-}})$ (here depicted sometimes as $\hat{\rho} = \ket{\psi}$ for short).
In order to include any initial quantum density state, the time dependent infidelity equation, given at Eq.~\ref{eq:fidelity}, generalizes to:
\begin{eqnarray}\label{eq:general_infidelity}
    P(t) = 1 - \left( tr \sqrt{\sqrt{\rho(0)} \rho(t) \sqrt{\rho(0)} }\right)^2
    \;.
\end{eqnarray}

\begin{figure*}
    \includegraphics[width=\linewidth,keepaspectratio]{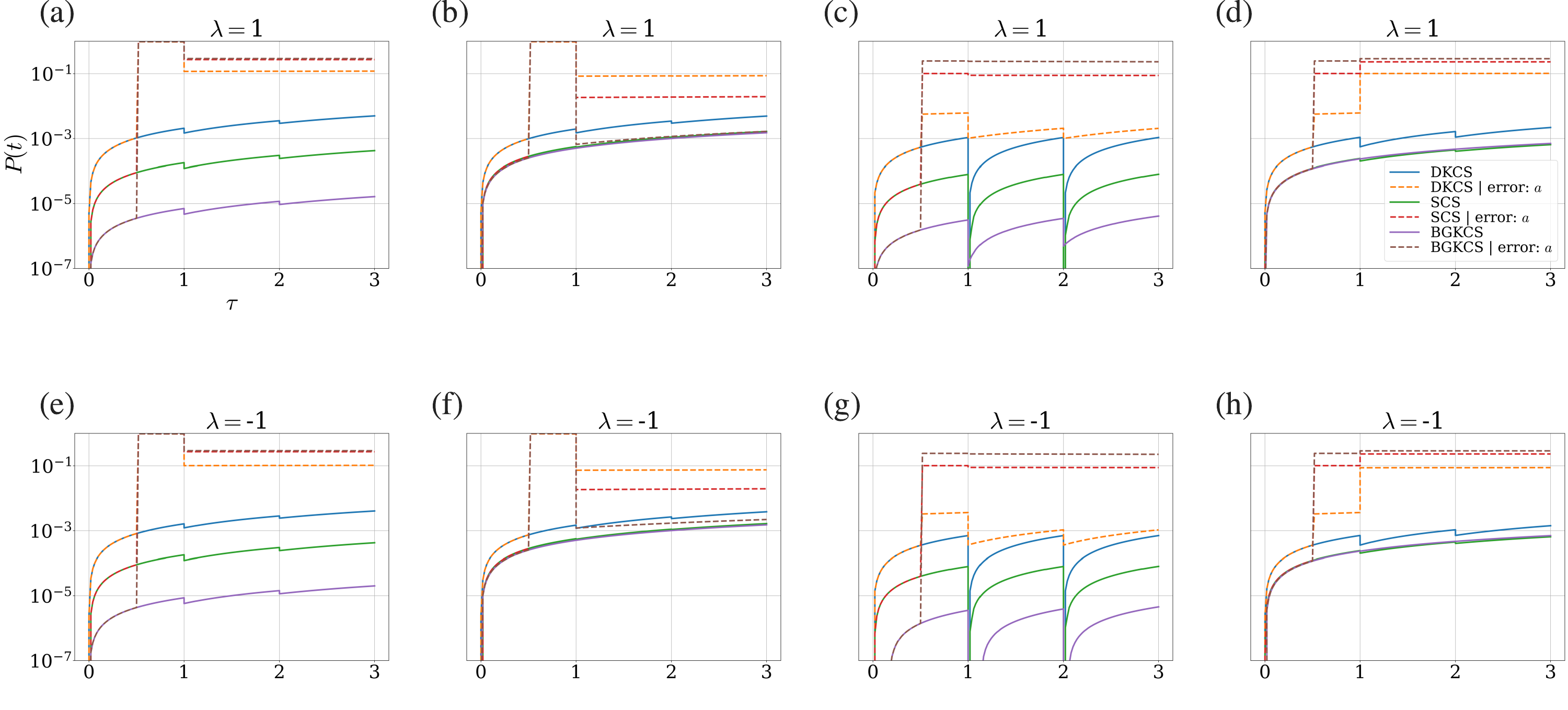}
    \caption{Evolution plot of the fidelity (logarithmic scale) of DKCS, BGKCS and SCS for values $j=5$, $|\alpha|=0.6$ across various input initial logical states: \eqref{eq:initial_cat_plus} for (a) and (e), \eqref{eq:initial_cat_minus} for (b) and (f), \eqref{eq:initial_max_mixed} for (c) and (g) and \eqref{eq:initial_pure_state} for (d) and (h). The states are prepared and then evolved using noise channel of Eq. (\ref{eq:noise_channel}) with rates $\kappa_l, \kappa_d = 10^{-3}$. In both evolution trajectories the optimized recovery channel of Eq. (\ref{eq:recovery_channel}) is performed to the state at time $\tau \in \{1, 2\}$ with $\tau = t/\kappa $. In dotted line evolutions a photon loss error is induced at $\tau= 0.5$, then measured and corrected at $\tau= 1$. plots (a)-(d) account for the case $\lambda = 1$ and subplots (e) - (h) correspond to $\lambda = -1$.}
    \label{fig:evolution_standard_noise_bigger_scope}
\end{figure*}

In Fig.  \ref{fig:evolution_standard_noise_bigger_scope} the dynamics of the quantum fidelity shows the difference for various initial states under the noise model.
Specifically for the both figures, Plots (a) through (d), represent the dynamics of SCS, DKCS and BGKCS for the positive Kerr parameter, $\lambda = 1$, starting at each respective initial state.
Meanwhile, plots (e) to (h) show analogously the corresponding dynamics for the negative Kerr parameter, $\lambda = -1$.
Each state is studied under the parameter $j=5$, with the exception of SCS which lacks of such degree of freedom.
A parallel scenario, portrayed in dotted line curve, corresponds to the case when a photon loss error is induced in the system.

Plot (a) shows a contrasting difference of the fidelity evolution between each positive cat state case, with BGKCS being overall one order of magnitude lower than SCS and two orders lower than DKCS. 
Once a photon loss error is induced in the system, the infidelities tend to all converge close together at the order of $10^{-1}$, with DKCS being slightly lower than the other two cases.

The negative cat state infidelity evolution can be observed in Plot (b).
In such plot, the BGKCS evolves quite similarly to the SCS, however DKCS traces a similar curve in shape, but shifted by half an order of magnitude higher. 
The dissimilarities emerge when considering the photon loss scenario, where BGKCS shows higher response to the QEC procedure by fully recovering the state from the photon loss error, while DKCS is the least restored state after the correction procedure.

In plot (c) the evolution of the logical maximally mixed state can be observed. 
The overall lowest infidelity corresponds to the BGKCS, by almost two orders of magnitude lower than the SCS and two and a half lower than the DKCS. 
This result is no longer true when considering a photon loss error, where DKCS shows to be much more resilient to the error than SCS and BGKCS.
Additionally, DKCS recovers substantially better than the other two Cat states from the photon loss error after performing the QEC procedure.
In contrast, BGKCS shows almost no alteration in their infidelities after the QEC, while SCS shows just a marginally small alteration.

The superposition pure state of positive Cat and negative Cat is shown in plot (d). The behavior of BGKCS's infidelity is similar to the one observed for the negative Cat state case when no photon loss is induced in the system, this is consequence of two factors.
First: considering finite dimensions and lower photon counts in the system induces an asymmetry in the infidelity evolution between positive and negative BGKCSs, since the leading orders of Eq.~\eqref{GK-def-pos} and Eq.~\eqref{GK-def-neg} are dominant at $n=0,1$. 
This is no longer the case as $|\alpha|\geq 1$ increases.
Second: the negative Cat state is dominant in this case, because it has higher infidelity than the positive case by two orders of magnitude.
It is notorious in this plot that the error correction fails to recover the state for the DKCS, which shifts the infidelity to higher values and closer to the Cat positive case, where the infidelity is the highest among the negative Cat state and the positive Cat state.
This effect can be attributed to the combination of classical photon loss, measurement and correction to induce a logical error in the code space, which in turn eludes the quantum recovery channel.
Overall we can observe that the evolution of such initial state gives the infidelity whose value is the highest among the infidelities of the positive Cat state and negative Cat state.

The behaviors of the infidelity evolution for the negative Kerr parameter case in plots (e) to (h) show no substantial difference from their positive Kerr parameter counterpart of plots (a) to (d).
Comparing all plots of Fig.  \ref{fig:evolution_standard_noise_bigger_scope}, we can observe that the quantum error correction procedure proves to be much more efficient for all systems where initial input is the maximally mixed state of their logical subspace.
Moreover, the systems are more resilient towards the quantum noise bath and the classically induced photon loss error for such initial state.
This result illustrates the variability of state evolution at different initial conditions.
An analogous study for the modified ladder operator noise can be followed in the Appendix \ref{subsec:dynamics_modified}

\subsection{Average infidelities for DKCS}\label{subsec:Biggerscope_DKCS}
\begin{figure*}
\includegraphics[width=\linewidth,keepaspectratio]{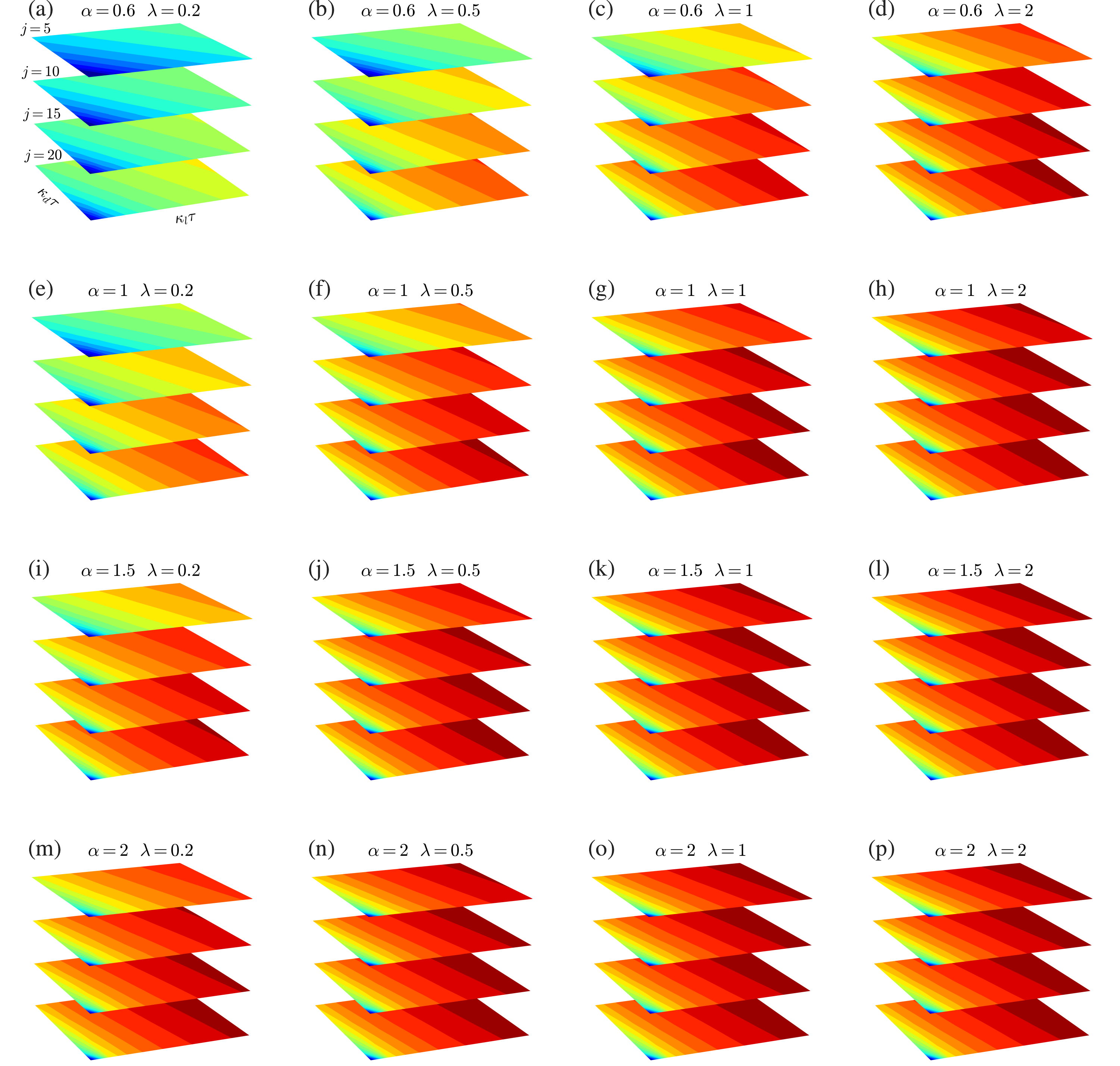}
\\
\includegraphics[width=0.7\linewidth]{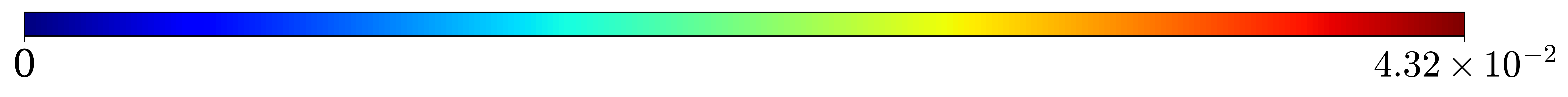}\\
    \caption{Stacked heatmaps of the average infidelity (logarithmic scale) for DKCS under combined dephasing and photon-loss channels. The error rates $\kappa_d \tau$ and $\kappa_l \tau$ are varied within $[0, 10^{-3}]$.}
    \label{fig:heatmaps_standard_noise_DKCS_pos}
\end{figure*}

\begin{figure*}
\includegraphics[width=\linewidth,keepaspectratio]{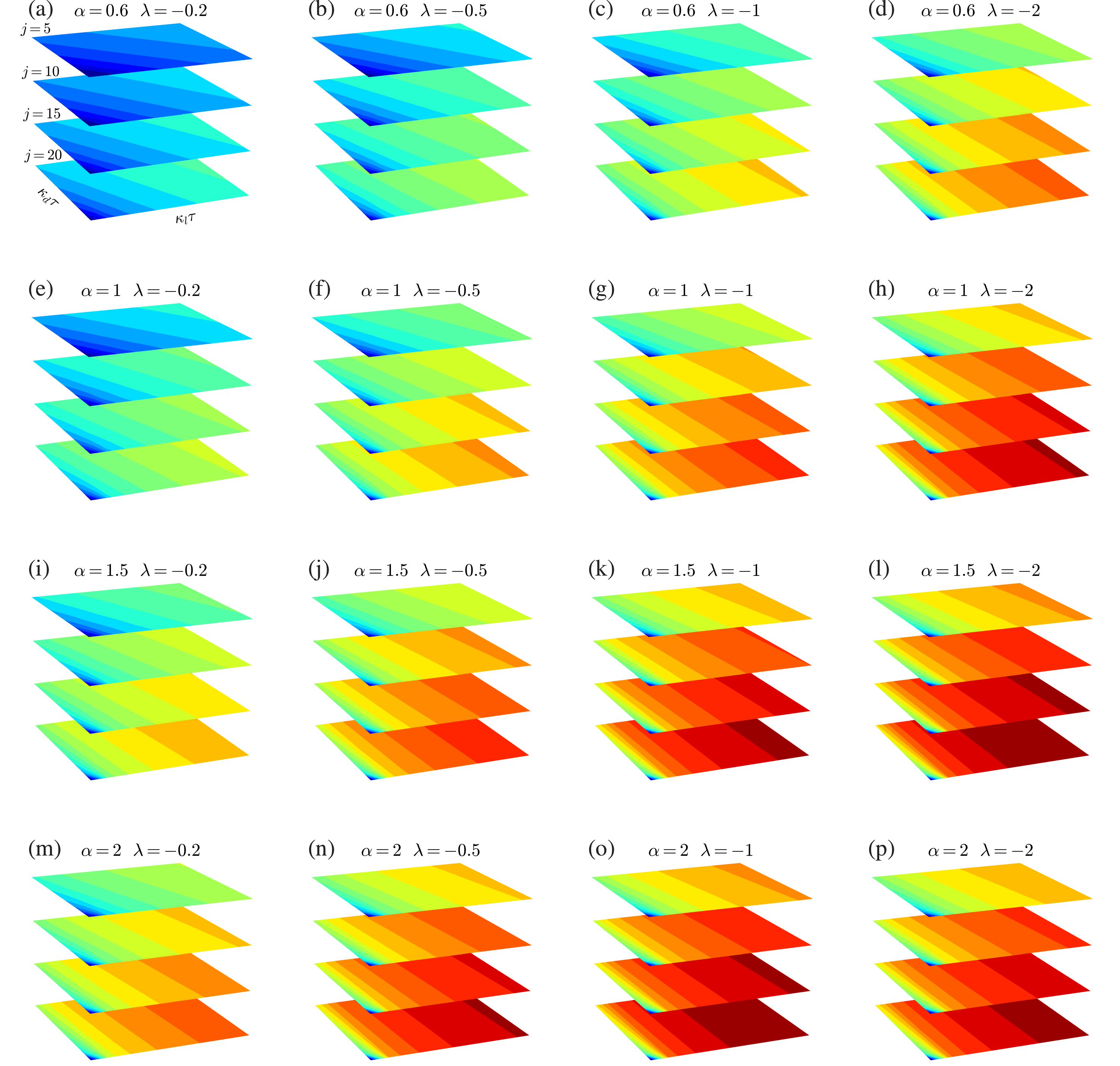}
\\
\includegraphics[width=0.7\linewidth]{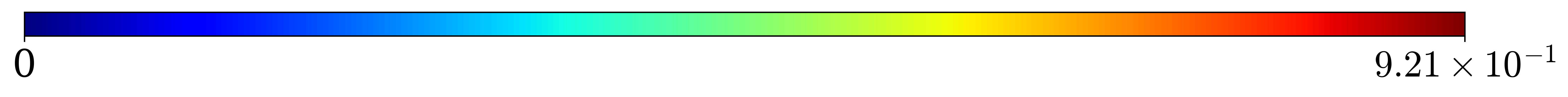}\\
    \caption{Stacked heatmaps of the average infidelity (logarithmic scale) for DKCS under combined dephasing and photon-loss channels. The error rates $\kappa_d \tau$ and $\kappa_l \tau$ are varied within $[0, 10^{-3}]$.}
    \label{fig:heatmaps_standard_noise_DKCS_neg}
\end{figure*}

Figs.  \ref{fig:heatmaps_standard_noise_DKCS_pos} and \ref{fig:heatmaps_standard_noise_DKCS_neg} illustrate an extension of the heatmaps presented in Section \ref{subsec:channel_fidelities} by showing a wider range of parameters. The subplots plots are ordered in increasing value of $|\lambda|$, from left to right in each row, and in increasing values of $|\alpha|$: from top to bottom in each column. 
Each plot contains vertically stacked heatmaps of the average infidelity of the the system under photon loss rates ($\kappa_l$) and dephasing noise rates ($\kappa_d$).
The stacked heatmaps of each plot are ordered in increasing order of $j$, from top to bottom. 
As a consequence of this arrangement, Figs. \ref{fig:heatmaps_standard_noise_DKCS_pos} and \ref{fig:heatmaps_standard_noise_DKCS_neg} show visibly the dependency of the average infidelity of DKCS across the 5 parameters: $\{\lambda,\alpha, \kappa_l, \kappa_d, j\}$.

In both figures it is noticeable that larger values of $|\lambda|$ increase the overall average infidelity of the system; diminishing the resilience of the DKCS under both noise effects.
The same behavior appears when increasing the value of $|\alpha|$, which in consequence leads to the lowest noise resilience for the highest values of $|\lambda|$ and $|\alpha|$, here studied ($|\lambda| = 2 = |\alpha|$).
This effect is more pronounced for $\lambda < 0$, as shown in Fig.  \ref{fig:heatmaps_standard_noise_DKCS_neg}, where the highest average infidelity reaches a value close to $0.921$ and substantially surpassing the maximum average infidelity of the case $\lambda > 0$.
Additionally, increasing $j$ leads to a slight increase in the overall average infidelity. 
The validity of this last observation is consistent through all plots of both figures.
Therefore DKCS reaches higher resilience to photon loss and dephasing noises at the low levels of nonlinearity ($|\lambda| \sim 0$) and lower photon count ($|\alpha| \sim 0$).

\begin{figure*}
\includegraphics[width=\linewidth,keepaspectratio]{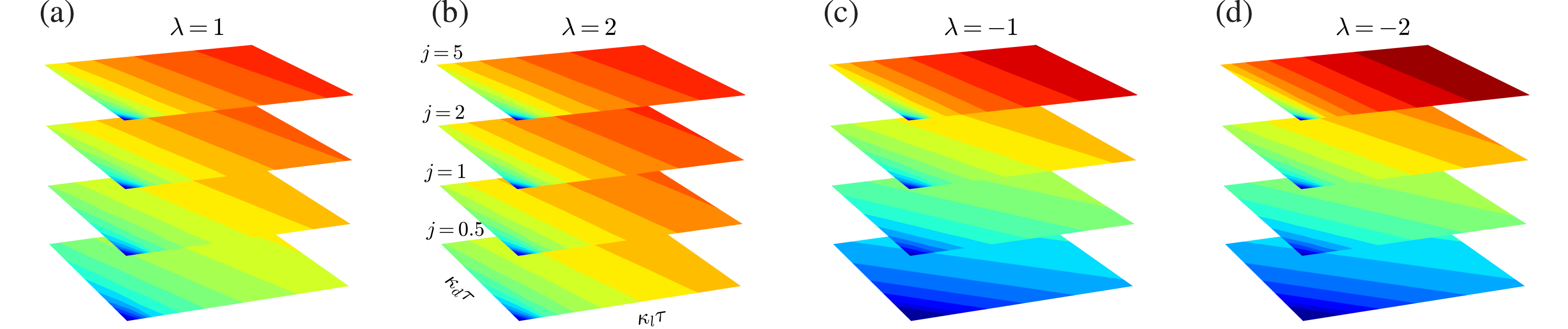}
\\
\includegraphics[width=0.7\linewidth]{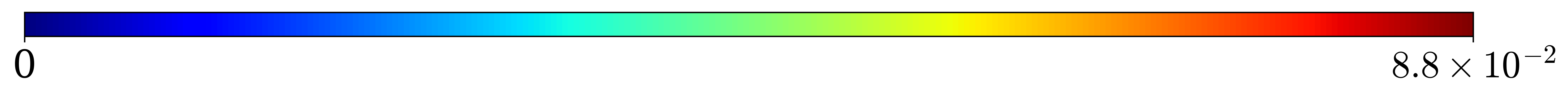}
    \caption{Stacked heatmaps of the average infidelity (logarithmic scale) for DKCS under combined dephasing and photon-loss channels. The error rates $\kappa_d \tau$ and $\kappa_l \tau$ are varied within $[0, 10^{-3}]$ and $|\alpha| = 1.5$.}
    \label{fig:heatmaps_DKCS_smaller_j}
\end{figure*}

It is also instructive to consider small values of $j$. In particular, for $\lambda = 2$ and $\lambda = -2$, the algebra reduces to $\mathfrak{su}(1,1)$ and $\mathfrak{su}(2)$, respectively, yielding the corresponding coherent states. The case $j = \tfrac{1}{2}$ is of special interest: the $\mathfrak{su}(1,1)$ coherent state becomes the squeezed vacuum state, whereas the $\mathfrak{su}(2)$ coherent state is defined on a two-dimensional Hilbert space. These limits connect the present formalism to well-established classes of quantum states. Fig.~\ref{fig:heatmaps_DKCS_smaller_j} illustrates the impact of the Kerr parameters and the parameter $j$. As shown, the average infidelity increases with increasing absolute values of the Kerr parameters as well as increasing the parameter $j$.

\subsection{Average infidelities for BGKCS}
\begin{figure*}
\includegraphics[width=\linewidth,keepaspectratio]{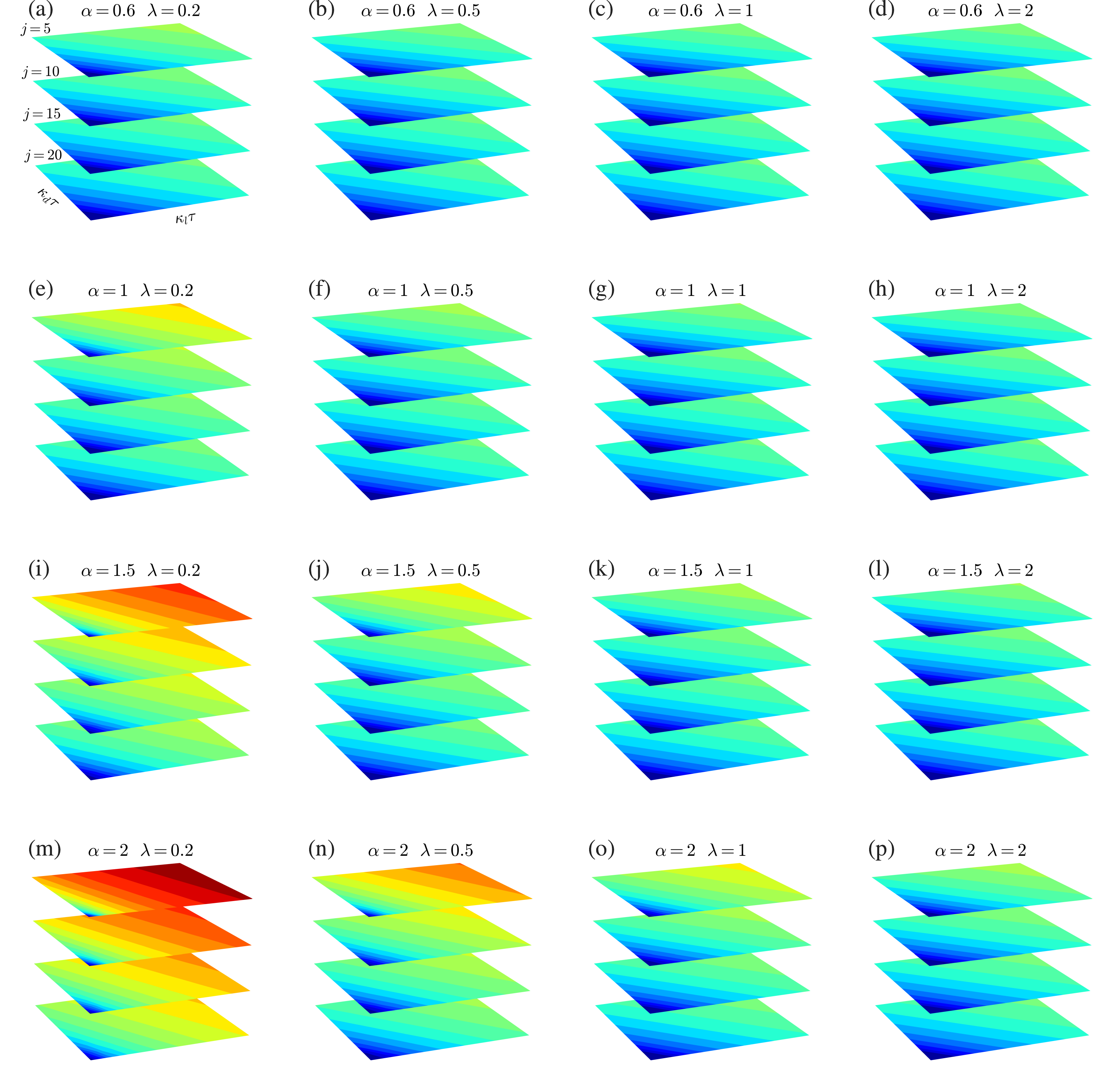}
\\
\includegraphics[width=0.7\linewidth]{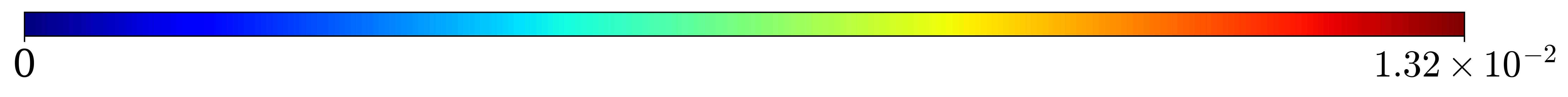}\\
    \caption{Stacked heatmaps of the average infidelity (logarithmic scale) for BGKCS under combined dephasing and photon-loss channels. The error rates $\kappa_d \tau$ and $\kappa_l \tau$ are varied within $[0, 10^{-3}]$.}
    \label{fig:heatmaps_standard_noise_BGKCS_pos}
\end{figure*}

\begin{figure*}
\includegraphics[width=\linewidth,keepaspectratio]{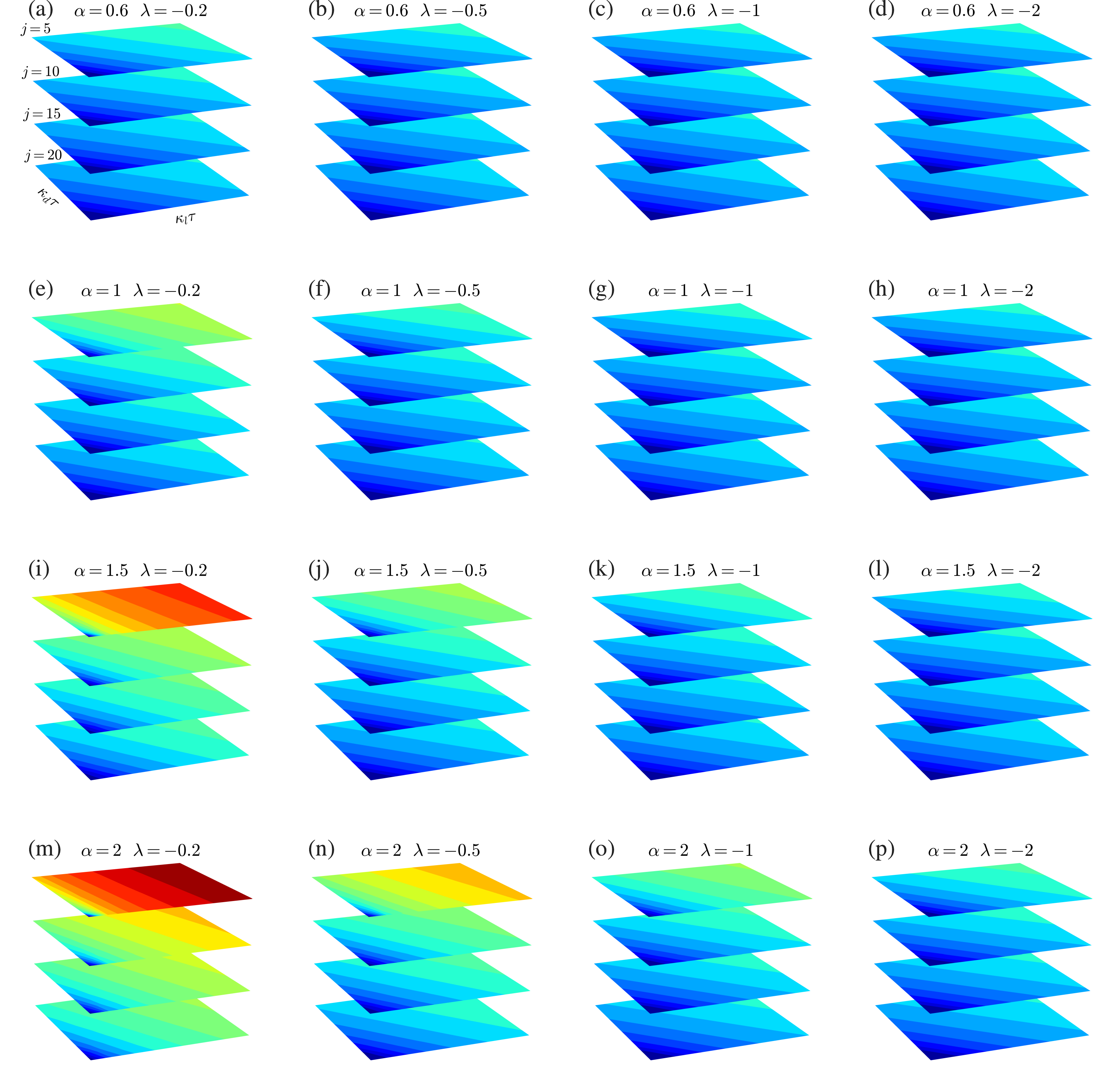}
\\
\includegraphics[width=0.7\linewidth]{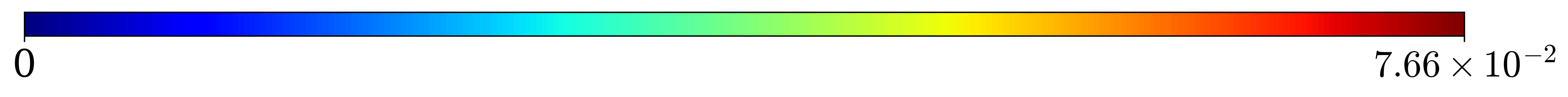}\\
    \caption{Stacked heatmaps of the average infidelity (logarithmic scale) for BGKCS under combined dephasing and photon-loss channels. The error rates $\kappa_d \tau$ and $\kappa_l \tau$ are varied within $[0, 10^{-3}]$.}
    \label{fig:heatmaps_standard_noise_BGKCS_neg}
\end{figure*}

Following the same ordering of the figures of subsection \ref{subsec:Biggerscope_DKCS}, Figs. \ref{fig:heatmaps_standard_noise_BGKCS_pos} and \ref{fig:heatmaps_standard_noise_BGKCS_neg} illustrate the dependency of the average infidelity of BGKCS across the 5 parameters: $\{\lambda,\alpha, \kappa_l, \kappa_d, j \}$.

In both figures, increasing $|\alpha|$ leads to a general increase in the average infidelity of BGKCS.
This effect is more pronounced when $\lambda < 0$ at $j = 5$, and becomes less significant as $j$ increases.
At higher Kerr nonlinearity, i.e. increasing values of $|\lambda|$, the resilience of the system to photon loss and dephasing noise becomes more prominent by reducing the overall average infidelity across the noise rates.
This result is persistent regardless of the sign of $\lambda$.
Additionally, higher Kerr nonlinearity substantially counteracts the increase in average infidelity produced at larger $|\alpha|$.

Under a more comprehensive perspective of all systems and parameters here studied, BGKCS shows consistently more stable resilience with less fluctuation across all the 5 studied parameters, $\{\lambda,\alpha, \kappa_l, \kappa_d, j \}$. 
Moreover, BGKCS systems present more resilience towards photon loss and dephasing noises than DKCS systems, by achieving a maximum average infidelity an order of magnitude lower than the maximum average infidelity of DKCS.

An analogous study that covers all systems, DKCS, BGKCS and SCS under the modified ladder operator noise can be followed in the Appendix \ref{sec:Error_correction_for_modified_noise}.

\section{Error correction for modified noise channel}\label{sec:Error_correction_for_modified_noise}
In this section we show the results of the analogous procedure as presented in Sections \ref{sec:Errors_and_their_corrections} and \ref{subsec:Dynamics} obtained when replacing the standard ladder operator in the noise model of Eq.~\eqref{eq:Lindbladian_standard} with the modified version of the ladder operator as in Eq.~\eqref{anh_pos}.
The dependency of the modified ladder operator $A$ on $\lambda$ and $j$ allows to extend the study of SCS resilience by including those parameters in the modified noise channel.
While it is natural to assume that operator $A$ for the noise model is the same for state preparation of DKCS and BGKCS systems, one could in general associate different modified ladder operators spanning from different parameters of $\lambda$ and $j$ on each scenario.
This general case is, however, outside of the scope of this research work and we will always assume that Equations \eqref{eq:DKCS_from_displacement_op}, \eqref{BG_def}, \eqref{eq:Lindbladian_modified} share the same operator of Eq.~\eqref{anh_pos}, originating from the same values of $\lambda$ and $j$.

\subsection{Cat states heatmaps under modified noise error} \label{subsec:heatmaps_modified}
\begin{figure*}
\includegraphics[width=\linewidth,keepaspectratio]{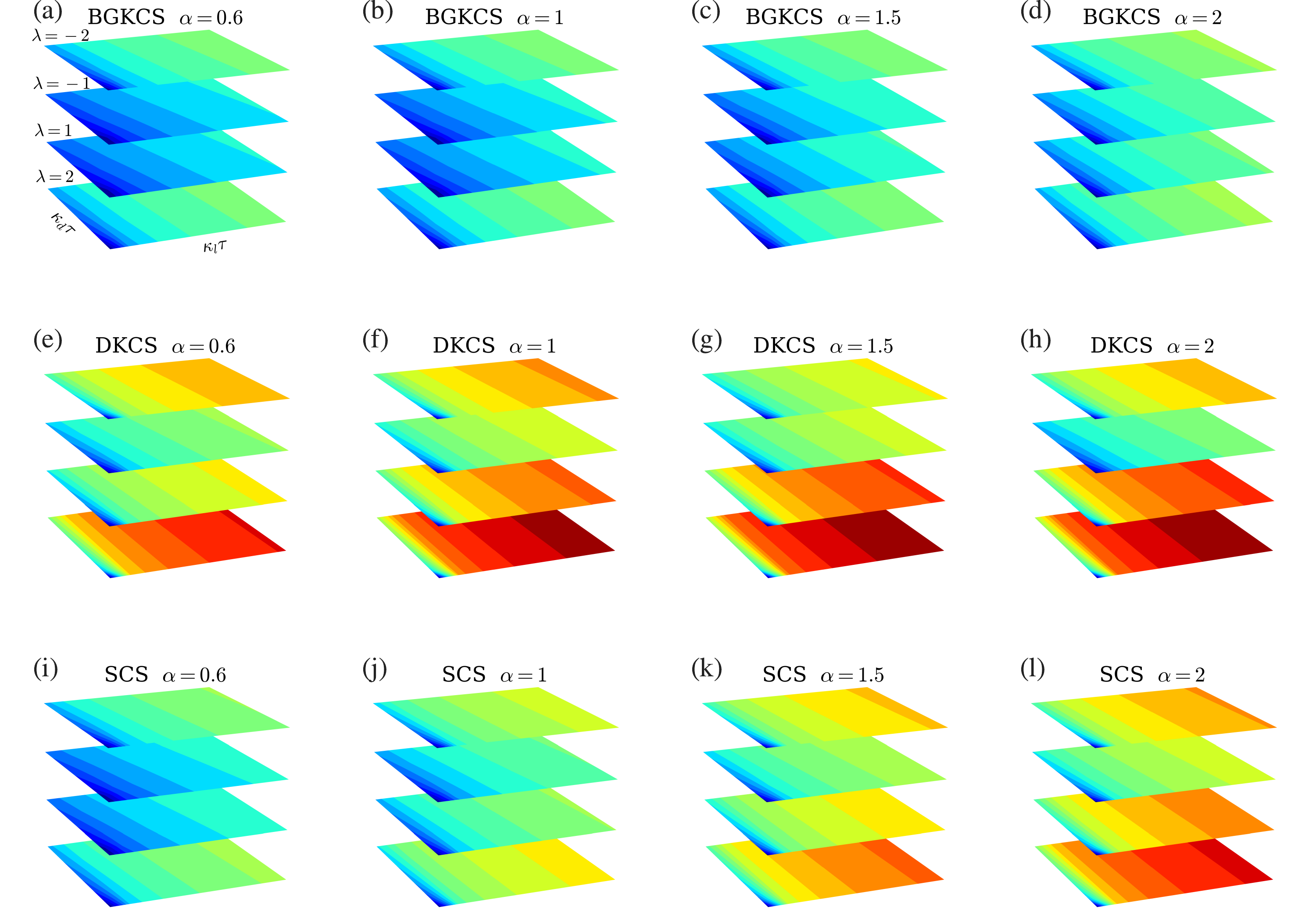}
\\
\includegraphics[width=0.7\linewidth]{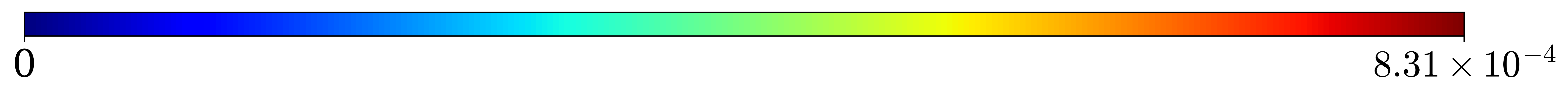}\\
    \caption{Stacked heatmaps of the average infidelity (logarithmic scale) DKCS, BGKCS and SCS under combined modified dephasing and photon-loss channels. The error rates $\kappa_d \tau$ and $\kappa_l \tau$ are varied within $[0, 10^{-7}]$ and $j=5$.}
    \label{fig:heatmaps_modified_noise}
\end{figure*}
In Fig.  \ref{fig:heatmaps_modified_noise} the average infidelities for BGKCS (plots (a) to (d)), DKCS (plots (e) to (h)) and SCS (plots (i) to (l)) are presented. 
All plots in a row are ordered with respect to increasing value of $|\alpha|$ from left to right.
Each plot contains vertically stacked heatmaps of the average infidelity across the photon loss rate ($\kappa_l$) and dephasing noise rate ($\kappa_d$) parameters.
The stacked heatmaps inside a plot are ordered by increasing values of $\lambda$ from top to bottom, where $\lambda \in \{-2,-1,1,2\}$.

In all the systems an increase in $|\alpha|$ manifests as an increase in the average infidelity, with BGKCS showing a slight increase, while the effect on SCS is more potent.
Additionally, in all cases, an increase in $|\lambda|$ generally reduces the system average fidelity.
This effect agrees with the fact that the dependency of the modified operator on $\lambda$ effectively translates as an increase of noise rate when $|\lambda|$ increases.
Such observation can be confirmed when considering the Kraus operators of noise channel of Eq.~\eqref{eq:Kraus_operators} and then absorbing the terms of orders of $|\lambda|$ into the noise rate coefficients.
For the case of DKCS we can observe that in general the average infidelity is higher than the BGKCS and SCS systems, reaching the value of $8.31\times 10^{-4}$ at $j = 5$, $\alpha = 2$, $\lambda =2$, the highest from among all the studied cases and systems.

\subsection{Cat states dynamics plots under modified noise error} \label{subsec:dynamics_modified}
\begin{figure*}
    \includegraphics[width=\linewidth,keepaspectratio]{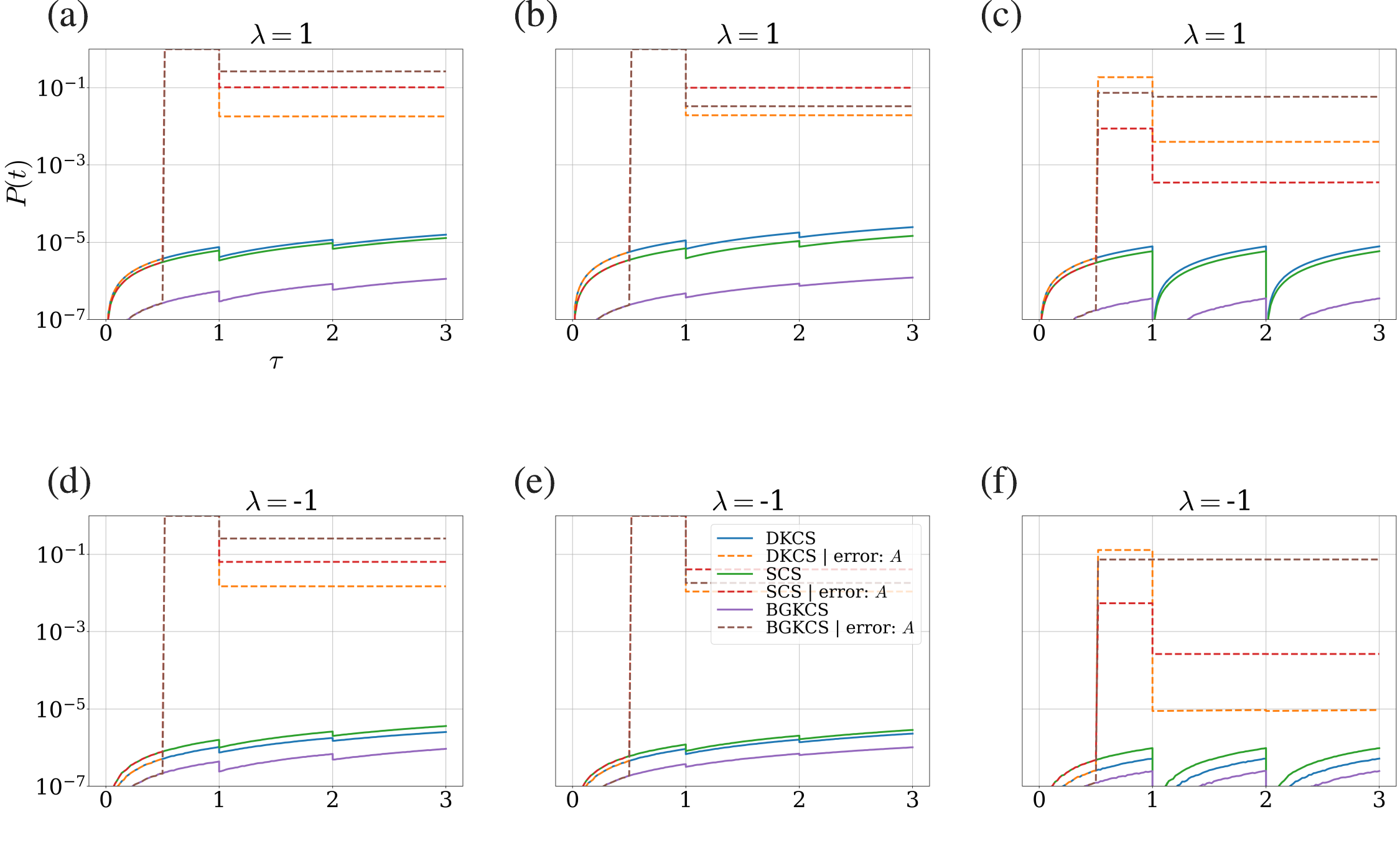}
    \caption{Evolution plot of the fidelity (logarithmic scale) of DKCS, BGKCS and SCS for values $j=5$ (not applicable to SCS) and $|\alpha|=1.5$. In plots row (a)-(c) the value $\lambda = 1$ is used, while in row (d)-(f) $\lambda = -1$ is used instead. The intial state: \eqref{eq:initial_cat_plus} was prepared in plots (a),(d),  \eqref{eq:initial_cat_minus} was prepared in plots (b),(e) and \eqref{eq:initial_max_mixed} was prepared in (c),(f), then evolved under the noise channel of Eq. (\ref{eq:noise_channel}) with error rates $\kappa_l, \kappa_d = 10^{-7}$. The same values of $j$ and $\lambda$ were used for the operators $A$ and $A^{\dagger}$ of the noise channel. In both evolution trajectories the optimized recovery channel of Eq. (\ref{eq:recovery_channel}) is performed to the state at time $\tau\in \{1, 2\}$ with $\tau = t/\kappa $. In dotted line evolutions a modified ladder, $\hat{A}$, error is induced at $\tau= 0.5$, then measured and corrected at $\tau= 1$.}
    \label{fig:dynamics_modified_noise}
\end{figure*}

Instead of plotting the averaged infidelity over all input states, Fig. \ref{fig:dynamics_modified_noise} shows the evolution of the infidelity at a given initial logical state for BGKCS, DKCS and SCS systems.
The dotted line represents the case when a modified ladder operator plots (a) to (c) shows the evolution of input logical states: $\rho = \{\ketbra{\mathcal{C}_{f}^{+}}{\mathcal{C}_{f}^{+}} , \ketbra{\mathcal{C}_{f}^{-}}{\mathcal{C}_{f}^{-}}, \frac{1}{2}I_l \}$ respectively, at the cases where $j = 5$, $\alpha = 1.5$, $\lambda = 1$, while plots (d) to (f) show the analogous cases when $\lambda = -1$.
In plot (a) the input states corresponds to the pure state $\ket{\mathcal{C}_{f}^{+}}$.
In the corresponding evolution curves, BGKCS is substantially more resilient to the quantum noise bath, by showing overall a lower infidelity evolution of an order of magnitude bellow.
However, the state shows the lowest resilience to the classical noise, since it recovers the least among DKCS and SCS after the error correction procedure.
The evolution of both systems, DKCS and SCS are similar with only SCS being slightly lower than DKCS.
The similarities between the two states cease to occur in the case of the classically induced error.
After the error correction procedure is performed in the system, DKCS show to have higher recovery than SCS by decreasing the infidelity of the state by close to an order of magnitude lower than SCS.

The pure state $\ket{\mathcal{C}_{f}^{-}}$ is studied in plot (b).
The states for DKCS and SCS show very similar behavior as their positive cat state counterpart shown in plot (a).
The difference between the two states is only apparent by being slightly separated with SCS being lower than DKCS.
The state BGKCS still remains substantially more resilient by remaining an order of magnitude lower than the evolutions of SCS and DKCS.
Its evolution curve is also comparably similar to its positive cat state counterpart of plot (a), however in this case the state shows to recover more from the classically induced noise, leaving its infidelity curve close to the lowest curve corresponding to DKCS state.
Therfore, SCS shows to recover less from the classically induced noise.

Plot (c) studies the case of the logical maximally mixed state as the input state.
In this case, all systems are benefited from the input state, since the evolution of the infidelity remains the lowest.
Additionally all of the systems improve in terms of recoverability, where the error correction manages to restore the initial state almost entirely.
Similarly as in plots (a) and (b), DKCS and SCS draw very close evolutions with SCS remaining slightly lower than DKCS.
The evolutions of the two states diverge once the classically induced noise case is considered with SCS's infidelity remaining an order of magnitude lower than DKCS's.
For the case of BGKCS, the state evolves with an infidelity of an order of magnitude lower than SCS and DKCS infidelities.
This is no longer true for the case of the classically induced noise in the system, where BGCS show to have the highest infidelity after the noise and error correction procedure are performed.

Plots (d) to (f) correspond to the cases when $\lambda = -1$.
The behavior of evolutions for this cases are similar as the $\lambda = 1$ counterpart scenario, with the exception that SCS and DKCS diminish in general their infidelity, making them closer to the consistently lower infidelity curve of BGKCS.
In general BGKCS curve remains almost identical as the $\lambda =1$ case, as well as its response to the error correction for the classically induced noise.
For the case of DKCS and SCS the recoverability of the state after the induced noise case remains the same as the $\lambda =1$ cases, however it substantially improves for the maximally mixed state input at plot (f).

Drawing a comparison of the previously discussed results of Fig.~\ref{fig:heatmaps_modified_noise} with Fig. \ref{fig:dynamics_modified_noise}, illustrates we observe that the evolution plots for all input states agree with what is shown in the averaged infidelities of the heatmaps.
In particular BGKCS remains consistently with higher resilience to the quantum noise, DKCS shows the lowest resilience and SCS remains in between but closer to DKCS.

\begin{widetext}
\section{Wigner distribution function}
By Considering the density matrix $\rho=\ket{\beta,f}\bra{\beta,f}$,  the Wigner characteristic function is obtained  as following  \cite{schleich2015quantum}:
\begin{eqnarray}
\chi(\eta)&=&Tr\big[e^{i(\eta a^{\dagger}-\eta^{\ast}a )}\rho\big]\nonumber\\
&=& \bra{\beta,f}D(\eta)\ket{\beta,f}\nonumber\\
&=&\sum_{m,n}^{\infty}\frac{\beta^{\ast  m}\beta^{n}}{\sqrt{m!n!}[f(m)]![f(n)]!}\bra{m}D(\eta)\ket{n}
\end{eqnarray}
in which for $m\geq n$,  
\begin{eqnarray}
\bra{m}D(\eta)\ket{n}=e^{-|\eta|^{2}/2}\sqrt{\frac{n!}{m!}}\eta^{m-n}\mathcal{L}_{n}^{m-n}(|\eta|^{2})
\end{eqnarray}
where $\mathcal{L}_{n}^{m-n}(|\eta|^{2})$ is an associated Laguerre polynomial.
The Wigner distribution is then given  as the Fourier transform of $\chi (\eta)$:
\begin{eqnarray}
W(\beta)&=&\frac{1}{\pi^{2}}\int d\eta^{2}  e^{\eta^{\ast} \beta - \eta \beta^{\ast} }\chi(\eta)\nonumber\\
&=& \sum_{m,n=0}^{\infty}W_{m,n}(\alpha) 
\end{eqnarray}
where
\begin{eqnarray}
W_{m,n}(\beta)=\frac{2(-1)^{n}}{\pi} \frac{2^{m-n} \beta^{\ast 2m} e^{-2|\beta|^{2}}}{m![f(n)]! [f(m)]!}  \mathcal{L}^{m-n}_{n} (4|\beta|^{2})\nonumber
\end{eqnarray}

\section{Useful calculations}\label{appendix_B}
\subsection{DSCSs with positive Kerr parameter}
By defining $z=e^{-i\varphi}\tanh\left[\sqrt{\frac{\lambda}{2}}|\alpha|\right]$, we have
\begin{eqnarray}
\sum_{n=0}^{\infty}\frac{\Gamma(2j+n)}{ \Gamma(2j)n!}|z|^{2n}=\left(1-|z|^{2}\right)^{-2j},
\end{eqnarray}
\begin{eqnarray}
_{D}\bra{\alpha,\lambda_{+}}\ket{-\alpha,\lambda^{
+}}_{D}=\left(\frac{1-|z|^{2}}{1+|z|^{2}}\right)^{2j}
\end{eqnarray}
\begin{eqnarray}
\hat{A}\ket{\alpha,\lambda_{+}}_{D}= \sqrt{\frac{\lambda}{2}} z\left[\left(1-|z|^{2}\right)^{j}\sum_{n=0}^{\infty}\sqrt{\frac{\Gamma(2j+n)}{\Gamma(2j)n!}}\, n z^{n}\ket{n}+2j\ket{\alpha,\lambda_{+}}_{D}\right]    
\end{eqnarray}
\begin{eqnarray}
_{D}\bra{\alpha,\lambda_{+}}\hat{A}\ket{\alpha,\lambda^{
+}}_{D}&=& \sqrt{2\lambda}\frac{j z}{1-|z|^{2}}
\end{eqnarray}
\begin{eqnarray}
\hat{A}\ket{-\alpha,\lambda_{+}}_{D}= \sqrt{\frac{\lambda}{2}} z\left[\left(1-|z|^{2}\right)^{-j}\sum_{n=0}^{\infty}\sqrt{\frac{\Gamma(2j+n)}{\Gamma(2j)n!}}\, n (-1)^{n+1}z^{n}\ket{n}+2j\ket{-\alpha,\lambda_{+}}_{D}\right]    
\end{eqnarray}

\begin{eqnarray}
_{D}\bra{\alpha,\lambda_{+}}\hat{A}\ket{-\alpha,\lambda^{
+}}_{D}&=&  \sqrt{2\lambda}\left[\frac{1-|z|^{2}}{1+|z|^{2}}\right]^{2j} \frac{j z}{1+|z|^{2}}
\end{eqnarray}

\begin{eqnarray}
\bra{\mathcal{C}^{\mp},\lambda_{+}}\hat{A}\ket{\mathcal{C}^{\pm},\lambda_{+}}&= \frac{\sqrt{2\lambda}}{\sqrt{1-\left[\frac{1-|z|^{2}}{1+|z|^{2}}\right]^{4j}}} \frac{z}{1-|z|^{2}}
\left[ 1 + \left[\frac{1-|z|^{2}}{1+|z|^{2}}\right]^{2j+1} \right]\\
&= j\sqrt{\frac{\lambda}{2}} \sinh\left[\sqrt{2\lambda}\, |\alpha|\right]\frac{1+\sech^{2j+1}\left[\sqrt{2\lambda}\, |\alpha|\right]}{1-\sech^{4j}\left[\sqrt{2\lambda}\, |\alpha|\right]}  
\end{eqnarray}
Note that for $j\gg 1$,
\begin{eqnarray}
    \frac{1+\sech^{2j+1}\left[\sqrt{2\lambda}\, |\alpha|\right]}{1-\sech^{4j}\left[\sqrt{2\lambda}\, |\alpha|\right]}\approx 1
\end{eqnarray}
Moreover, in the situation in which $j\ll 1$ and $\sqrt{\lambda} |\alpha| \ll 1 $, leads into
\begin{eqnarray}
    \frac{1+\sech^{2j+1}\left[\sqrt{2\lambda}\, |\alpha|\right]}{1-\sech^{4j}\left[\sqrt{2\lambda}\, |\alpha|\right]}\approx \frac{1+e^{-j\lambda |\alpha|^{2}}}{1-e^{-2j\lambda |\alpha|^{2}}} \rightarrow \frac{1}{2j\lambda |\alpha|^{2}}
\end{eqnarray}
Therefore, under these conditions, we have 
\begin{eqnarray}
\bra{\mathcal{C}^{-},\lambda_{+}}\hat{A}\ket{\mathcal{C}^{+},\lambda_{+}}\approx \frac{1}{|\alpha|^{2}}  
\end{eqnarray}

\begin{eqnarray}
_{D}\bra{\alpha,\lambda_{+}}\hat{A}^{\dagger}\hat{A}\ket{\alpha,\lambda^{
+}}_{D}&=&  \left[\frac{\lambda j |z|^{2}(2j+|z|^{2})}{(1-|z|^{2})^{2}}\right]=\lambda j\sinh^{2}\!\left(\sqrt{\frac{\lambda}{2}}\,|\alpha|\right)
\left[
2j\,\cosh^{2}\!\left(\sqrt{\frac{\lambda}{2}}\,|\alpha|\right)
+
\sinh^{2}\!\left(\sqrt{\frac{\lambda}{2}}\,|\alpha|\right)
\right]
\end{eqnarray}
\begin{eqnarray}
_{D}\bra{\alpha,\lambda_{+}}\hat{A}^{\dagger}\hat{A}\ket{-\alpha,\lambda^{
+}}_{D}
&=& -j\lambda \frac{|z|^{2}}{1+|z|^{2}} \left[\frac{1-|z|^{2}}{1+|z|^{2}}\right]^{2j}
\end{eqnarray}

\begin{eqnarray}
\bra{\mathcal{C}^{+},\lambda_{+}}\hat{A}^{\dagger}\hat{A}\ket{\mathcal{C}^{+},\lambda_{+}}= 2j\lambda \sinh^{2}\left[\sqrt{\frac{\lambda}{2}} |\alpha|\right] \frac{1-\sech^{2j+1}\left[\sqrt{2\lambda}\, |\alpha|\right]}{1+\sech^{2j}\left[\sqrt{2\lambda}\, |\alpha|\right]}
\end{eqnarray}

\begin{eqnarray}
\bra{\mathcal{C}^{-},\lambda_{+}}\hat{A}^{\dagger}\hat{A}\ket{\mathcal{C}^{-},\lambda_{+}}= 2j\lambda \sinh^{2}\left[\sqrt{\frac{\lambda}{2}} |\alpha|\right] \frac{1+\sech^{2j+1}\left[\sqrt{2\lambda}\, |\alpha|\right]}{1-\sech^{2j}\left[\sqrt{2\lambda}\, |\alpha|\right]}
\end{eqnarray}
\begin{eqnarray}
\hat{a}\ket{\alpha,\lambda_{+}}_{D}=  z\left[\left(1-|z|^{2}\right)^{j}\sum_{n=0}^{\infty}\sqrt{\frac{\Gamma(2j+n)} {\Gamma(2j)n!}} z^{n} \sqrt{2j+n} \ket{n}\right]    
\end{eqnarray}
\begin{eqnarray}\label{B_17}
_{D}\bra{\mathcal{C}^{\pm},\lambda_{+}}\hat{a}\ket{\mathcal{C}^{\mp},\lambda_{+}}_{D} &=&  |z|^{2}\left(1-|z|^{2}\right)^{2j}\sum_{n=0}^{\infty}\frac{\Gamma(2j+n)} {\Gamma(2j)n!} |z|^{2n} \sqrt{2j+n} \left[1\pm(-1)^{n}\right]\nonumber\\
&\geq &
|z|^{2}\left(1-|z|^{2}\right)^{2j}\sum_{n=0}^{\infty}\frac{\Gamma(2j+n)} {\Gamma(2j)n!} |z|^{2n} \sqrt{2j} \left[1+(-1)^{n}\right]=\sqrt{2j} |z|^{2} \left[1\pm\left(\frac{1-|z|^{2}}{1+|z|^{2}}\right)^{2j}\right]
\end{eqnarray}
Hence, if we assume $j\gg 1$, we have  
\begin{align}\label{B_18}
    _{D}\bra{\mathcal{C}^{\pm},\lambda_{+}}\hat{a}\ket{\mathcal{C}^{\mp},\lambda_{+}}_{D} \geq \sqrt{2j} \tanh^{2} \left[\sqrt{\frac{\lambda}{2}}|\alpha|\right]
\end{align}

\begin{eqnarray}
_{D}\bra{\alpha,\lambda_{+}}\hat{a}^{\dagger}\hat{a}\ket{\alpha,\lambda^{
+}}_{D}&=& 2j \frac{|z|^{2}}{1-|z|^{2}}
\end{eqnarray}
\begin{eqnarray}
_{D}\bra{\alpha,\lambda_{+}}\hat{a}^{\dagger}\hat{a}\ket{-\alpha,\lambda^{
+}}_{D}
&=& -2j \frac{|z|^{2}}{1+|z|^{2}} \left[\frac{1-|z|^{2}}{1+|z|^{2}}\right]^{2j}
\end{eqnarray}

\begin{align}
\bra{\alpha;j,\lambda_{+}} a^{2} \ket{\alpha;j,\lambda_{+}}
=
\frac{j(2j+1)}{2}e^{-2i\phi}\,
\sinh^{2}\left[\sqrt{2\lambda}\,|\alpha|\right]
\end{align}

\begin{eqnarray}
\bra{\mathcal{C}^{+},\lambda_{+}}\hat{a}^{\dagger}\hat{a}\ket{\mathcal{C}^{+},\lambda_{+}}= 4j \sinh^{2}\left[\sqrt{\frac{\lambda}{2}} |\alpha|\right] \frac{1-\sech^{2j+1}\left[\sqrt{2\lambda}\, |\alpha|\right]}{1+\sech^{2j}\left[\sqrt{2\lambda}\, |\alpha|\right]}
\end{eqnarray}

\begin{eqnarray}
\bra{\mathcal{C}^{-},\lambda_{+}}\hat{a}^{\dagger}\hat{a}\ket{\mathcal{C}^{-},\lambda_{+}}= 4j \sinh^{2}\left[\sqrt{\frac{\lambda}{2}} |\alpha|\right] \frac{1+\sech^{2j+1}\left[\sqrt{2\lambda}\, |\alpha|\right]}{1-\sech^{2j}\left[\sqrt{2\lambda}\, |\alpha|\right]}
\end{eqnarray}

\begin{eqnarray}
_{D}\bra{\alpha,\lambda_{+}}(\hat{A}^{\dagger}\hat{A})^{2}\ket{\alpha,\lambda^{
+}}_{D}&=& j\lambda^{2} \left[\frac{z}{1-z^{2}}\right]^{2} \Bigg[
jz^{6}+2\left(1+2j+3j^{2}\right)z^{4}\nonumber\\
&\times & \left[1+2j\left(3+2j+2j^{2}\right)\right]z^{2}
 +2j^{2}\Bigg]
\end{eqnarray}

\begin{eqnarray}
_{D}\bra{\alpha,\lambda_{+}}(\hat{A}^{\dagger}\hat{A})^{2}\ket{-\alpha,\lambda^{
+}}_{D}&=& j\lambda^{2} \left[\frac{z}{1-z^{2}}\right]^{2}\left[\frac{1-z^{2}}{1+z^{2}}\right]^{2j}
 \Bigg[
jz^{6}-2\left(1+2j+3j^{2}\right)z^{4}\nonumber\\
&\times & \left[1+2j\left(3+2j+2j^{2}\right)\right]z^{2}- 2j^{2}\Bigg]
\end{eqnarray}

\begin{eqnarray}
_{D}\bra{\alpha,\lambda_{+}}(\hat{A}^{\dagger}\hat{A})\hat{A}\ket{\alpha,\lambda^{
+}}_{D}&=& 4j (2j+1)\left[\frac{\lambda}{2}\right]^{3/2} \left[\frac{z}{1-z^{2}}\right]^{3}\left[z^{2}+j\right]
\end{eqnarray}

\begin{eqnarray}
_{D}\bra{\alpha,\lambda_{+}}(\hat{A}^{\dagger}\hat{A})\hat{A}\ket{-\alpha,\lambda^{
+}}_{D}&=& 4j (2j+1)\left[\frac{\lambda}{2}\right]^{3/2}\left[\frac{1-z^{2}}{1+z^{2}}\right]^{2j} \left[\frac{z}{1-z^{2}}\right]^{3}\left[z^{2}-j\right]
\end{eqnarray}

\begin{eqnarray}
_{D}\bra{\alpha,\lambda_{+}}(\hat{A}^{\dagger}\hat{A})^{2}\hat{A}\ket{\alpha,\lambda^{
+}}_{D}&=& 8j  (2j+1)\left[\frac{\lambda}{2}\right]^{5/2}\frac{z^{2}}{\left[1-z^{2}\right]^{5}}
\Bigg[(2j+1)z^{6}+(5j^{2}+6j+4)z^{4}\nonumber\\
&+& (2j^{3}+3j^{2}+5j+1)z^{2}+j^{2}
\Bigg]
\end{eqnarray}

\begin{eqnarray}
_{D}\bra{\alpha,\lambda_{+}}(\hat{A}^{\dagger}\hat{A})^{2}\hat{A}\ket{-\alpha,\lambda^{
+}}_{D}&=& 8j  (2j+1)\left[\frac{\lambda}{2}\right]^{5/2} \frac{z^{2}}{\left[1+z^{2}\right]^{5}} 
\left[\frac{1-z^{2}}{1+z^{2}}\right]^{2j}
\Bigg[(2j+1)z^{6}-(5j^{2}+6j+4)z^{4}\nonumber\\
&+& (2j^{3}+3j^{2}+5j+1)z^{2}-j^{2}
\Bigg]
\end{eqnarray}

\begin{eqnarray}
_{D}\bra{\alpha,\lambda_{+}}(\hat{A}^{\dagger}\hat{A})^{3}\ket{\alpha,\lambda^{
+}}_{D}&=& 8j   \left[\frac{\lambda}{2}\right]^{3} \frac{z^{2}}{\left[1+z^{2}\right]^{6}}
\Bigg[(j^{2}z^{10}
+(14j^{2}+18j+2)z^{8}
+ (24j^{4}+56j^{3}+90j^{2}+59j+15)z^{6}\nonumber\\
&+& (8j^{5}+24j^{4}+78j^{3}+92j^{2}+58j+12)z^{4}+ (12j^{4}+20j^{3}+24j^{2}+8j+1)z^{2}+2j^{3}
\Bigg]\nonumber
\end{eqnarray}

\begin{eqnarray}
_{D}\bra{\alpha,\lambda_{+}}(\hat{A}^{\dagger}\hat{A})^{3}\ket{-\alpha,\lambda^{
+}}_{D}&=& 8j   \left[\frac{\lambda}{2}\right]^{3} 
\frac{z^{2}}{\left[1+z^{2}\right]^{6}} 
\left[\frac{1-z^{2}}{1+z^{2}}\right]^{2j}\Bigg[(j^{2}z^{10}+(14j^{2}+18j+2)z^{8}\nonumber\\
&+& (24j^{4}+56j^{3}+90j^{2}+59j+15)z^{6}+ (8j^{5}+24j^{4}+78j^{3}+92j^{2}+58j+12)z^{4}\nonumber\\
&+& (12j^{4}+20j^{3}+24j^{2}+8j+1)z^{2}+2j^{3}
\Bigg]
\end{eqnarray}
\begin{eqnarray}
\hat{a}\ket{\alpha,\lambda^{
+}}_{D}&=& (1-z^{2})^{-2j} z e^{-i\varphi} \sum_{n=0}^{\infty} \sqrt{\frac{\Gamma(2j+n)}{\Gamma(2j)n!}} \sqrt{2j+n} \, e^{-i n\varphi} z^{n}\ket{n}     
\end{eqnarray}
\subsection{DSCSs with negative Kerr parameter}
By defining $z=e^{-i\varphi}\tan\left[\sqrt{\frac{|\lambda|}{2}} |\alpha|\right]$, we have 
\begin{eqnarray}
    \hat{A}\ket{\alpha,\lambda_{-}}=\sqrt{\frac{\lambda}{2}}\frac{z}{(1+|z|^{2})^{j}}\sum_{n=0}^{2j}\sqrt{\frac{(2j)!}{(2j-n)!n!}}\, (2j-n)z^{n}\ket{n}
\end{eqnarray}
\begin{eqnarray}
_{D}\bra{\alpha,\lambda_{-}}\ket{-\alpha,\lambda^{
-}}_{D}=\left(\frac{1-|z|^{2}}{1+|z|^{2}}\right)^{2j}=\cos^{2j}\left(\sqrt{2|\lambda|} |\alpha|\right)
\end{eqnarray}
\begin{eqnarray}
_{D}\bra{\alpha,\lambda_{-}}A\ket{\alpha,\lambda^{
-}}_{D}=\sqrt{\frac{|\lambda|}{2}}\frac{2jz}{1+|z|^{2}}
\end{eqnarray}
\begin{align}
_{D}\bra{\alpha,\lambda_{-}}\hat{a}^{2}\ket{-\alpha,\lambda^{
-}}_{D}=2j(2j-1)\frac{z^{2}}{\left(1+|z|^{2}\right)^{2}}=\frac{j(2j-1)}{2} e^{-2i\phi}
\sin^{2}\left[
\sqrt{2\lambda}\,|\alpha|
\right]
\end{align}

\begin{eqnarray}
_{D}\bra{\alpha,\lambda_{-}}A\ket{-\alpha,\lambda^{
-}}_{D}=\sqrt{\frac{|\lambda|}{2}}\frac{2jz}{1-|z|^{2}}\left[\frac{1-|z|^{2}}{1+|z|^{2}}\right]^{2j}
\end{eqnarray}
\begin{eqnarray}
\bra{\mathcal{C}^{\mp},\lambda_{-}}\hat{A}\ket{\mathcal{C}^{\pm},\lambda_{-}}&= \frac{\sqrt{2\lambda}}{\sqrt{1-\left[\frac{1-|z|^{2}}{1+|z|^{2}}\right]^{4j}}} \frac{z}{1+|z|^{2}}
\left[ 1 + \left[\frac{1-|z|^{2}}{1+|z|^{2}}\right]^{2j-1} \right]\\
&= j\sqrt{\frac{\lambda}{2}} \sin\left[\sqrt{2\lambda}\, |\alpha|\right]\frac{1+\cos^{2j+1}\left[\sqrt{2\lambda}\, |\alpha|\right]}{1-\cos^{4j}\left[\sqrt{2\lambda}\, |\alpha|\right]}  
\end{eqnarray}
Note that for $j\gg 1$,
\begin{eqnarray}
    \frac{1+\cos^{2j-1}\left[\sqrt{2\lambda}\, |\alpha|\right]}{1-\cos^{4j}\left[\sqrt{2\lambda}\, |\alpha|\right]}\approx 1
\end{eqnarray}
Moreover, in the situation in which $j\gg 1$ and $\sqrt{\lambda} |\alpha| \ll 1 $, we have
\begin{eqnarray}
\bra{\mathcal{C}^{\mp},\lambda_{-}}\hat{A}\ket{\mathcal{C}^{\pm},\lambda_{-}}\approx \frac{1}{|\alpha|^{2}}  
\end{eqnarray}
In addition by adjust $\sqrt{2\lambda}|\alpha|=s\pi/2$, with $s\in \mathbb{Z}$, we have 
\begin{eqnarray}
\bra{\mathcal{C}^{\mp},\lambda_{-}}\hat{A}\ket{\mathcal{C}^{\pm},\lambda_{-}}= j\sqrt{\frac{\lambda}{2}}
\end{eqnarray}
\begin{eqnarray}
_{D}\bra{\alpha,\lambda_{-}}A^{\dagger}A\ket{\alpha,\lambda^{
-}}_{D}=  \frac{|\lambda| j |z|^{2}(2j+|z|^{2})}{(1+|z|^{2})^{2}}= |\lambda|\, j\sin^{2}\!\left(\sqrt{\frac{\lambda}{2}}\,|\alpha|\right)
\left[
2j\,\cos^{2}\!\left(\sqrt{\frac{\lambda}{2}}\,|\alpha|\right)
+
\sin^{2}\!\left(\sqrt{\frac{\lambda}{2}}\,|\alpha|\right)
\right]
\end{eqnarray}
\begin{eqnarray}
_{D}\bra{\alpha,\lambda_{-}}A^{\dagger}A\ket{-\alpha,\lambda^{
-}}_{D}= \frac{|\lambda| j z^{2}(z^{2}-2j)}{(1-z^{2})^{2}}\left[\frac{z^{2}-1}{z^{2}+1}\right]^{2j}
\end{eqnarray}

\begin{eqnarray}
_{D}\bra{\alpha,\lambda_{-}}(A^{\dagger}A)A\ket{\alpha,\lambda^{
-}}_{D}=\left[\frac{|\lambda|}{2}\right]^{3/2}z^{3}
\end{eqnarray}

\begin{eqnarray}
_{D}\bra{\alpha,\lambda_{-}}\left(A^{\dagger}A\right)^{2}\ket{\alpha,\lambda^{
-}}_{D}=j\frac{|\lambda|^{2}}{2}z^{2}\frac{1+2j z^{2}}{1+z^{2}}
\end{eqnarray}

\begin{eqnarray}
_{D}\bra{\alpha,\lambda_{-}}(A^{\dagger}A)^{2}\ket{-\alpha,\lambda^{
-}}_{D}=j\frac{|\lambda|^{2}}{2}z^{2}\frac{1-2j z^{2}}{1-z^{2}}\left(\frac{1-z^{2}}{1+z^{2}}\right)^{2j}\nonumber\\
\end{eqnarray}

\begin{eqnarray}
_{D}\bra{\alpha,\lambda_{-}}\left(A^{\dagger}A\right)^{2}A\ket{\alpha,\lambda^{
-}}_{D}=2j\left[\frac{|\lambda|}{2}\right]^{5/2}z^{3}\frac{1+2j z^{2}}{1+z^{2}}\nonumber\\
\end{eqnarray}

\begin{eqnarray}
_{D}\bra{\alpha,\lambda_{-}}(A^{\dagger}A)^{2}A\ket{-\alpha,\lambda^{
-}}_{D}&=&-2j\left[\frac{|\lambda|}{2}\right]^{5/2}z^{3}\frac{2j z^{2}-1}{1-z^{2}} \left(\frac{1-z^{2}}{1+z^{2}}\right)^{2j}
\end{eqnarray}

\begin{eqnarray}
_{D}\bra{\alpha,\lambda_{-}}\left(A^{\dagger}A\right)^{3}\ket{\alpha,\lambda^{
-}}_{D}=z^{2}\frac{4 j b}{(z^{2}+1)^4}
\end{eqnarray}
where
\begin{eqnarray}
b=4 j^3 z^{4}+2 j^2 z^{2} \left(\eta^{4}-2 z^{2}+3\right)+j \left(6 z^{4}-4 z^{2}+1\right)-(z^{2}-2) z^{2}\nonumber
\end{eqnarray}

\begin{eqnarray}
_{D}\bra{\alpha,\lambda_{-}}\left(A^{\dagger}A\right)^{3}\ket{-\alpha,\lambda^{
-}}_{D}=\frac{-4 z^{2} j b^{\prime}}{(1-z^{2})^4}\left(\frac{1-z^{2}}{1+z^{2}}\right)^{2j}
\end{eqnarray}
where
\begin{eqnarray}
b^{\prime}=4 j^3 z^4 - z^{2} (2 + z^{2}) - 2 j^2 z^{2} (3 + 2 z^{2} + z^4) + 
 j (1 + 4 z^{2} + 6 z^4)\nonumber
\end{eqnarray}
\subsection{BGCSs with positive Kerr parameter}
\begin{align}
{}_{BG}\braket{\alpha;j,\lambda_{+}}{\alpha;j,\lambda_{+}}_{BG}={}_{0}F_{1}(,2j,\frac{2}{\lambda}|\alpha|^{2})
\end{align}
\begin{align}
{}_{BG}\braket{\alpha;j,\lambda_{+}}{-\alpha;j,\lambda_{+}}_{BG}={}_{0}F_{1}(,2j,-\frac{2}{\lambda}|\alpha|^{2})
\end{align}
\begin{align}
\hat{a}\ket{\alpha;j,\lambda_{+}}_{BG}=\frac{1}{\sqrt{{}_{0}F_{1}(,2j,\frac{2}{\lambda}|\alpha|^{2})}}\sqrt{\frac{2}{|\lambda|}}|\alpha|\sum_{n=0}^{\infty} \frac{1}{\sqrt{2j+n}}\sqrt{\frac{\Gamma(2j)}{\Gamma(2j+n) n!}} \left[\sqrt{\frac{2}{\lambda}}\alpha\right]^{n}\ket{n}
\end{align}
\begin{eqnarray}
_{BG}\bra{\mathcal{C}^{\mp},\lambda_{\pm}}\hat{a}\ket{\mathcal{C}^{\pm},\lambda_{\pm}}_{BG}=\frac{\sqrt{\frac{2}{|\lambda|}}|\alpha|}{\sqrt{1 - {}_{0}F_{1}^{2}(,2j,-|z|^{2})}} \sum_{n=0}^{\infty} \frac{1}{\sqrt{2j+n}}\frac{\Gamma(2j)}{\Gamma(2j+n) n!} \left[\frac{2}{\lambda}|\alpha|^{2}\right]^{n}
\left[ 1 + (-1)^{n}
\right]\nonumber\\
\end{eqnarray}

\begin{eqnarray}
\frac{1}{(2j)^{3/2}} \frac{\sqrt{\frac{2}{|\lambda|}}|\alpha|}{\sqrt{ 1 - {}_{0}F_{1}^{2}(,2j,-|z|^{2})}} \frac{{}_{0}F_{1}(,2j,|z|^{2})+ {}_{0}F_{1}(,2j,-|z|^{2})}{{}_{0}F_{1}(,2j,|z|^{2})}
&=& \frac{\sqrt{\frac{2}{|\lambda|}}|\alpha|}{\sqrt{1 - {}_{0}F_{1}^{2}(,2j,-|z|^{2})}} \frac{1}{{}_{0}F_{1}(,2j,|z|^{2})}\nonumber\\
&\times&  \sum_{n=0}^{\infty} \frac{1}{\sqrt{2j}}\frac{\Gamma(2j)}{\Gamma(2j+n) n!} \left[\frac{2}{\lambda}|\alpha|^{2}\right]^{n}
\left[ 1 + (-1)^{n}
\right]\nonumber\\
&\ge &_{BG}\bra{\mathcal{C}^{\pm},\lambda_{\pm}}\hat{a}\ket{\mathcal{C}^{\pm},\lambda_{\pm}}_{BG}
\end{eqnarray}
where 
\begin{align}
    \frac{1}{(2j)^{3/2}} \frac{\sqrt{\frac{2}{|\lambda|}}|\alpha|}{\sqrt{ 1- {}_{0}F_{1}^{2}(,2j,-|z|^{2})}} \frac{{}_{0}F_{1}(,2j,|z|^{2})+ {}_{0}F_{1}(,2j,-|z|^{2})}{{}_{0}F_{1}(,2j,|z|^{2})}\approx \frac{\sqrt{2}}{2 j} \left[ 1 - \frac{5 |\alpha|^2}{8 \lambda\, j} + \frac{5 |\alpha|^4}{8 \lambda^2\, j^2} \right]
\end{align}
\begin{align}
_{BG}\bra{\mathcal{C}^{\pm},\lambda_{+}}\hat{A}^{\dagger}\hat{A}\ket{\mathcal{C}^{\pm},\lambda_{+}}_{BG}= \frac{2}{j\lambda}\frac{1}{1\pm {}_{0}F_{1}(,2j,-|z|^{2})} \left[1\pm\frac{{}_{0}F_{1}(,2j,-|z|^{2})}{{}_{0}F_{1}(,2j,|z|^{2})}\right]   
\end{align}
which can be expanded up to $\mathcal{O}(j^{-2})$ as
\begin{align}
_{BG}\bra{\mathcal{C}^{+},\lambda_{+}}\hat{A}^{\dagger}\hat{A}\ket{\mathcal{C}^{+},\lambda_{+}}_{BG} &\approx \frac{2}{j\lambda} \left[1-\frac{|\alpha|^{2}}{2j\lambda}\right]\\
_{BG}\bra{\mathcal{C}^{-},\lambda_{+}}\hat{A}^{\dagger}\hat{A}\ket{\mathcal{C}^{-},\lambda_{+}}_{BG} &\approx \frac{2}{j\lambda} \left[2-\frac{|\alpha|^{2}}{j\lambda}\right]
\end{align}

\begin{align}
_{BG}\bra{\mathcal{C}^{\pm},\lambda_{+}}\hat{a}^{\dagger}\hat{a}\ket{\mathcal{C}^{\pm},\lambda_{+}}_{BG}= 
 \frac{|\alpha|^{2}}{j\lambda}\frac{{}_{0}F_{1}(,2j,|z|^{2})\pm {}_{0}F_{1}(,2j,|z|^{2})}{1\pm {}_{0}F_{1}(,2j,-|z|^{2})}    
\end{align}
which can be expanded for $j\gg 1$, as
\begin{align}
_{BG}\bra{\mathcal{C}^{+},\lambda_{+}}\hat{a}^{\dagger}\hat{a}\ket{\mathcal{C}^{+},\lambda_{+}}_{BG} &\approx \frac{|\alpha|^{2}}{j\lambda}(1+\frac{|\alpha|^{2}}{2j\lambda})\\
_{BG}\bra{\mathcal{C}^{-},\lambda_{+}}\hat{a}^{\dagger}\hat{a}\ket{\mathcal{C}^{-},\lambda_{+}}_{BG} &\approx \frac{|\alpha|^{2}}{j\lambda}(2+\frac{|\alpha|^{2}}{j\lambda})
\end{align}
and for $\lambda\ll 1$ and $\alpha\gg 1$,
\begin{align}
    _{BG}\bra{\mathcal{C}^{\pm},\lambda_{+}}\hat{a}^{\dagger}\hat{a}\ket{\mathcal{C}^{\pm},\lambda_{+}}_{BG} \approx \frac{|\alpha|^{2}}{j\lambda}. 
\end{align}
\begin{align}
{}_{BG}\!\bra{\alpha,\lambda_{+}}\hat{a}^{2}
\ket{\alpha,\lambda_{+}}_{BG}
&=
\frac{2\alpha^2}{\lambda}\,
\frac{\Gamma(2j)}{
{}_0F_1\!\left(;2j,\frac{2|\alpha|^{2}}{\lambda}\right)
}
\sum_{n=0}^{\infty}
\frac{
\left[\frac{2|\alpha|^{2}}{\lambda}\right]^{n}
}{
n!\,\Gamma(2j+n)\sqrt{(2j+n)(2j+n+1)}
}
\end{align}

\begin{align}\label{B63}
{}_{BG}\!\bra{\alpha,\lambda_{+}}\hat{a}^{2}
\ket{\alpha,\lambda_{+}}_{BG}
=
\frac{2\alpha^2}{\lambda\sqrt{(2j)(2j+1)}} 
+\frac{2\alpha^2|\alpha|^2}{j\sqrt{2j+1}\lambda^2}
\left[
\frac{1}{\sqrt{2j+2}}
-\frac{1}{\sqrt{2j}}
\right] 
+ \mathcal{O}\!\left(\frac{|\alpha|^6}{\lambda^4}\right).
\end{align}

\subsection{BGCSs with negative Kerr parameter}

\begin{align}
\hat{a}\ket{\alpha;j,\lambda_{-}}_{BG}=\sqrt{\frac{2}{|\lambda|}}|\alpha|\frac{1}{\sqrt{{}_{0}F_{1}(,-2j,-\frac{2}{\lambda}|\alpha|^{2})}}\sum_{n=0}^{2j} \frac{1}{\sqrt{2j-n}}\sqrt{\frac{(2j)!}{(2j-n)! n!}} \left[\sqrt{\frac{2}{\lambda}}\alpha\right]^{n}\ket{n}
\end{align}

\begin{align}
_{BG}\bra{\mathcal{C}^{\mp},\lambda_{-}}\hat{a}\ket{\mathcal{C}^{\pm},\lambda_{-}}_{BG}=\frac{\sqrt{\frac{2}{|\lambda|}}|\alpha|}\sum_{n=0}^{\infty} \frac{1}{\sqrt{2j+n}}\frac{\Gamma(2j)}{\Gamma(2j+n) n!} \left[\frac{2}{\lambda}|\alpha|^{2}\right]^{n}
\left[ 1 + (-1)^{n}
\right]\nonumber\\
\end{align}

\begin{eqnarray}
\frac{1}{2j}\frac{\sqrt{\frac{1}{j|\lambda|}}|\alpha|\left[{}_{0}F_{1}(,-2j+1,-|z|^{2})+{}_{0}F_{1}(,-2j+1,-|z|^{2})\right]}{\sqrt{1-{}_{0}F_{1}^{2}(,-2j,|z|^{2})}}&=& \frac{\sqrt{\frac{|\alpha|^{2}}{2j^{3}|\lambda|}}}{\sqrt{1-{}_{0}F_{1}^{2}(,-2j,|z|^{2})}} \sum_{n=0}^{\infty} \frac{(2j)!}{(2j-n)! n!} \left[\frac{2}{\lambda}|\alpha|^{2}\right]^{n}
\left[ 1 + (-1)^{n}
\right]\nonumber\\ 
&\geq &_{BG}\bra{\mathcal{C}^{-},\lambda_{\pm}}\hat{a}\ket{\mathcal{C}^{\pm},\lambda_{-}}_{BG}
\end{eqnarray}
where 
\begin{align}
\sqrt{\frac{|\alpha|^{2}}{j^{3}|\lambda|}}\frac{\left[{}_{0}F_{1}(,-2j+1,-|z|^{2})+{}_{0}F_{1}(,-2j+1,|z|^{2})\right]}{\sqrt{1-{}_{0}F_{1}^{2}(,-2j,|z|^{2})}}\approx \frac{\sqrt{2}}{j}
\left[
1+\frac{|\alpha|^2}{2j|\lambda|}
+\frac{|\alpha|^4}{24j^2|\lambda|^2}
-\frac{|\alpha|^6}{48j^3|\lambda|^3}
+O\!\left(j^{-4}\right)
\right]
\end{align}
\begin{align}
_{BG}\bra{\mathcal{C}^{\pm},\lambda_{-}}\hat{a}^{\dagger}\hat{a}\ket{\mathcal{C}^{\pm},\lambda_{-}}_{BG}= 
\frac{|\alpha|^{2}}{j\lambda}\frac{{}_{0}F_{1}(,-2j,|z|^{2})\pm {}_{0}F_{1}(,-2j,-|z|^{2})}{1\pm {}_{0}F_{1}(,-2j,-|z|^{2})}    
\end{align}
\begin{align}
{}_{BG}\!\bra{\alpha,\lambda_{-}}\hat{a}^{2}
\ket{\alpha,\lambda_{-}}_{BG}
&=
\frac{2\alpha^2}{\lambda}\,
\frac{(2j)!}{
{}_0F_1\!\left(;-2j,-\frac{2|\alpha|^{2}}{\lambda}\right)
}
\sum_{n=0}^{2j}
\frac{
\left[\frac{2|\alpha|^{2}}{\lambda}\right]^{n}
}{
n!\,(2j-n)!\sqrt{(2j-n)(2j-n+1)}
}
\end{align}
\begin{align}
{}_{BG}\!\bra{\alpha,\lambda_{-}}\hat a^{2}
\ket{\alpha,\lambda_{-}}_{BG}
=
\frac{2\alpha^2}{\lambda}
\frac{1}{\sqrt{(2j)(2j+1)}} +
\frac{\sqrt{2}|\alpha|^2 \alpha^{2}}{j^{3/2}\lambda^{2}}\left(
\frac{4j^{2}}{\sqrt{2j-1}}
-\frac{1}{\sqrt{2j+1}}
\right)
+ \mathcal{O}(\frac{|\alpha|^{6}}{\lambda^{4}}).
\end{align}

\end{widetext}
\end{document}